\documentclass[a4paper,14pt]{article}
\usepackage[utf8]{inputenc}
\usepackage{tabularx}
\usepackage{amssymb}
\usepackage{amsthm}
\usepackage{amsmath}
\usepackage{setspace}
\usepackage{enumitem}

\usepackage[pdftex,bookmarks,colorlinks]{hyperref}
\usepackage[pdftex]{graphicx}
\usepackage[section]{placeins}
\usepackage[utf8]{inputenc}
\usepackage{wrapfig}
\usepackage{epstopdf}
\usepackage[toc,page]{appendix}

\newtheorem{thm}{Theorem}

\newtheorem{defi}{Definition}

\newcommand{\guv}{g_{\mu\nu}}

\newcommand{\beq}{\begin{equation}}
\newcommand{\eeq}{\end{equation}}

\newcommand{\gvu}{g^{\nu\mu}}
\newcommand{\gab}{g^{\alpha\beta}}
\newcommand{\ogab}{{}^{(0)}g^{\alpha\beta}}
\newcommand{\sdg}{\sqrt{|g|}}

\newcommand{\delij}{\delta_{ij}}

\newcommand{\pari}{\partial_i}
\newcommand{\Tuv}{T_{\mu\nu}}
\newcommand{\Ruv}{R_{\mu\nu}}
\newcommand{\dphi}{\dot{\varphi}}
\newcommand{\Vp}{V(\varphi)}
\newcommand{\Vphi}{V_{,\varphi}}
\newcommand{\intkpi}{\int \frac{d^3 \mathbf{k}}{(2\pi)^{3/2}}}
\newcommand{\eikx}{e^{i\mathbf{k\cdot x}}}
\newcommand{\emikx}{e^{-i\mathbf{k\cdot x}}}
\newcommand{\apl}{\hat{a}_k^+}
\newcommand{\aplp}{\hat{a}_{k'}^+}
\newcommand{\ami}{\hat{a}_k^-}
\newcommand{\amip}{\hat{a}_{k'}^-}
\newcommand{\vac}{|0\rangle_t}
\newcommand{\cav}{\langle 0|_t}
\newcommand{\xip}{\xi^i_{\perp}}
\newcommand{\intd}{\int d^Dx\sqrt{|g|}}

\newcommand{\HL}{H_{\Lambda}}
\newcommand{\vacuum}{|0\rangle}
\newcommand{\Dmuo}{\left(\partial_{\mu}-\frac{i}{2}g_1B_{\mu} + W_{\mu}\right)}
\newcommand{\Dmut}{\left(\partial_{\mu}+\frac{i}{2}g_1B_{\mu} + W_{\mu}\right)}
\newcommand{\TR}{\mathrm{Tr}(\phi\phi^{\ast})}

\title{Conformal standard model and inflation.} 
\author{Jan Kwapisz, MISMaP UW }
\begin{document}
\maketitle
\author
\begin{abstract}
This thesis presents a possible inflation scenario as a consequence of non-minimal gravitational couplings in the Conformal Standard Model. That model consists, in comparison to the SM, of additional right-chiral neutrinos and a complex scalar sextet which, in distinction to the Higgs boson, is not coupled to the SM particles. The inflation is driven by two non-minimally coupled fields. One of those is the SU(2) Higgs-like doublet and the second is associated with the trace of the sextet. These two fields are related to the Higgs particle and ``shadow'', heavier Higgs via the mass mixing matrix. Since these type of models, with non-minimal coupling(s), can match the observational data related to inflation, they are of high interest. For this particular model, as it turned out, the tensor to scalar ratio and spectral tilt match the current data for a wide range of parameters. Also the unitary issue was adressed.
\end{abstract}
\setcounter{page}{1}
\tableofcontents
\renewcommand{\theequation}{\arabic{section}.\arabic{equation}}
\section{Preface}
Inflation started its career in the 80ties when it turned out that classical cosmology picture struggles with fine tuning of matter density i.e. very uniform distribution of matter at the time of last scattering in circa $10^{84}$ disconnected regions. The Universe must have started its history with very a special initial state. Starobinsky \cite{STAROBINSKY} and Guth \cite{Guthinflation} argued that the possible solution to this problem exists without fine-tuning the initial conditions. If one assumes that at the very early stage of expansion of the Universe there was an accelerated expansion era, which eventually went into the decelerated FLRW epoch, then the fine tuning problem is solved. This solution of the problem is called inflation but it introduced an issue of its own initial conditions what is under active investigation nowadays. \\
There were many mechanisms proposed and discussed to generate the $\ddot{a}>0$ epoch just after the Big Bang. In general they can be divided into two groups: one modifies the structure of gravity, while another proposes a particle physics mechanism ensuring inflation.  Following Starobisky (first) approach, accelerated expansion originates from the addition of $R^2$ term to the gravity lagrangian. However, original article of Starobinsky discussed quite a different issue, namely the initial singularity at the Big-Bang, and the $R^2$ term came from the anomaly rather than from the modification of gravity but it was later recognised as a successful inflation scenario. Other proposal is to extend general relativity to Einstein-Cartan theory, where the torsion tensor doesn't vanish. Then that tensor gives strong repulsive behaviour of matter at the very beginning of the Universe and one can derive the accelerated expansion within this theory \cite{Desai:2015haa,Poplawski:2010kb}. However this attempt is far from being complete, it might have some advantages. On the other hand accelerated expansion can be provided by scalar field(s) satisfying certain conditions what will be described in section \ref{Slow-roll}. If these fields are in addition non-minimally coupled to gravity then it is an equivalent mechanism to Starobinksy approach discussed in section \ref{Starobinskyinflation}. The recent WMAP3 and PLANCK measurements discarded all the other approaches except for Starobinsky, the equivalent ones and natural inflation. \\
Since the only observed (so far) fundamental scalar field in nature is the Higgs field it is then a natural candidate for inflaton. However, the scenario with minimally coupled Higgs driving inflation predicts much smaller self-coupling than it was measured in LHC so the non-minimally coupled scenario with Higgs particle as inflaton was investigated in \cite{Bezrukov}. It turned out that the current CMB measurements agree with predictions of this model. From the particle physics theory point of view Standard Model should be extended since it suffers from numerous problems like the hierarchy problem. Then, even though pure Higgs non-minimally coupled gives a successful inflation scenario, it should be anyway supplemented by high energy SM extension to give a complete description.\\
There were many extension of Standard Model proposed, most popular being supersymmetric ones, but there is also another direction, namely the Conformal Standard Model \cite{CSMtwo,Latosinski2015} and the latter is the subject of this thesis. This extension not only solves hierarchy problem, using SBCS mechanism \cite{CSMtwo} but also provides possible candidates for Dark Matter and gives natural explanation to the baryogenesis via resonant leptogenesis \cite{PILAFTSIS2004303}. Moreover, it preserves the general structure of SM and can be valid up to Planck scale. It possesses two Higgs-like particles mixed by the mass mixing matrix. They are natural candidates to be non-minimally coupled to gravity and to serve as inflatons. There arises a question whether those candidates can provide a successful scenario with parameters which don't contradict the ones coming from CSM. \\
 The purpose of this thesis is to investigate this model. The model is presented in chapter \ref{CSMinflation}. We show there that inflation driven by those particles can give a successful scenario. Moreover, as far as the author is informed, there was no previous analysis concerning non-minimal couplings to gravity in minimally extended SM where the inflation was discussed on a concrete extended Standard Model candidate -- most of the scenarios concerning multi-Higgs-like scalars don't involve a detailed extension of SM. Also, the natural way of solving unitary issues and broad range of parameters for which inflation can occur are advantages of this model. The mechanism of decoupling described in section \ref{taubehaviour} haven't been discussed that way either. It seems also that this kind of reasoning can be justified in presence of more inflatons with different mass scales. Therefore CSM supplemented by the inflation model, described below, can be a complete theory up to the Planck scale, however it obviously has to be tested experimentally.\\
The thesis is organised in the following way. In chapter \ref{Mainideas} the main ideas of cosmology and general relativity are presented. The purpose of this chapter, besides discussing those topics, is to adjust notation. Also the drawbacks of standard $\Lambda$-CDM cosmology are presented in this chapter. The inflation hypothesis is presented in chapter \ref{inflation}. Slow-roll conditions are discussed. The origin and spectrum of perturbations is addressed. The possible mechanism of causing inflation are described with emphasis on Starobinsky inflation, as the most relevant scenario for us. The following chapter \ref{TheSM} describes the Standard Model (SM) and its problems. Higgs mechanism and Yang-Mills theory are described. The discussion of Conformal Standard Model as a successful extension of SM can be found in chapter \ref{ConformalSM}. The chapter \ref{HiggsInflaton} discusses the possibility of driving inflation by non-minimally coupled Standard Model Higgs particle. Main results of the thesis are presented in the section \ref{CSMinflation}. Here the non-minimally multi inflaton scenario in CSM framework is discussed. The role of chapter \ref{Summary} is to summarize the thesis and outline possible extensions of the work done here. Also another possible mechanisms of causing inflation, which are already incorporated (or naturally could be) in CSM, are outlined. In appendices reader can find extensions of topics raised in the thesis but not related to the main parts, like $f(R)$ gravity equations or the discussion of Quantum Field Theory in curved spacetime. 
\subsection{Conventions}
\noindent
\textbf{Units}\\
We work in particle physics units, namely: $\hbar =c=1$. For discussion of gravity and in context of Inflation, we will use the following normalisation: $M_P^2=\frac{1}{8\pi G}=1/\kappa \equiv 1$, where $M_P^2$ is called reduced Planck mass.\\
\textbf{Three and four vectors}\\
Four vectors will be written in cursive style ${x}_{\mu}$, have greek indices like $\emph{x} = (t, \vec{x})$, and $\mu \in \{0,1,2,3\}$. Three vectors will be bolded \textbf{p} and spatial coordinates has latin indices, so $p_i$ is the $i$-th coordinate, $i \in \{1,2,3\}$.\\
\textbf{Signature}\\
Tensor $\eta_{\mu\nu}$ is defined as: $\eta_{\mu\nu} = diag(+1,-1,-1,-1)$. The same is for signature of $g_{\mu\nu}$. Then $\det{(g_{\mu\nu})} \geq 0$. However we will use $\sqrt{|g|}$ as square root of determinant, to fit both conventions for signature of $g_{\mu\nu}$.\\
\noindent\textbf{Scale factor} \\*
For scale factor in FLRW metric we will use letter $a$.
We introduce conformal time as: 
\beq
\label{Conformaltime}
\eta = \int \frac{dt}{a}.
\eeq 
Then differentiation with respect to physical time will be noted as $\dot{a}$, and with respect to conformal: $a'$.\\
\textbf{Metric}\\
For the spatial part of metric $h_{ij}$ symbol will be used, and $|h|$ as its determinant. Further, additional notation is discussed in detail in the following chapters when needed.
\section{Main ideas of Cosmology and General Relativity}
\label{Mainideas}
\setcounter{equation}{0}
This chapter is dedicated to presenting the main ideas of General Relativity and Cosmology. Here the most important concepts and equations of these theories are presented. The subject is obviously too broad to be presented in detail so we restrain only to fundamental ideas. Much more details can be found in specialistic books, for example \cite{Gravity,hawking1973large,Fieldtheory,Mukhanov}. Derivations of geodesic equation or Einstein equations from Einstein - Hilbert action, also for $f(R)$ gravity, are discussed in Appendix \ref{GRappendix}.
\subsection{Lorentzian Manifolds and Einstein equations}
In this paragraph we will show how to construct General Relativity as theory of Lorentzian manifolds satisfying certain postulates, following the famous Hawking Ellis book \cite{hawking1973large}. The level of precision, used in this book, exceeds much from what we need to describe inflation, where only Einstein equations are needed. However, the purpose is to give possibly wide background. Also the formalism used by Hawking-Ellis gives a very simple and straightforward derivation of GR as a theory with some well justified axioms, while the notation and definitions of tensors are given in Appendix \ref{GRappendix}. We start with some definitions:
\begin{defi}\textbf{Lorentzian manifold} is a pair $(\mathcal{M},g)$ where M is $4$-dimensional smooth Haussdorff manifold and g is a metric with Lorentz signature. We require that $g$ depends smoothly on points of manifold M.  
\end{defi}
\noindent Two models $(\mathcal{M},g)$  and $(\mathcal{M}',g')$ are called equivalent when there exists diffeomorphism $\theta: \mathcal{M} \to \mathcal{M'}$ such that: $ \theta_{\star} g = g'$. So we are considering a whole class of equivalence of manifolds and metrics as our physical spacetime.
\begin{defi} Lorentzian manifold is \textbf{time-orientable}, if there exists a continuous timelike vector field $\eta$ on M. Threesome $(M,g,\eta)$ is called \textbf{time-oriented} Lorentzian manifold / spacetime.
\end{defi}
\noindent For a given metric $\guv$ we calculate infinitesimal length squared as:
\beq
ds^2 = g_{\mu\nu}dx^{\mu}dx^{\nu}
\eeq
\begin{defi} \textbf{Path} is a smooth map between an interval and manifold.\\
 \mbox{$\mu: I \to \mathcal{M}$} and is called \textbf{regular}, if it has everywhere non-vanishing derivative. A \textbf{curve} is an image of equivalence class of paths due to reparametrisation.
\end{defi}
\noindent The curve $\gamma(t)$ is future/past/not oriented if: $\forall_t$ $\dot{\gamma}^{\mu}(t)$ is future/past timelike or null, ie.:
\begin{itemize}
\item \textbf{timelike}: $\guv u^{\mu}u^{\nu}>0$,
\item \textbf{zero (null)}: $\guv u^{\mu}u^{\nu}=0$,
\item \textbf{spacelike}:  $\guv u^{\mu}u^{\nu}<0$,
\end{itemize} 
where: 
\beq
\dot{\gamma}^{\alpha} = \left(\frac{dx^{\alpha}}{d\lambda}\right) = u^{\alpha}.
\eeq
\textbf{Postulate 0: Extremal path principle}\\*
Particles obey the extremal path principle, they choose the curve $\gamma$ such that its length is extremal:
\beq
\delta \int_{\gamma} ds =0,
\eeq
the equation following from this principle is called geodesic equation:
\beq
\frac{d^2x^{\mu}}{ds^2} + \Gamma^{\mu}_{\nu\rho}\frac{dx^{\rho}}{ds}\frac{dx^{\nu}}{ds} =0,
\eeq
and the derivation can be found in appendix \ref{Geodesic}. In our investigation we separate gravity and matter, where we call matter everything which has non-gravitational origin.\\
\textbf{Postulate 1: Local causality}\\*
\noindent
We postulate that, if $\mathcal{U}$ is a convex normal neighbourhood and $p,q \in \mathcal{U}$, then a light signal can be sent between $p$ and $q$ if and only if $p$ and $q$ can be joined by by a $C^1$ curve $\gamma$ lying entirely in $\mathcal{U}$ and $\gamma$ mustn't vanish along the path, moreover it mustn't be \textbf{spacelike}.
Equivalently let $\{x_i\}_{i=0}^3$ be normal coordinates around $p$ ($p$ as origin) in $\mathcal{U}$, then only points satisfying:
$$
(x_0)^2 - (x_1)^2 -(x_2)^2 - (x_3)^2 \geq 0,
$$ 
can comunicate and this subset is called lightcone for point $p$. The structure of the lightcone is presented on the figure below:
\FloatBarrier
\begin{figure}[h!]
\begin{center}
\includegraphics[scale=0.4]{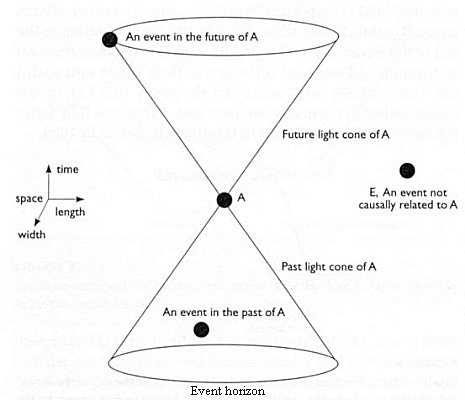}
\caption{Structure of a lightcone}
\label{fig:Light_cone}
\end{center}
\end{figure}
\FloatBarrier
\noindent
The Figure was taken from: \url{https://physics.stackexchange.com/questions/25460/can-we-see-all-of-the-observable-universe}.\\\\
Thus the boundary of the hyperspace is called null-cone $N_p$ and moreover when $N_p$ is known, then the metric $g$ is determined up to a conformal factor. The further discussion of global properties of spacetime can be found in appendices, which concerns Killing vectors and Cauchy surfaces.\\
\textbf{Postulate 2: Local energy and momentum conservation}\\*
The matter fields are characterised by symmetric energy - momentum tensor: $T_{\mu\nu}$, which depends on fields, derivatives of those and $g$. It has to satisfy following properties:
\begin{enumerate}[label=(\alph*)]
\item $T_{\mu\nu}$ vanishes on $\mathcal{U}$ if and only if all the matter fields vanish on $\mathcal{U}$,
\item The covariant derivative of $T_{\mu\nu}$ is zero - local conservation of energy and momentum:
$$
\nabla^{\mu} T_{\mu\nu} = 0.
$$
\end{enumerate}
Then conformal factor may be determined by the conservation of $T_{\mu\nu}$ equation.
One can argue that postulates (a) don't determine how to build $T_{\mu\nu}$, neither say whether energy-momentum tensor is unique. However, if we can derive equation of motion from a Lagrangian density ($\mathcal{L}$), then symmetric energy-momentum tensor in uniquely defined. Let us assume that Lagrangian depends on the field $\phi$ and its first derivatives. Then the action is the following:
\beq
S = \int d^4x \sqrt{|g|} \mathcal{L}(\phi, \partial_{\mu} \phi),
\eeq
If we require that the action be stationary then we obtain Euler-Lagrange equations:
\beq
\label{ELequations}
\frac{\delta S}{\delta \phi} = \partial_{\mu}\frac{\partial \mathcal{L}}{\partial\left(\partial_{\mu} \phi\right)} - \frac{\partial \mathcal{L}}{\partial \phi}=0.
\eeq
We also obtain symmetric $T_{\mu\nu}$ using the expression:
\beq
\label{Tuv}
\frac{\delta S}{\delta g_{\mu\nu}} = \int d^4x T^{\mu\nu} \delta g_{\mu\nu}.
\eeq
The $T_{\mu\nu}$ can contain any kind of matter eg. scalar field, electromagnetic field or any other types of ``matter''.\\
\textbf{Postulate 3: Field Equations}\\*
 So far we haven't said anything about $g$ metric relation to $T_{\mu\nu}$. In Special Theory of Relativity we take $\eta_{\mu\nu} = diag(+1,-1,-1,-1)$. Since the light rays are deflected in the presence of huge masses we cannot introduce gravity as additional field keeping metric conformally flat which was Nordstr\"om's original idea. Then we are left with one option, we need to introduce curvature of spacetime. Now the problem arises how to relate metric to $T_{\mu\nu}$. Curvature should only depend on the distribution of energy and momenta rather than the type of matter. One can either discuss this problem as in \cite{hawking1973large} using physical arguments, or simply write down Einstein-Hilbert action:
\beq
S_{EH}= \int d^4x \sqrt{|g|}\left(R + \Lambda + \mathcal{L}_{m}\right),
\eeq 
where $\Lambda$ is dark energy term, which we will drop in the whole text (except for discussion of de-Sitter solution) and $\mathcal{L}_m$ is the matter part lagrangian density. By variation for $g_{\mu\nu}$ of the action we obtain Einstein field equations using (\ref{Tuv}):
\beq
\label{Einsteinequations}
R_{\mu\nu} + \frac{1}{2}Rg_{\mu\nu} + \Lambda g_{\mu\nu} = T_{\mu\nu},
\eeq
where $R$ is Ricci scalar and $R_{\mu\nu}$ is Ricci tensor, the details and derivation are presented in the appendix \ref{GRappendix}. We formulate the following postulate:
$$
 g_{\mu\nu} \textbf{    } \mathbf{satisfies}\textbf{ }  \mathbf{Einstein} \textbf{ } \mathbf{equations}.
$$ 
\noindent The left and right hand side of equations are completely divergenceless, which means that  energy is conserved for gravity in absence of matter and for non-zero $T_{\mu\nu}$. There arises a question, whether EH action is unique and whether it is the only well defined action for gravity. The answer is given by \emph{Lovelock theorem} and is discussed in section \ref{EHaction}.  Now we will describe solutions, which are fundamental from the view of cosmology, namely the ones which satisfiy Cosmological Principle.
\subsection{Friedmann-Lema{\^i}tre-Robertson-Walker cosmology}
Cosmology is a part of physics which applies fundamental theories, mainly General Relativity and various theories of particles, to understand how the Universe was born, how it developed and how will it end. All the cosmological models follow from one fundamental assumption called the Copernicus Principle or Cosmological Principle, which says: \emph{The Universe is homogeneous and isotropic in large scales}. Homogeneity means that the Universe has the same matter distribution everywhere. Isotropic property says that Universe looks the same in every direction. Cosmic microwave background shows that these assumptions are satisfied in general up to very small deviations. The origin of such homogenous Universe and generation of tiny anisotropies is still under investigation. The main theory explaining it is inflation which will be described in chapter \ref{inflation}. Most of the Universe energy density (around $70\%$) is now in the form of cosmological constant. Cosmological constant is probably equivalent to dark energy in a form of perfect fluid with negative pressure spreading across the Universe. The rest thirty percent is matter but again almost $80\%$ of it is in a form of so called dark matter which doesn't (or almost doesn't) react with electromagnetic field. We actually don't know what this dark matter or dark energy actually is. These are questions among ones of the biggest puzzles of cosmology nowadays. Still on large distances we can approximate all of them as fluids with certain relation between pressure and energy which is homogeneous and isotropic. So we can analyse their gravitational behaviour (for ordinary matter, dark matter or cosmological constant) with the advantage of the Cosmological Principle and apply it to the metric -- then instead of 10 equations we get two. To satisfy this principle we have to assume that our metric is of the form:
\beq
\label{FLRWmet}
ds^2 = \left[ dt^2 - a^2(t)h_{ij} dx^idx^j \right], 
\eeq
and in isotropic, homogenous models the only choices of non-vanishing components of $h_{ij}$ (in spherical frame with $dR$, $d\theta$, $d\varphi$ as basis) are: 
\beq
\label{hij}
\begin{array}{lcr}
h_{11} = \frac{1}{1-kR^2},&  h_{22} = R^2, & h_{33} = R^2\sin^2\theta,
\end{array}
\eeq 
where $k = \pm 1$, $0$ is curvature of space related to $h_{ij}$ and $|h|=\det (h_{ij})$. If we change the variables using: $d\chi^2 = \frac{dR^2}{1 - kR^2}$, we can easily write down the equations for the 3d spatial metric, expressing them using $k$:
\beq 
ds_{3d}^2= a^2(d\chi^2 + \Phi^2(\chi)d\Omega^2) \equiv a^2 \left[ d \chi^2 + \left(
\begin{array}{c}
\sinh^2 \chi \\
\chi^2 \\
\sin^2 \chi \\
\end{array} \right)
d\Omega^2 \right] 
\begin{array}{l} 
k = -1; \\
k =  0; \\
k = +1, \\
\end{array}
\label{metryka}
\eeq
where: $d\Omega^2 = (d\theta^2 + \sin^2\theta d\varphi^2)$. However, the curvature of Universe is almost negligible.  The reason why is it so is a puzzling issue, it is one of the problems of initial conditions in cosmology solved by inflation. The energy-momentum tensor $\Tuv$ takes perfect fluid form characterised by: energy density $\varepsilon$, pressure $p$ and 4-velocity $u_{\mu}$, where relation $p(\varepsilon)$ needs to be specified in a concrete model:
\beq
\label{TuvFLRW}
T^{\mu}_{\nu} = \left( \varepsilon + p\right) u^{\mu}u_{\nu} - p\delta^{\mu}_{\nu},
\eeq
for (\ref{FLRWmet}) the only non-vanishing parts of metric are:
\beq
\begin{array}{lccr}
g_{00}=1,& g_{ij} = -a^2(t)h_{ij},& g^{00} = 1,& g^{ij} = -\frac{1}{a^2(t)}h^{ij}.
\end{array}
\eeq
Then the Christoffel symbols of second kind are (\ref{FLRWmet}, \ref{hij}):
\begin{align}
\Gamma^{i}_{0j} =& \frac{1}{2}g^{ik}\frac{\partial g_{jk}}{\partial t} = \frac{\dot{a}}{a}\delta_{ij},\\
\Gamma^{0}_{ij} =& a\dot{a}h_{ij}.
\end{align}
Ricci tensor has following components:
\begin{align}
\label{Riccizero}
R_{00} =& -3\frac{\ddot{a}}{a},\\
R_{0i} =& 0, \\
R_{ij} =& (\ddot{a}a +2\dot{a}^2 + 2k)h_{ij},
\end{align}
and Ricci scalar is:
\beq
R =g^{\mu\nu}R_{\mu\nu} = R_{00} - \frac{1}{a^2}\delta^{ij}R_{ij} = -6\left(\frac{\ddot{a}}{a} + \frac{\dot{a}^2}{a^2} +\frac{k}{a^2}\right).
\eeq
If we rewrite the Einstein equations in another form:
\beq
R_{\nu}^{\mu} = \kappa T^{\mu}_{\nu} - \frac{1}{2}T\delta_{\nu}^{\mu},
\eeq
where $T = \varepsilon -3p$, finally we get two Friedmann equations:
\begin{align}
\label{IFriedeq}
- 3\frac{\ddot a}{a} &= 8\pi G \cdot \frac{1}{2}(\varepsilon+ 3p), \\
\label{IIFried}
- \frac{2{\dot a}^2 + a\ddot a + 2k}{a^2} &= -8\pi G \cdot \frac{1}{2}(p - \varepsilon).
\end{align}
In FLRW metric we can introduce Hubble parameter as: 
\beq 
\label{Hubble}
H = \frac{\dot{a}}{a}.
\eeq
Then Ricci scalar can be rewritten in the form:
\beq
\label{RicciHubble}
R = -6\left(\dot{H}+ 2H^2+ \frac{k}{a^2} \right).
\eeq
and for $k=0$ the time derivative reads:
\beq
\label{RicciHubbledot}
\dot{R} = -6\left(4H\dot{H} +\ddot{H}\right).
\eeq
Hubble parameter can be constant or can vary in time and Hubble law connects velocity with space interval between the observer and particle:
\beq
v_i = H r_i.
\eeq
Current Hubble parameter value is about \cite{Mukhanov} 65 - 80 km s$^{-1}$ Mpc$^{-1}$. Then the frequency of a photon will drop while the Universe expands:
$$
\omega(t) = \frac{a_0}{a(t)}\omega_0,
$$
so we have the following relation: 
\beq 
T \sim 1/a,
\eeq
for isotropic and relatively cold Universe. This is why the Universe is becoming colder and the spectral lines are redshifted. If we plug (\ref{IFriedeq}) into (\ref{IIFried}) and use the notion of Hubble parameter, we obtain more familiar form of Friedman equations:
\begin{align}
\label{IFried}
\frac{\ddot{a}}{a} =\dot{H} + H^2 =  -\frac{8\pi G}{6}\left(\varepsilon + 3p\right),\\
\label{IIFriedeq}
H^2  = \frac{8\pi G}{3}\varepsilon - \frac{k}{a^2}.
\end{align}
 We can also use (\ref{TuvFLRW}) to obtain energy - momentum conservation equation:
\beq
\frac{d\varepsilon}{dt} + 3H(\varepsilon +p) = 0,
\eeq
so if equation of state is $p \sim \varepsilon$, with a constant $w$:
\beq
p = w \varepsilon,
\eeq
we get from energy conservation \cite{Senatore}:
\beq
\varepsilon \varpropto a^{-3(1+w)},
\eeq
and
\beq
\label{atw}
a(t) \varpropto \left\{
\begin{array}{cc}
t^{\frac{2}{3(1+w)}} & w \neq -1,\\
e^{Ht} & w=-1.
\end{array}
\right.
\eeq
Moreover, if we define for each component fractional quantities: $\varepsilon_i$ and $p_i$, then $\varepsilon$ and $p$ are sums over fractions. We would like to define critical density $\varepsilon_{cr}$ as the density of a fraction required to obtain Hubble parameter $H$ without any other contributions from other fractions. Then we define relative scale-dependent density as:
\beq
\Omega_{i}(a) = \frac{\varepsilon^i}{\varepsilon_{cr}},
\eeq
and measure of relative curvature contribution as:
\beq
\label{omegak}
\Omega_{k}(a) = - \frac{k}{a^2H^2(a)},
\eeq
at present time ($t_0$) we set $a(t_0) =a_0 =1$. We know from astronomical data that: 
\beq
\label{critden}
\sum_i \Omega_{i}(a_0) + \Omega_{k}(a_0) =1,
\eeq
then the Friedmann equations becomes:
\beq
\Omega_k(a) = 1 - \sum_i \Omega_i(a).
\eeq
When strong energy condition: 
\beq
\label{strongenergy}
\varepsilon + 3p > 0; 
\eeq
is satisfied we observe, from the Friedmann equations, that gravity is attractive. This is not the case for de-Sitter space where the cosmological constant term causes accelerated expansion. The paradigm where the Universe evolution is described by FLRW metric with Friedmann equations coupled to matter and dark energy is called $\Lambda$-CDM cosmology, where $\Lambda$ comes from dark-energy and CDM states for cold dark matter. Matter is divided into radiation ($v_i \sim c$) and cold (normal) matter for which $v_i \ll c$. Notice that for matter and radiation: $\varepsilon_{matter} \varpropto a^{-3}, \varepsilon_{radiation} \varpropto a^{-4}$. There are many pieces of the evolution of the Universe which are not understood with the $\Lambda$CDM model alone. For example, one can ask about the fundamental fact that the critical density equation (\ref{critden}) is almost perfectly satisfied. This problem is solved in inflation scenario which predicts accelerated expansion at the very beginning of the Universe. \\


\subsection{Propagation of light and De-Sitter spacetime}
\label{Conformaltransformations}
\label{de-Sitter}
The light geodesic \cite{Mukhanov} can be described by the condition that the spacetime interval is equal to zero:
\beq
ds^2 =0
\eeq
To get symmetry between space and time we can introduce conformal time (\ref{Conformaltime}): 
\beq
\eta = \int \frac{dt}{a},
\eeq
then the FLRW becomes:
\beq
\label{ConformalFLRW}
ds^2 = a^2(\eta)\left(d\eta^2 -d\chi^2 - \Phi^2(\chi)(d\theta^2 + \sin^2\theta d\varphi^2)\right),
\eeq
with (\ref{metryka}):
\beq
\Phi^2(\chi)=\left(
\begin{array}{c}
\sinh^2 \chi \\
\chi^2 \\
\sin^2 \chi \\
\end{array} \right)
\begin{array}{l} 
k = -1; \\
k =  0; \\
k = +1, \\
\end{array}
\eeq
One can observe that: $\theta, \varphi =$ const is a geodesic. Then we obtain
\beq
ds^2 = +d\eta^2 - d\chi^2 =0.
\eeq
Hence radial, light geodesics are straight lines described by
\beq
\chi(\eta) = \pm \eta + \textrm{const}
\eeq
Now we provide some useful definitions of horizons.
\begin{defi}\textbf{Particle horizon} The lightcone structure in general relativity divides, for each point, spacetime into three parts: future, past and elsewhere (\ref{development}, \ref{Lightcone}). The set $V^-(p)$, so the boundary of of domain from which point $p$ can receive information is so called \textbf{particle horizon}. The maximum comoving distance the light can propagate is:
\beq
\chi_p(\eta) = \eta - \eta_i = \int_{t_i}^t \frac{dt}{a}.
\eeq
\end{defi}
\begin{defi}
The complement of the particle horizon is so called \textbf{event horizon}. The event horizon is the closure of the set of points which can sent signal \emph{at time $\eta$} such that they will be never received by an observer in the \emph{future} i.e.: 
\beq
\chi > \chi_e(\eta) = \int_{\eta}^{\eta_{max}}d\eta = \eta_{max} - \eta, 
\eeq
where ``max'' is related to the final moment of time eg. $t_{max} = \infty$. 
\end{defi}
\noindent Physical quantities are related to the conformal ones via $a(t)$ factor: 
\begin{align}
d_e(t) = a(t)\int_t^{t_{max}} \frac{dt}{a}, \\
d_p(t) = a(t)\int^t_{t_i} \frac{dt}{a}.
\end{align}
Then the causal structure of the Universe can be described by two dimensional diagram, where each point is a 2-sphere. Those diagrams are called \emph{Conformal/Penrose diagrams} \cite{Mukhanov, DiagPenrose}. Moreover, in principle, for a metric with spherical symmetry \cite{Mukhanov}, there is a coordinate transformation, such that:
\beq
\begin{array}{lcr}
 \widetilde{g}_{00} = \widetilde{g}_{11} = a^2(\eta, \chi),& & \widetilde{g}_{01} = 0.
 \end{array}
\eeq
A priori the transformation can be hard to find. However, in cosmological cases, the metric is already in the desired form. We will illustrate these quantities on one of the most important solutions, namely de-Sitter spacetime.
\subsubsection{De-Sitter spacetime}
De-Sitter spacetime is solution for positive cosmological constant and in absence of matter and moreover we assume that solution in is in FLRW metric form. It is one of most famous solutions of the Einstein equations. As we have said cosmological constant is equivalent to a ``perfect fluid'' with equation of state $p_{\Lambda} = - \varepsilon_{\Lambda}$. Then from first Friedmann equation (\ref{IFriedeq}) we get:
\beq
\ddot{a} - H_{\Lambda}^2 a =0,
\eeq
where: $H_{\Lambda} = (8\pi G \varepsilon_{\Lambda}/3)^{1/2}$.
General solution of this equation is in the form:
\beq
a = C_1 \exp(\HL t) +C_2 \exp(-\HL t),
\eeq
and the constants of integration are constrained by the second Friedmann equation:
\beq
4\HL^2 C_1C_2 = k,
\eeq
where $k \in \{-1,0,1\}$ is the curvature. The solution can be summarised as:
\beq
ds^2= dt^2 -\left(\begin{array}{c}
\sinh^2(\HL t) \\ 
\exp(2\HL t) \\
\cosh^2(\HL t)
\end{array}\right)
\left[ d \chi^2 + \left(
\begin{array}{c}
\sinh^2 \chi \\
\chi^2 \\
\sin^2 \chi \\
\end{array} \right)
d\Omega^2 \right] 
\begin{array}{l} 
k = -1; \\
k =  0; \\
k = +1. \\
\end{array}
\eeq
The solutions exists of all values of $\varepsilon_{\Lambda}$ so they all describe the same spacetime in different coordinate systems. For flat De-Sitter spacetime horizons reads as:
\begin{align}
d_p(t) &= \exp(\HL t) \int_{t{_i}}^t \exp(-\HL s) ds = \HL^{-1}(\exp(\HL(t-t_i))-1) \\
d_e(t)  &= \exp(\HL t) \int_t^{\infty} \exp(-\HL s) ds = \HL^{-1}
\end{align}
This means that not all parts of Universe are causally connected. This will be crucial for 
De-Sitter spacetime which can be viewed as the 0-th order approximation to the inflation scenario. In De-Sitter, where $\varepsilon_{\Lambda} = \textrm{const}$ the energy density is conserved, so does Hubble constant. For inflationary models the energy density varies during the inflation stage and the evolution of the Friedmann Universe depends on its behaviour during the accelerated expansion. On the other hand it seems that our Universe is dominated by dark energy nowadays, so on large scales De-Sitter solution approximately describes our Universe now.

\subsection{Perturbations in General Relativity}
\label{Perturbations}
\label{Gravperturbations}
Einstein equations can be exactly solved in a very limited class of problems. Small inhomogeneities in the distribution of matter cause perturbations which can be analysed in the linear approximation. They can be also classified by gauge(coordinate)-invariant variables. In the following paragraph we analyse main results for gauge independent perturbations in the case of hydrodynamical $T_{\mu\nu}$; much more details can be found in \cite{Mukhanov}.
Flat Friedmann metric with linear perturbation can be written as:
\beq
ds^2 = \left[ {}^{(0)}g_{\alpha\beta} + \delta g_{\alpha\beta}(x^{\gamma})\right]dx^{\alpha}dx^{\beta},
\eeq
where $|\delta \gab| \ll |\ogab|$, and $\ogab$ is FLRW metric in conformal coordinates (\ref{ConformalFLRW}). For simplicity we take the curvature equal to zero.
General form of $\delta g_{\alpha\beta}$ can be divided into three cases.
\begin{itemize} 
\item The $\delta g_{00}$ component:
\beq
\label{scalarper}
\delta g_{00} = 2a^2 \phi,
\eeq
where $\phi$ is scalar function.
\item  The $\delta g_{0i}$ component:
\beq
\label{vectorper}
\delta g_{0i} = a^2 \left(B_{,i} + S_i\right),
\eeq
where $B_{,i} = \pari B$ derivative of some scalar function and vector part $S_i$ is a divergenceless. Each vector can be written in such a way by the virtue of Helmholtz theorem \cite{pracaAmsterdam}.
\item  And the $\delta g_{ij}$ component:
\beq
\label{tensorper}
\delta g_{ij} = a^2\left(2 \psi \delta_{ij} + 2E_{ij} + F_{i^{'}j} + F_{j^{'}i} + h_{ij}\right), 
\eeq
 is decomposed into irreducible components \cite{Mukhanov}, where:
 $\psi$, $E$ are scalar functions, $F_i$ is divergenceless vector, and $h_{ij}$ is tensor satisfying: 
\beq
h_i^i =0, \textbf{   } h^{i}_{j,i}=0.
\eeq
\end{itemize}
So in general the perturbations can be divided into three types depending on their transformation laws (bearing in mind that differentiation rises the covariant /contravariant tensor rank by one):
\begin{itemize}
\item \textbf{Scalar perturbations} are described by $\phi, \psi, B, E$. Inhomogeneities of energy density entails scalar perturbations.
\item \textbf{Vector perturbations} are related to the fluid rotational motions, and are characterised by $S_i$ and $F_i$.
\item  The origin of \textbf{tensor perturbations} $h_{ij}$ are gravitational waves, which are explicitly degrees of freedom of the Einstein theory.
\end{itemize} 
Since all types of perturbations propagate separately, we can study them apart of each other. Now we will consider a coordinate transformation, to see how the inhomogeneities will transform:
\beq
x^{\alpha} \to \tilde{x}^{\alpha} = x^{\alpha} + \xi^{\alpha},
\eeq
where $\xi^{\alpha}$ are infinitesimal translations in space and time. Since we can again divide metric into background and perturbation, we obtain following transformation laws (up to linear term in $\xi$):
\begin{align}
\ogab(x^{\sigma}) &\approx \ogab(\tilde{x}^{\sigma}) - \ogab_{,\gamma}\xi^{\gamma} \\
\delta \gab &\to \delta \tilde{\gab} -\gab_{,\gamma}\xi^{\gamma} - g_{\gamma\beta}\xi^{\gamma}_{,\alpha} - g_{\alpha\delta}\xi^{\delta}_{,\beta}.
\end{align}
We can write $\xi^{\alpha} = (\xi^{0}, \xi^i)$ and the spatial part can be divided as \cite{pracaAmsterdam, Mukhanov}:
\beq
\xi^i = \xi^i_{\perp} + \zeta^{'i},
\eeq
 where $\zeta$ is a scalar function and $\xip$ is divergentless vector, plugging into (\ref{ConformalFLRW}) we obtain the following:
 \begin{align}
 \label{deltaguv}
 \delta \tilde{g}_{00} &= \delta g_{00} - 2a\left(a\xi^0\right)^{'}, \nonumber \\
 \delta \tilde{g}_{0i} &= \delta g_{0i} + a^2\left[\xi_{\perp i}^{'} + \left(\zeta^{'} - \xi^0\right)_{,i}\right], \\
 \delta \tilde{g}_{ij} &= \delta g_{\ij} + a^2\left[2\frac{a'}{a}\delta_{ij}\xi^{0} + 2\zeta_{,ij} + \left(\xi_{\perp i,j} + \xi_{\perp j,i}\right)\right]. \nonumber
 \end{align}
 \noindent
Now we would like to build gauge invariant functions, to provide a coordinate free framework. We again treat each type of perturbations separately.\\
\textbf{Scalar perturbations} \\
If we compare  scalar functions and their proper derivatives from (\ref{deltaguv}) with scalar originated perturbations from (\ref{scalarper}, \ref{vectorper}, \ref{tensorper}) we obtain such coordinate transformations:
\beq
\begin{array}{lcr} 
\label{scalartransformation}
\phi \to \tilde{\phi} = \phi - \frac{1}{a}\left(a\xi^0\right)', & B \to \tilde{B} = B +\zeta^{'} - \xi^0, \\
&\\
\psi \to \tilde{\psi} = \psi + \frac{a^{'}}{a} \xi^0, & E \to \tilde{E} = E + \zeta.
\end{array}
\eeq
As we see only $\xi^0$ and $\zeta$ contribute to transformation of scalar perturbations. Then we choose them in such a manner to get rid (set them to zero) of two functions from: $E,B,\psi,\phi$. The most common choice is \emph{Newtonian (longitudinal) gauge} where $B_l=E_l=0$, then $\psi_l = \phi_l= \Phi=\Psi$ and there remain one variable describing the scalar perturbations. After fixing the gauge there is no coordinate freedom since any change of frame spoils this gauge conditions. On the other hand \emph{synchronous gauge} is the gauge where $\delta g_{0\alpha} =0$, then $\phi_l = B_l =0$, however this assumption doesn't fix the coordinates \cite{Mukhanov} and there exists a whole class of coordinates systems which share that property.\\
The simplest gauge invariant functions made from those four are:
\beq
\begin{array}{lcr}
\Phi = \phi - \frac{1}{a}\left(a\left(B - E^{'}\right)\right)', & &\Psi = \psi - \frac{a'}{a}\left(B-E'\right).
\end{array}
\eeq
\textbf{Vector and tensor perturbations}\\
For vector perturbations we obtain a metric:
\beq
\label{Vector perturbations}
ds^2 =a^2\left[d\eta^2 +2S_idx^i d\eta - \left(\delta_{ij} - F_{i,j} - F_{j,i}\right)dx^idx^j\right]
\eeq
and $S_i$ and $F_i$ have the following transformations:
\beq
\begin{array}{lcr}
S_i \to \tilde{S_i} = S_i + \xi^{'}_{\perp i},& & F_i \to \tilde{F_i} = F_i + \xi_{\perp i}.
\end{array}
\eeq
So the gauge invariant vector is
\beq
\overline{V}_i = S_i - F_i^{'}
\eeq
\emph{Tensor perturbation} $h_{ij}$ (\ref{tensorper}), \cite{Mukhanov} is already gauge-invariant tensor describing gravitational waves.\\\\
\noindent
\textbf{Energy momentum tensor and scalar perturbations}\\
If we  derive transformations for $\delta T_{\beta}^{\alpha}$ and $G_{\beta}^{\alpha}$ on FLRW background using the fact that ${}^{(0)}T_j^i \varpropto
\delta_{ij}$ and from it construct the gauge invariant quantities, in the same manner as we do for metric. Then we obtain a coordinate invariant expression \cite{Mukhanov}:
\beq
\label{PerGaugeInv}
\overline{\delta G}_{\beta}^{\alpha} = 8\pi G \overline{\delta T}_{\beta}^{\alpha},
\eeq
with following energy density and 4 velocities $(u^{\alpha})$ variations \cite{Mukhanov}:
\begin{align}
\label{varepspert}
 \overline{\delta\varepsilon} = \delta \varepsilon - \varepsilon_0^{'}\left(B - E^{'}\right) \\
\overline{\delta u_{\alpha}} = \delta u_{\alpha} \left[ a\left(B - E^{'}\right)\right]_{\alpha}
\end{align}
\\
\noindent From now on we consider the longitudinal gauge, ie $E_l,B_l=0$. For pure scalar perturbations we have:
 \beq
 ds^2 = a^2 \left[ (1 + 2\psi) d\eta^2 - \delta_{ij}\left(1-2\phi\right)dx^idx^j\right].
 \eeq
If we calculate the perturbed Einstein tensor $\delta G_{\mu\nu}$ for those perturbations it will allow us to write general coordinate independent equations for $\Psi$ and $\Phi$, with the write hand side (\ref{PerGaugeInv}) $\overline{\delta T}_{\mu\nu}$. Since we work in Newtonian gauge we have: $\Psi = \Phi$ \cite{Eingorn, Mukhanov}. We get:
\begin{align}
\label{Scalarperturbations}
\Delta \Psi - 3H\left(\Psi' + H \Psi\right) &= 4\pi G a^2 \overline{\delta T}_0^0, \\
\left(\Psi' + H\Psi\right)_{,i} &= 4\pi G a^2  \overline{T}_i^0, \\
\left[ \Psi'' +3H\Psi' + 2\left(H' +H^2\right)\Psi \right]\delta_{ij} &= -4\pi G a^2 \overline{\delta T}_j^i.
\end{align}
This can be extended also to a vector and tensor perturbation analysis. Here we focus only on hydrodynamical perturbations as is the case for inflation where energy-momentum tensor (\ref{TuvFLRW}) has the form:
\beq
\label{tuvhydro}
T_{\nu}^{\mu} = (\varepsilon + p)u^{\mu}u_{\nu} - p \delta_{\nu}^{\mu},
\eeq
where $u_{\nu}$ are four-velocities. We can calculate gauge-invariant perturbations as
\beq
\label{perturbationtuv}
\begin{array}{lcccr}
\overline{\delta T}_0^0 = \overline{\delta \varepsilon}, & & \overline{\delta T}_i^0 = \frac{1}{a}\left(\varepsilon_0 + p_0\right) \left(\overline{\delta u}_{\parallel i} + \delta u_{\perp i}\right), & & \overline{\delta T}_j^i = - \overline{\delta p} \delta_j^i,
\end{array}
\eeq
where $ \overline{\delta u}_{\mu}, \overline{\delta \varepsilon}$ are defined as in \cite{Mukhanov}, (\ref{varepspert}).
\noindent Then gauge invariant perturbations can be expressed as 
\begin{align}
\Delta \Psi - 3H\left(\Psi' + H \Psi\right) &= 4\pi G a^2 \overline{\delta \varepsilon}, \\
\left(\Psi' +H\Psi\right)_{,i} &= 4\pi G a^2 \left(\varepsilon_0 + p_0\right)  \overline{\delta u}_{\parallel i}, \label{Toi} \\
\left[ \Psi'' +3H\Psi' + 2\left(H' +H^2\right)\Psi \right] &= 4\pi G a^2 \overline{\delta p}.
\end{align}
For non-expanding Universe Hubble parameter is equal to $0$ then the first equation reproduces the known formula for Newtonian potential, namely Poisson equation. Second equation shows that $(a\Psi)^{'}$ behaves as velocity potential. Pressure fluctuations can be expressed using equation of state for $p(\varepsilon, S)$, where $S$ stands for entropy:
\beq
\overline{\delta p} =  c_s^2\overline{\delta \varepsilon} + \tau \delta S,
\eeq
with $c_s^2 =(\partial p/\partial \varepsilon)_S$ and $\tau = (\partial p / \partial S)_{\varepsilon} $. Then combining the first and third equations we obtain a closed form equation:
\beq
\label{SPerturbations}
\Psi'' + 3\left(1 +c^2_s\right)H\Psi' - c_s^2\Delta \Psi + \left(2H' + \left(1 + 3c_s^2\right)H^2\right) \Psi = 4\pi G a^2 \tau \delta S
\eeq
One can consider many cases like \cite{Mukhanov}:
\begin{itemize}
\item non-relativistic matter with $p=0$,
\item relativistic matter with $p = \omega \varepsilon$,
\item general form of adiabatic perturbations $\delta S=0$ for general $p(\varepsilon)$,
\item entropy perturbations.
\end{itemize}
\noindent
The notion of inhomogeneities and perturbations is crucial for the understanding of Planck and WMAP data. Their presence and properties can be explained in inflation paradigm rather than in the standard $\Lambda$-CDM cosmology.
\subsection{Why $\Lambda$ - CDM model is incomplete?}
\label{Incomplete}
We would like to know the evolution of the Universe from the very beginning. Of course we cannot start our scenario from initial singularity or even from Planck scale because of effects of quantum gravity that we do not know. Let us start the description just below Planck scale, so as initial data we will take some temperature below QG scale $T_i$, lets say $T_i = 10^{17}$ GeV with some initial $t_i$ and $a_i$ \cite{Guthinflation}. $\Lambda$-CDM model is one most popular and broadly accepted model in cosmology as it explains many things like CMB anisotropies, creation of Galaxies, accelerating expansion and other features. However, one has to address fine tuning of initial conditions problems, namely \cite{Guthinflation,Senatore} and \cite{Mukhanov}:
\begin{enumerate}
\item homogenity/horizon problem,
\item initial velocities/flatness problem, 
\item initial perturbation problem.
\end{enumerate}
The first two problems will be described below and perturbation problem will be presented in section \ref{Inhomogeneouslimit}. \\
\textbf{Homogeinity/horizon problem}\\
As we have mentioned, currently our Universe is homogenous on large scales where the domain scale is comparable to the present horizon scale, $l_0 = ct_0 \sim 10^{26}$ m according to \cite{Mukhanov}. The sizes of domains change with time according to the scale factor, $a_i/a_0$. Then from the Planck time $t_i \sim 10^{-43}$, the Universe  growth is roughly:
$$
l_i \sim l_0 \frac{a_i}{a_0}
$$
It is natural to  compare it to a size of a causal region $l_c \sim ct_i$: 
$$
\frac{l_i}{l_c} \sim \frac{t_0}{t_i}\frac{a_i}{a_0},
$$
if we use the relation $a = 1/T$
and use:
$$
a_i/a_0 \sim T_0/T_{P} \sim 10^{-32},
$$
we  obtain a rough estimate:
\beq
\label{flatness}
\frac{l_i}{l_c}\sim \frac{\dot{a}_i}{\dot{a}_0} \sim 10^{28}
\eeq
Then on Planck scale there should be around: $(10^{28})^3 = 10^{84}$ causally disconnected regions, with energy variation: $\frac{\delta \varepsilon}{\varepsilon} \sim 10^{-4}$ \cite{Mukhanov}. It is an extreme fine tuned value. Moreover, if we approximate: $a/t \sim \dot{a}$ we will get:
\beq
\frac{l_i}{l_c} \sim \frac{\dot{a}_i}{\dot{a}_0}, 
\eeq
therefore we see that homogeneity scale in $\Lambda$-CDM has always been larger than causality scale. The fact that the Universe remains homogenous despite being causally disconnected is called the \emph{horizon} or \emph{homogeinity} problem. However, initial velocities problem is even more puzzling. Let us assume for a moment that matter is distributed in an observed way, bearing in mind previous paragraph, how about initial velocities? \\
\textbf{Initial velocities/flatness problem}\\
Initial velocities should obey Hubble law (\ref{Hubble}), otherwise homogeneity will be no longer a realistic assumption. One can show using very rough arguments that matter must be distributed very precisely.
Since energy is conserved ($E^k$ is kinetic energy and $E^p$ is potential energy):
$$
E^{tot} = E^k_i + E^p_i = E^k_0 + E^p_0.
$$
We know that:
$$
E^k_i = E^k_0\left(\dot{a}_i/\dot{a}_0\right)^2,
$$
so we get:
\beq
\frac{E^{tot}_i}{E^k_i} = \frac{E^k_i + E^p_i}{E^k_i} = \frac{E^k_0 + E^p_0}{E^k_0}\left(\dot{a}_i/\dot{a}_0\right)^2.
\eeq
Assuming that \cite{Mukhanov} $E^k_0 \sim |E^p_0|$ and $\frac{\dot{a_0}}{\dot{a_i}} \le 10^{-28}$, we find:
\beq
\label{initialvelocities}
\frac{E^{tot}_i}{E^k_i} \le 10^{-56}.
\eeq
So the error in the initial velocities bigger than $10^{-54}\%$ gives completely different cosmology picture than we live in today. One can rephrase the problem also in terms of relative curvature and densities \cite{Senatore}. We recall the equation for curvature (\ref{omegak}) and assuming that the matter with equation of state connected to $w$ dominates the expansion with $a$ given by (\ref{atw}) we can write:
\beq
\begin{array}{lcr}
\dot{\Omega}_k = H\Omega_k(1+3w), & & \frac{\partial \Omega_k}{\partial \log a} = \Omega_k(1+ 3w),
\end{array}
\eeq
for $w> -1/3$ the $\Omega_k=0$ is an unstable fixed point. The current value of $\Omega_k$ is less than $10^{-2}$. The curvature redshifts as $a^{-2}$, much slower than ordinary matter or radiation. Taking to account the fact that it is almost negligible nowadays it should be even smaller at the Planck scale. So since $\Omega_k =0$ is unstable in decelerating expansion the Universe should either start from $k=0$, which is a very fine tuned value or, conversely, there could be a special moment of accelerated expansion when the curvature drops to a very small value and this process is called inflation. 
\section{Inflation}
\setcounter{equation}{0}
\label{inflation}
\subsection{General View}
In the paragraph \ref{Incomplete} we have described the problems of $\Lambda$-CDM cosmology. In this chapter we will inspect how the inflation theory explains problems mentioned there. Roughly speaking inflation is a stage of accelerated expansion of the Universe at its beginning, when gravity was repulsive. In inflation cosmology model there is a moment when acceleration solution turns into decelerated Friedmann solution and that moment is called the graceful exit. We will show that if we start with small causally connected spacetime which grows rapidly, it will give us, after the stage of inflation, causally disconnected very homogeneous Universe we observe today. But what if Universe was highly non homogeneous before the inflation stage, can this scenario make it homogeneous and isotropic? It turns out the answer is yes. One has to stress that inflation has itself many problems and there are many puzzles which haven't been solved and understood. Among them there is a problem of the Planck initial state \cite{Dimopoulos:2016yep}. Another, mentioned in the preface, problem is of the fundamental mechanism which generates inflation. Currently there are two: one is ``natural'' inflation, associated to shift symmetry and pseudo-Goldstone bosons and another one is Starobinsky modified gravity scenario and its descendants. The third problem is the reheating process which is responsible for production of the particles we observe today. However, quoting Mukhanov: \emph{The theory of reheating is far from complete. Not only the details, but even the over- all picture of inflaton decay depend crucially on the underlying particle physics theory beyond the Standard Model.} Despite those, and unmentioned, problems inflation is most successful in providing the explanation of $\Lambda$-CDM problems. In this thesis the second one is fully addressed in chapter \ref{CSMinflation}. The first problem might be solved by proto-inflation era and decoupling of heavy scalar degrees of freedom and can be investigated for CSM inflation it is not clear whether this is the right solution to the issue mentioned in \cite{Dimopoulos:2016yep}. The third one will be not addressed in this thesis, besides paragraph \ref{examplem2}.\\
So in this chapter we will give a general picture of inflation scenario, regardless of the mechanism driving it. If we analyse (\ref{IFriedeq}) for the $k=0$ case:
\beq
\ddot{a} = - \frac{4\pi}{3}G(\varepsilon +3p)a,
\eeq
we see that $\ddot{a}>0$ occurs if and only if $(\varepsilon +3p)<0$, i.e. when the strong energy condition (\ref{strongenergy}) is violated. We know de-Sitter solution doesn't have a natural time, when the acceleration stops. That is why De-Sitter can serve only as in general 0-th approximation for inflation scenario. If we look at the left hand side of the equation, we see that we have to allow Hubble parameter to vary in time:
\beq
\frac{\ddot{a}}{a} = H^2 + \dot{H},
\eeq
and $\dot{H}<0$. The graceful exit takes place roughly when $\dot{H} \sim H^2$. So if we assume that $H^2$ changes faster than the $\dot{H}$ we can roughly estimate the time for duration of inflation as:
\beq
t_f \sim H_i/|\dot{H}_i|,
\eeq
where $i$ refers to the beginning of inflation. Moreover, CMB observations require that $\dot{a}_i/\dot{a}_0 <10^{-5}$:
\beq
\frac{\dot{a}_i}{\dot{a}_f}\frac{\dot{a}_f}{\dot{a}_0} = \frac{a_i}{a_0}\frac{H_i}{H_f}\frac{\dot{a}_f}{\dot{a}_0} < 10^{-5},
\eeq
since $\dot{a}_f/\dot{a}_0$ should be larger than $10^{28}$, we obtain that inflation is successful if:
\beq
\frac{a_f}{a_i} > 10^{33} \frac{H_i}{H_f},
\eeq
so we roughly estimate:
\beq
a_f/a_i \sim \exp(H_it_f) >10^{33},
\eeq
Hence only if $t_f > 75H_i^{-1}$, so the inflation last longer than 75 e-folds then it can solve the horizon problem. However, this is a very rough estimate, most of the papers suggests that 50-60 e-folds \cite{Remmen:2014mia} is enough to solve the problem. It obviously depends on the value of the initial scalar field/fields which can cause the inflation scenario. \\
\subsection{Slow roll regime}
\label{Slow-roll}
Natural candidate for a source term in FLRW equations which can give us the desired graceful-exit is a scalar field. It could be Higgs field or fermionic condensate acting like ``effective'' scalar field or more exotic proposition. The section \ref{Menu} is an overview of inflation mechanism models. The particle giving rise to the inflation is called the Inflaton. In this paragraph we stay on classical / semi-classical level of description. Namely, we will assume homogeneous distribution of the fields which is the case for inflation at the classical level. Classical homogenous scalar field is characterised by the action (we neglect spatial derivatives term):
\beq
\mathcal{S} =\int_{\Omega} d^4x \sdg \left(\frac{1}{2}\dot{\varphi}^2 - V(\varphi)\right),
\eeq
For the scalar field we have 1-dim Klein-Gordon equation in FLRW background:
\beq
\label{KGin}
\ddot{\varphi} + 3H\dphi + \Vphi = 0. 
\eeq
Additionally we have Friedman equation as a constraint:
\beq
\label{inflationII}
H^2  = \frac{1}{3} \left(\frac{1}{2}\dphi^2 + \Vp\right),
\eeq
 where we use $M_P^2 =1$. So if we introduce energy density: 
\beq
\label{inflenergy}
\varepsilon = \frac{1}{2}\dphi^2 + \Vp,
\eeq
and pressure:
\beq
\label{inflpressure}
p = \frac{1}{2}\dphi^2 - \Vp,
\eeq
then the equation of state is:
\beq
w = \frac{p}{\varepsilon} = \frac{\frac{1}{2}\dot{\varphi}^2 - V(\phi)}{\frac{1}{2}\dot{\varphi}^2 + \Vp},
\eeq
For $\dphi^2 \ll \Vp$, we get $\varepsilon \approx -p$ and $w \approx -1< -1/3$, which is the desired value for accelerated expansion. Therefore we can give the first inflation condition: \\
\textbf{Slow roll condition 1:} $\dot{\varphi}^2 \ll V(\phi)$, we obtain:\\
\beq
\label{Hslow}
H = \frac{d\ln a}{dt} = \frac{d\ln a}{d \varphi}\dot{\varphi} \backsimeq \sqrt{\frac{1}{3}\Vp},
\eeq
and a related parameter:
\beq
\epsilon = -\frac{\dot{H}}{H^2} \simeq \frac{1}{2}\left(\frac{\Vphi}{V}\right)^2 \ll 1.
\eeq
Since Klein-Gordon equation has an attractor solution for large friction term: $\dot{\varphi}\simeq \Vphi/(3H)$, which is the case for Inflation, then being on this trajectory gives exactly:\\
\noindent \textbf{Slow roll condition 2:} $\ddot{\varphi} \ll 3H\dot{\varphi}$. From this condition we get:
\beq
\label{KGslow}
3H\dot{\varphi} + \Vphi \simeq 0,
\eeq
Then the second slow-roll parameter is
\beq
\eta = \frac{\ddot{\varphi}}{H\dot{\varphi}} \ll 1,
\eeq
so for both of them the conditions can be replaced as a condition for the potential:
\begin{align}
\label{Inflconditions}
\epsilon &\simeq \frac{M_P^2}{2} \left(\frac{\Vphi}{V}\right)^2, \\
\eta  & \simeq M_P^2 \frac{V_{,\varphi\varphi}}{V} \ll 1,
\end{align}
and we have reintroduced $M_P^2$ for later purposes. The inflation ends when  approximately: $\epsilon \simeq 1$. However, there is also another suitable choice of slow roll parameters ($\epsilon_H, \eta_H$), using the Hubble constant as the measure of inflation \cite{Liddle:1994dx}, then we have precisely that $\ddot{a} >0$ equivalent to $\epsilon_H <1$.\\
Because the second time derivative term in KG equation is suppressed by the Hubble term we called this part of evolution slow-roll. As we can see slow-roll regime is sufficient for inflation to take place, doesn't matter what drives it. So here arises a question.\\
\noindent
\textbf{How long does Inflation last?}\\*
The measure for time of accelerated expansion is number of e-folds, one e-fold is the era, when the Universe grow by $e$. In slow-roll regime we can calculate the number of e-folds (duration of inflation) by its definition:
\beq
dN = Hdt,
\eeq
so the number of e-folds is given by:
\beq
\label{Numberefolds}
N = \int_{t_i}^{t_{end}} Hdt = \int_{\varphi_i}^{\varphi_{end}} \frac{H}{\dot{\varphi}} d\varphi \simeq \int^{\varphi_i}_{\varphi_{end}} \frac{V(\varphi)}{V_{,\varphi}(\varphi)}d\varphi,
\eeq
where in the last equality we took advantage of slow-roll conditions.
So the equation for the scale factor can be solved in the integral form:
\beq
a(\varphi) \simeq a_i \exp\left(8\pi \int_{\varphi}^{\varphi_i} \frac{V}{\Vphi} d\varphi \right)
\eeq
For example the power-law potential $V= (1/n)\lambda \varphi^n$, which satisfies both conditions for $|\varphi| \gg 1$, we obtain the following scale factor:
\beq
\label{power-lawscale}
a(\varphi(t)) = a_i \exp\left(\frac{4\pi}{n}\left(\varphi_i^2 - \varphi^2(t)\right)\right).
\eeq


\subsection{Case study: $V= \frac{1}{2}m^2\varphi^2$}
\label{examplem2}
\begin{figure}
\begin{center}
\includegraphics[scale=0.6]{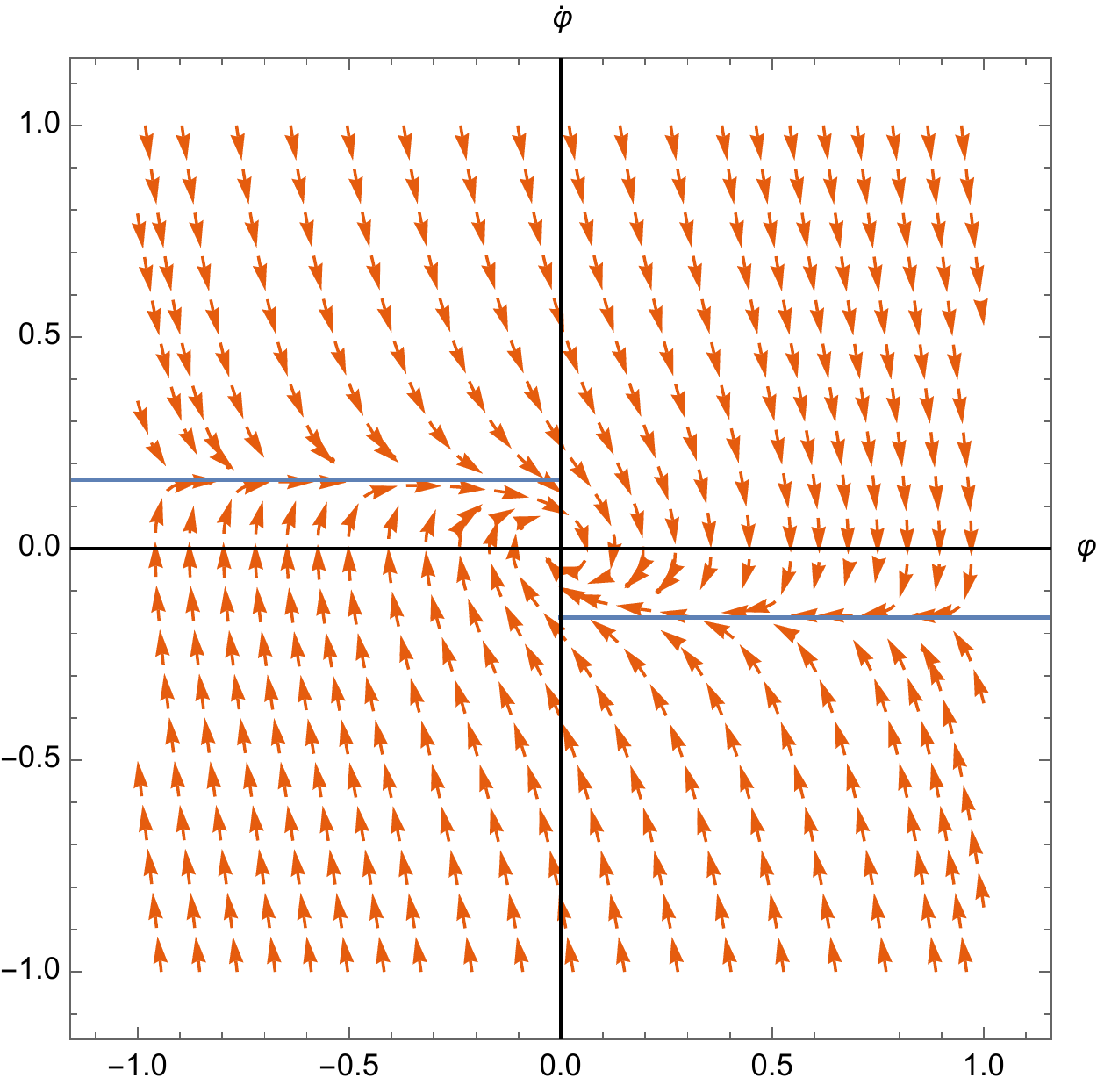}
\caption{Attractor for $m^2\varphi^2$ potential}
\label{fig:m2phi2potential}
\end{center}
\end{figure}

As an example we will analyze one of the simplest models: a massive scalar field without interactions. From (\ref{IIFriedeq}) and (\ref{KGin}) we get one equation:
\beq
\ddot{\varphi} + \sqrt{12\pi}\left(\dphi^2 + m^2\varphi^2\right)^{1/2} \dphi + m^2\varphi = 0.
\eeq
If we use relation:
$$
\ddot{\varphi} = \dphi \frac{d\dphi}{d\varphi},
$$
we obtain: 
\beq
\label{Attractorm2}
\dphi \frac{d\dphi}{d\varphi} = - \sqrt{12\pi}\left(\dphi^2 + m^2\varphi^2\right)^{1/2} \dphi + m^2\varphi.
\eeq
We want to analyse it by drawing the phase diagram and point out the attractors. If we change second-order ODE to a system of two first-order ODE's, using substitution $\dot{\varphi}=\varphi_1$, we get:
$$
\left\{\begin{array}{ccl}
\dot{\varphi} &=& \varphi_1,  \\
\dot{\varphi}_1 &=&  -\sqrt{12\pi} \left[\varphi_1^2 + m^2\varphi^2\right]^{1/2}  \varphi_1 - m^2\varphi.
\end{array} \right.
$$
We will use Lapunov theorem to prove stability of point $(0,0)$, but first we would like to introduce Lapunov function.
\begin{defi}{\textbf{Lapunov function}} of vector field $V(x)$, ie system of ODE defined as: $\dot{\mathbf{x}} =V(\mathbf{x})$ on domain $\Omega$, for point $x_{e}$, where $V(x_{e})=0$ is satisfied, is a scalar function:
$$
\mathcal{L}: \Omega \to \mathbb{R},
$$
and there exists $U$, which is a non-empty neighbourhood of $x$, where $\forall_{x \in U}$ $\mathcal{L}(x)$ satisfies:
\begin{enumerate}
\item $ \mathcal{L}(x) > 0$, 
\item $\mathcal{L}(x)=0 \Leftrightarrow x=x_e$,
\item $\dot{\mathcal{L}}(x)= \langle \nabla(\mathcal{L}(x)),V(x)\rangle < 0$.
\end{enumerate}
Lapunov theorem says that, when a system posses a Lapunov function for point $x_e$, this point is asymptotically stable. It means that the neighbouring trajectories approach this point.
\end{defi}
\noindent It is easy to show that for our system Lapunov function, around $(0,0)$, is simply: $\mathcal{L}(x) = \frac{1}{2}m^2 \varphi^2 + \frac{1}{2}\varphi_1^2 =  \frac{1}{2}m^2 \varphi^2 + \frac{1}{2}\dot{\varphi}^2$. Hence the point $(0,0)$ is stable and the trajectories of equation (\ref{Attractorm2}) converge to this point. We can see it on the plot below Figure \ref{fig:m2phi2potential}. Moreover, the divergence of our system is strictly negative for $\mathbb{R}^2 \backslash (0,0)$:
$$
\nabla \cdot V = 0 - \sqrt{\frac{12\pi}{\varphi_1^2+m^2\varphi^2}}\left(2\varphi_1^2 +m^2\varphi^2\right) < 0,
$$
hence eventually all the trajectories drops to $(0,0)$.
\FloatBarrier
\FloatBarrier
\noindent
The picture was made in  Mathematica programme (with a very simple code), for this concrete diagram I took $m=1$. The blue lines represents $\dot{\varphi}=\pm\frac{m}{\sqrt{12\pi}}$. On the phase diagram we can see attractors to which all solutions converge. Depending on a region we have two options either $|\dot{\varphi}|\gg m|\varphi|$ - \emph{ultra hard equation of state} $p \approx \varepsilon$ \cite{Mukhanov}. Or we have inflationary conditions, namely (\ref{Inflconditions}). 
Now we can use slow roll approximation to analyze attractor solution, so (\ref{Hslow}) is exactly $H = \pm \sqrt{\frac{8\pi G}{6}} m \varphi(t)$, from this we get exact solution for $\varphi$ \cite{pracaAmsterdam, Mukhanov}:
\beq
\label{phim2}
\varphi \simeq \varphi_0 - \frac{m}{\sqrt{12\pi}}(t-t_i) \simeq \frac{m}{\sqrt{12\pi}}(t_f -t),
\eeq
and from (\ref{IIFriedeq}):
\beq
a(t) \simeq a_i\exp\left(\frac{H_i +H(t)}{2}(t-t_i)\right).
\eeq
Since Hubble parameter decreases linearly with the field, we obtain time of inflation period (\ref{Hslow}, \ref{phim2}):
\beq
\Delta t \simeq t_f - t_i \simeq \sqrt{12\pi} \varphi_i/m.
\eeq
During this period the scale factor increases:
\beq
\frac{a_f}{a_i} \simeq \exp(2\pi\varphi_i^2).
\eeq
This result is in good agreement with previous estimates: (\ref{flatness}, \ref{initialvelocities}). The initial value $\varphi_i$ has to be four times bigger than the Planckian value to last more than 75-efolds. We can observe that Hubble constant decreases much slower than the scale factor which grows exponentially:
$$
\frac{H_i}{H_f} \ll \frac{a_f}{a_i}.
$$
\textbf{Graceful exit and afterwards.} After the field drops below the Planckian value the oscillation begins. To analyze this part of evolution we investigate (\ref{inflationII}). If we define new variables, namely:
\beq
\label{variableangle}
\begin{array}{lcr}
\dot{\phi} = \sqrt{\frac{3}{4\pi}} H \sin\theta,& & m\varphi = \sqrt{\frac{3}{4\pi}} H \cos\theta.
\end{array}
\eeq
then the equation (\ref{Attractorm2}) will reduce to a system of differential equations:
\begin{align}
\dot{H} &= - 3H^2 \sin^2\theta, \\
\dot{\theta} &= -m - \frac{3}{2}H\sin 2\theta.
\end{align}
If we neglect second term in equation for theta we obtain:
$$
\theta \simeq -mt +\alpha,
$$
where $\alpha$ can be set as zero. Then we obtain a closed form solution for $H(t)$:
\beq
H(t) \simeq \frac{2}{3t}\left(1-\frac{\sin(2mt)}{2mt}\right)^{-1}.
\eeq
So plugging in (\ref{variableangle}):
\beq
\phi(t) \simeq \frac{cos(mt)}{\sqrt{3\pi}mt}\left(1+\frac{\sin(2mt)}{2mt}\right) + \emph{O}\left((mt)^{-3}\right).
\eeq
Using above results we obtain scalar curvature as:
$$
R \simeq - \frac{4}{3t^2}\left(1+3\cos(2mt) + \emph{O}\left((mt)^{-1}\right)\right),
$$
in comparison with matter dominated Universe, where $R = - \frac{4}{3t^2}$. Graceful exit occurs naturally in models with massive scalar fields. So now we move on to inhomogeneities generated by inflation.
\subsection{Generation of primordial inhomogeneities}
\label{Inhomogeneouslimit}
We have already discussed the fact that inflation preserves the homogeneity of the Universe in a sense that when we assumed homogeneity at the beginning of evolution no inhomogeneous terms appear. Moreover, we will show that inflation can generate inhomogeneities, originating from quantum fluctuations, are in agreement with cosmological microwave background (CMB). These CMB data presented on the picture below (taken from \cite{Rubakov}) shows that Universe is almost homogeneous with $\delta T/T \sim 10^{-5}$. The red regions represents hotter temperature by $\delta T/T \sim 10^{-5}$, the blue one represent the colder by the same amount. 
\FloatBarrier
\begin{figure}[!h]
\begin{center}
\includegraphics[scale=0.15]{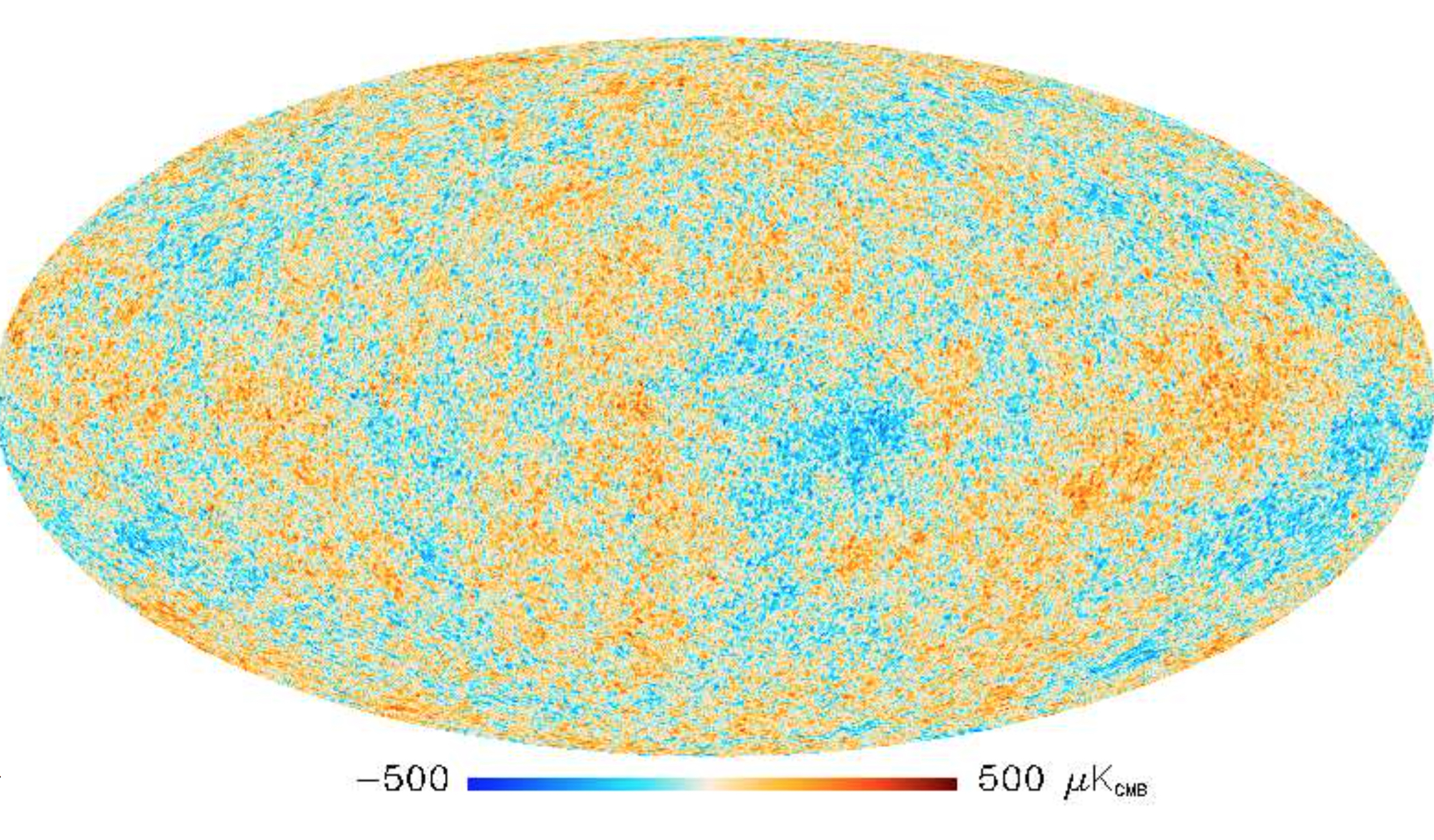}
\caption{CMB data}
\label{CMBdata}
\end{center}
\end{figure}
\FloatBarrier
\noindent
Inflationary cosmology allows us to predict their origin and calculate their spectrum while in $\Lambda$-CDM model they can only be postulated as initial conditions. According to inflation primordial perturbations originated from quantum fluctuations. These fluctuations started to arise on scales close to Planckian length. Inflation cause them to stretch to huge (galactic) scales with almost unchanged amplitudes. So this is exactly content of problem 3, we mentioned in chapter \ref{Mainideas}, how to explain their origin, spectra and statistical properties? Since the number of hotter and colder regions, where the small inhomogeneities occurs is huge, then it is convinient to treat their configurations as random fields, namely we assume that each of the Fourier components has random Gaussian \cite{Mukhanov} distribution with variance:
\beq
\sigma_k^2 \equiv |\Phi_k|^2,
\eeq
which is also the amplitude of a two point function:
\beq
\langle f_k f_k'\rangle = \sigma_k^2 \delta(k+k'),
\eeq
and we define power spectrum as:
\beq
\mathcal{P}_s = \frac{|\Phi_k|^2 k^3}{2\pi^2}.
\eeq
If we introduce the spectral index:
\beq
n_s - 1 = \frac{d \ln \mathcal{P}_s}{d\ln k}, 
\eeq 
we see that the flat spectrum gives $n_s =1$ \cite{Mukhanov}. Before we move on one has to stress that gaussianity of inhomogeneities is an assumption coming from experiment. However, the higher point functions can be also discussed, they originate especially in multi-inflaton models but they will be not discussed in this thesis. Moreover, there is an upper limit on their amplitudes, see \cite{1475-7516-2011-11-040,Two-Higgs-doublet,Senatore}, then gaussianity is a reasonable assumption. Then we have to calculate two point functions for metric perturbations and relate them to scalar perturbations. Here we will develop a systematic way of treating inflation inhomogeneities. The inflation is also source of primordial gravitational waves.
\subsubsection{Scalar field perturbations}
Following Garriga and Mukhanov \cite{GARRIGA} we will consider the most general local action for a scalar field:
\beq
\label{actioninhomo}
S = \frac{1}{16\pi}\int d^4x\sqrt{|g|} p(X,\varphi), 
\eeq
where $X = \frac{1}{2}g^{\mu\nu}\partial_{\mu}\varphi\partial_{\nu}\varphi$.
As we can see $p$ plays the role of pressure, so if we derive the energy momentum tensor we see that it is already in the hydrodynamical form:
\beq
T_{\mu\nu} = (\varepsilon +p)u_{\mu}u_{\nu} - p\guv,
\eeq
where:
\beq
u_{\mu} = \frac{\varphi_{,\mu}}{(2X)^{1/2}},
\eeq
and the energy density given by:
\beq
\varepsilon = 2Xp_{,X} - p.
\eeq
The Lagrangian (\ref{actioninhomo}) can be used to describe not only the scalar field but also motion of hydrodynamical fluid.\\
\textbf{Background}
As our background we consider FLRW metric:
\beq
ds^2 = [dt^2 -a^2(t) \delij dx^idx^j],
\eeq
in our case with spatial curvature $k=0$, then we can write two equations for background variables $\varphi_0(t)$ and $a(t)$ as:
\begin{align}
H^2 = \frac{8\pi G}{3}\varepsilon, \\
\label{epsH}
\dot{\varepsilon} = -3H(\varepsilon +p),
\end{align}
where first is (\ref{IIFriedeq}) and second is $\nabla^{\mu}T_{0\mu} =0$. We get by substitution:
\beq
\dot{H} = 4\pi G(\varepsilon +p).
\eeq 
 We will write the full derivative for energy density $\dot{\varepsilon} = \varepsilon_{\,X}\dot{X} + \varepsilon_{,\varphi}\dot{\varphi}$ and the same for the pressure:
\beq
\dot{p} = p_{\,X}\dot{X} + p_{,\varphi}\dot{\varphi} = -3c_s^2H(\varepsilon +p) + \dot{\varphi}(p_{,\varphi} -c_s^2\varepsilon_{,\varphi}),
\eeq
with :
\beq
c_s^2 = \frac{p_{,X}}{\varepsilon_{,X}} = \frac{\varepsilon +p}{2X\varepsilon_{,X}},
\eeq
where $c_s^2$ is called the speed of sound for perturbations, for canonical kinetic term we have: $c_s=1$. \\
\textbf{Perturbations} Now we introduce small inhomogeneities in the scalar field:
\beq
\varphi(t,x) = \varphi_0(t) + \delta\varphi(t,x),
\eeq
which will cause the following perturbations in metric:
\beq
ds^2 = [dt^2(1 +2\Phi) - a^2(t)\delij(1-2\Phi) dx^idx^k],
\eeq 
then to linear order we get that:
\beq
\delta X = \frac{1}{2}\delta g^{00} \varphi_0'{}^{2} + g^{00}\varphi'\delta\varphi' = 2X_0\left(-\Phi + \frac{\delta \varphi'}{\varphi_0'}\right).
\eeq
Keeping in mind that $\delta T^i_k \varpropto \delta_k^i$ and applying (\ref{epsH}) we obtain:
\beq
\delta T_0^0 = \varepsilon_{\,X}\delta X + \varepsilon_{,\varphi}\delta\varphi = \frac{\varepsilon +p}{c_s^2} \left[\left(\frac{\delta \varphi}{\dot{\varphi}_0}\right)^{.} - \Phi\right] - 3H(\varepsilon +p)\frac{\delta \varphi}{\dot{\varphi}_0},
\eeq
and:
\beq
\delta T_i^0 = (\varepsilon +p)\left(\frac{\delta \varphi}{\dot{\varphi}}\right)_{,i}.
\eeq
Applying the formalism from section \ref{Gravperturbations}, we obtain the following equations:
\beq
\Delta \Phi = \frac{4\pi G a^2(\varepsilon +p)}{c_s^2H}\left(H\frac{\overline{\delta \varphi}}{\varphi_0'} + \Phi\right)',
\eeq
and:
\beq
\left(a^2\frac{\Phi}{H}\right)' = \frac{4\pi G a^4(\varepsilon +p)}{H^2}\left(H\frac{\overline{\delta \varphi}}{\varphi_0'} + \Phi\right),
\eeq
if we will recast the equations to the physical time we will obtain \cite{GARRIGA}:
\begin{align}
\left(\frac{\overline{\delta \varphi}}{\dot{\varphi}_0}\right)^{\cdotp} =& \left(1 + \frac{c_s^2\Delta}{4\pi G a^2 (\varepsilon +p)} \right)\Phi\\
\left(a\Phi\right)^{\cdotp} =& 4\pi G a(\varepsilon +p)\frac{\overline{\delta \varphi}}{\dot{\varphi}_0},
\end{align}
similar derivation in conformal time can be found in \cite{Mukhanov}.
From now on we will skip $0$ in $\varphi_0$ and bar in $\overline{\delta\varphi}$.
Let us define new variables $\xi$ and $\zeta$ via the following equations:
\beq
\Phi a = 4\pi G H \xi,
\eeq
and:
\beq
\frac{\delta\varphi}{\dot{\varphi}} = \frac{\zeta}{H} - \left(\frac{4\pi G}{a}\right)\xi.
\eeq
Then we obtain the linear equations for both of them:
\begin{align}
\dot{\xi} &= \frac{a(\varepsilon +p)}{H^2}\zeta, \\
\dot{\zeta} &= \frac{c_s^2 H^2}{a^3(\varepsilon +p)}\Delta\xi.
\end{align}
\textbf{Action} Now we introduce the action which reproduces the equations of motion:
\beq
S = \int dtd^3x \left[\xi \Delta\dot{\zeta} - \frac{1}{2}\frac{H^2c_s^2}{a^3(\varepsilon +p)}\xi \Delta\xi + \frac{1}{2} \frac{a(\varepsilon + p)}{H^2}\zeta\Delta\zeta\right].
\eeq
If we propose a new variable:
\beq
z = \frac{a(\varepsilon +p)^{1/2}}{c_sH}, 
\eeq
and we introduce 
\beq 
v= z\zeta,
\eeq
it turns out that $v$ is a canonical quantisation variable, then in conformal time we obtain:
\beq
\label{etaaction}
S = \frac{1}{2}\int\left[v'{}^2 + c_s^2v\Delta{v} + \frac{z''}{z}v^2\right]d\eta d^3x.
\eeq
The corresponding Klein-Gordon equation takes a simple form:
\beq
v'' - c_s^2\Delta v - \frac{z''}{z} v =0,
\eeq
which has a long-wavelength solution $v \varpropto z$.\\
\textbf{Power spectra} We are interested in calculating the spectrum of $\zeta$, which is related to the gravitational perturbation via the formula:
\beq
\zeta = \frac{5\varepsilon +3p}{3\varepsilon +p}\Phi + \frac{2}{3}\frac{\varepsilon}{\varepsilon +p}\frac{\dot{\Phi}}{H},
\eeq
let us note \cite{GARRIGA} that when $p/\varepsilon$ is constant for long wavelengths the second term drops out. The quantisation of the action (\ref{etaaction}) is standard and is described in the section \ref{QFTcurved}. Then we define the spectral density as:
\beq
\label{Pzeta}
\mathcal{P}_{\zeta} = \frac{1}{2\pi^2}|\zeta_k|^2k^3 = \frac{k^3}{2\pi^2}\frac{|v_k|^2}{z^2},
\eeq
where $v_k$ is the solution of the equation for Fourier modes:
\beq
v_k'' + \left(c_s^2k^2 - \frac{z''}{z}\right)v_k =0,
\eeq
with the wavelength defined as: $\lambda = \frac{2\pi}{k}$. During ``slow roll'' the changes of the scale factor are much faster than following quantities: the Hubble rate, $c_s$ and $(\varepsilon +p)$. Thus: $z''/z \approx 2(Ha)^2$ and the equation takes the form:
\beq
\label{vkshort}
\begin{array}{lcr}
v_k \approx \frac{e^{\pm ikc_s \eta}}{(2c_sk)^{1/2}}, & & (aH \ll c_s k),
\end{array}
\eeq
while for long wavelength the equation simplifies to:
\beq
\begin{array}{lcr}
v_k'' - \left(\frac{z''}{z}\right)v_k =0, & & (aH \gg c_s k),
\end{array}
\eeq
then the solution is:
\beq
\begin{array}{lcr}
v_k \approx C_k z, & &(aH \gg c_s k),
\end{array}
\eeq
and the $C_k$ is the constant, found by matching the solutions. In the transition region $(aH = c_s k)$ we find that: $|C_k|^2 = (2c_s k z^2_s)^{-1}$, where $z_s$ is the value of $z$ for horizon crossing, then the power spectrum is given by the formula:
\beq
\mathcal{P}_{\zeta} =\left. \frac{16}{9} \frac{\varepsilon}{c_s(1+p/\varepsilon)}\right|_{aH \simeq c_s k}.
\eeq
With the definition of spectral-index we obtain:
\beq
n_s  - 1 := \frac{\ln P_{\zeta}}{d \ln k} \simeq - 3 \left(1 + \frac{p}{\varepsilon}\right) - \frac{1}{H} \left(\ln\left(1 + \frac{p}{\varepsilon}\right)\right)^{\mathbf{.}} - \frac{(\ln{c_s})^{\mathbf{.}}}{H} + \ldots
\eeq  
One can also show that, when $c_s \equiv 1$, using (\ref{inflenergy}, \ref{inflpressure}, \ref{Inflconditions}),  we obtain:
\beq
\label{nsepsiloneta}
n_s - 1 =\left(-6\epsilon + 2\eta\right),
\eeq
where first term gives: $-2\epsilon$, and the second: $2\eta - 4 \epsilon$. 
There arises a question whether the quantum fluctuations during the Planck era can provide us the required amplitude we observe in CMB? The only way to obtain observed: $\Phi \sim 10^{-5}$ is to stretch a very short-wavelength fluctuations otherwise the metric fluctuations would be too small. Without inflation stage: $\ddot{a} <0$ so $H^{-1}$ grows faster than $\lambda_{ph}$, and since the Universe is expanding they remain inside the horizon and according to (\ref{Pzeta}, \ref{vkshort}) decay. The perturbations with wavelength slightly larger than the horizon will eventually enter it and also decay. For inflation scenario this is not the case however and the perturbations leave the horizon and it results in growth of the amplitude. Moreover, the inflation mechanism doesn't affect the statistical properties of the fluctuations rather then their power spectrum so the fluctuations remains Gaussian during the inflation stage, as we observe them today. \\
\subsubsection{Tensorial perturbations}
\label{Tensor to scalar ratio}
The theory of quantum gravity is far from being complete, but one of the prediction of inflation is the existence of such a theory. Namely PLANCK and WMAP data can be described by inflation scenario with help of gravitational waves generated during this period. If we assume that we can quantise gravitational waves then we can calculate so called tensor to scalar ratio $r$ and compare it with the data, see \cite{pracaAmsterdam, Mukhanov} and section \ref{Experimental} for details. If we take traceless tensor perturbations $h_{ik}$, we will obtain the following action:
\beq
\label{Gravwavesaction}
S = \frac{1}{64 \pi G} \int a^2 \left( h^{i'}_j  h^{j'}_i - h^i_{j,l}h_i^{j,l} \right) d\eta d^3x,
\eeq
if we expand it into Fourier components we obtain:
\beq
h_j^i(\mathbf{x},\eta) = \int h_{\mathbf{k}}(\eta)e^i_j(\mathbf{k})e^{i\mathbf{k}\mathbf{x}}\frac{d^3 k}{(2\pi)^{3/2}}, 
\eeq
where $e^i_j(\mathbf{k})$ are polarisation tensors. With the help of a new variable:
\beq
v_k = \sqrt{\frac{e_j^ie^j_i}{32\pi G}}ah_{\mathbf{k}},
\eeq
we obtain the action (\ref{Gravwavesaction}) as:
\beq
S = \frac{1}{2} \int \left(v_{\mathbf{k}}'v'{}_{-\mathbf{k}} - \left(k^2 - \frac{a''}{a}\right)v_{\mathbf{k}}v_{-\mathbf{k}}\right)d\eta d^3k,
\eeq
the equations of motion are then: 
\beq
\begin{array}{lcr}
v_{\mathbf{k}}'' + \omega_k^2(\eta)v_{\mathbf{k}}=0 , & & \omega_k^2(\eta) = k^2 - a''/a,
\end{array}
\eeq
and the two point function is the following:
\beq
\langle 0| h_i^j(\eta,\mathbf{x})h_j^i(\eta,\mathbf{y})| 0 \rangle = \frac{8}{\pi a^2}\int |v_k|^2k^3\frac{\sin kr}{kr} \frac{dk}{k}.
\eeq
The power spectrum can be calculated as:
\beq
\mathcal{P}_h = \frac{16|v_{\mathbf{k}}|^2k^3}{\pi a^2},
\eeq
 where we have multiplied by two, due to summation over polarisations. If we use the 0-th order approximation for inflation scenario (gravitational waves don't depend much on the equation of state) and put $H = H_{\Lambda} = \textrm{const}$, then $a= - (H_{\Lambda}\eta)^{-1}$. Then we obtain according to Mukhanov \cite{Mukhanov}:
\beq
v_{\mathbf{k}}(\eta) = \frac{1}{\sqrt{k}}\left(1 + \frac{i}{k\eta}\right)\exp(ik(\eta - \eta_i)),
\eeq
and:
\beq
\mathcal{P}_{h} =\frac{8H^2_{\Lambda}}{\pi}[1 + (k\eta)^2].
\eeq
Finally we obtain for long wavelengths:
\beq
\mathcal{P}_h \simeq \frac{16H^2_{k\simeq Ha}}{\pi} \simeq \frac{128}{3}\varepsilon_{c_sk\simeq Ha},
\eeq
and tensor spectral index as:
\beq
n_T  =  \frac{d\ln\mathcal{P}_h}{d \ln k} \simeq -3\left(1 + \frac{p}{\varepsilon}\right)_{c_sk \simeq Ha}.
\eeq
We introduce tensor to scalar ratio as:
\beq
r= \frac{\mathcal{P}_h}{\mathcal{P}_s} \simeq 24\left[c_s\left(1 + \frac{p}{\varepsilon}\right)\right]_{c_sk\simeq Ha},
\eeq
for a scalar field we have: $c_s=1$ and for so called k-inflation, can be in principle: $c_s \ll 1$. This gives a possibility to distinguish both theories phenomenologically. The consistency relation is
\beq
r = - 8n_T, 
\eeq
The relation breaks down if we modify the structure of gravity like in Gauss-Bonnet gravity \cite{pracaAmsterdam} or when we have non-canonical kinetic terms.
We also have an expression for $r$ in terms of slow-roll parameter:
\beq
\label{repsilon}
r = 16 \epsilon.
\eeq
Now we briefly discuss possible particle physics scenarios which can give rise of inflation.
\subsection{Menu of scenarios}
\label{Menu}
The scalar field satisfying slow-roll condition is the only thing we require for inflation scenario. This means that there are many mechanisms which can give such a conditions and obviously one has to show that they fit the data for $n_s$ and $r$. This can be fundamental scalar field or a fermionic condensate described effectively by an effective scalar field. Let us then inspect some of the possible scenarios.\\
\textbf{Scalar field} First proposal is  obviously the scalar field with a potential with satisfying slow roll conditions. For example it can be:
\beq
V(\varphi) = \frac{A}{n}\varphi^n,
\eeq
with $A$ being some constant. We have checked in section \ref{examplem2} that for a massive potential slow-roll conditions are satisfied for a large value of the field. However, in so called \textbf{old inflation} scenarios the potential has two minima, one is meta-stable and the second one is global one and at first the Universe supercools to the the meta-stable minima and the accelerated expansion is obtained via tunneling to the global one. This scenario has however problems with the graceful exit. \\
\textbf{New inflation} is based on another concept, namely the potential has a maximum at $\varphi =0 $. Then the field explicitly rolls down to the minimum which has a barrier on its right so the field cannot escape. The potential escapes the maximum due to the quantum fluctuations rather than by tunneling. This model however has problem with initial conditions since one has to take the thermal initial state what is quite unlikely. That's why it is popular to think that the Universe might be in a ``self-reproducing'' regime \cite{Mukhanov}.\\
\textbf{Chaotic inflation} is the name, coming from the fact, that the initial conditions might be taken \emph{chaotic} with the only requirement that the field must be beyond the Planck scale. Even though it can provide a large inhomogeneity the quantum fluctuations can provide self-reproduction of the Universe and induce a very complicated \emph{global} structure of the Universe.\\
Another proposal for lagrangian are non-canonical kinetic terms which are the origin of the inflation. We recall (\ref{actioninhomo}):
\beq
S = \int p(X,\varphi) \sqrt{|g|}d^4x,
\eeq
with 
\beq
\varepsilon = 2X\frac{\partial p}{\partial X} -p,
\eeq
If $p$ satisfies the condition $X\partial p/\partial X \ll p$ for some range of $X$ then we have $p \approx - \varepsilon$ and we have inflationary solution. In this approach $c_s$ is not equal to one and can adopt many values to avoid fine-tuning. This is also a prediction of this kind of theories. This scenario is called $k$-inflation and is discussed in chapter \ref{Extensions}.\\
There is also another proposal, namely modified gravity. The \emph{Lovelock theorem} guaranties that the E-H action can be extended to higher powers of $R$ or contractions of two Ricci or Riemann tensors, see \cite{pracaAmsterdam} and appendix \ref{EHaction} for details. Starobinsky inflation type is such a case and is currently of great interest because it turns out that it is consistent with the Planck data.
\subsection{Experimental tests of inflation}
\label{Experimental}
The probes WMAP and PLANCK can actually measure the spectral tilt and tensor to scalar ratio. The picture below shows the constraints of $n_s$ and $r$ from Planck data \cite{Ade:2015lrj, pracaAmsterdam}.
\FloatBarrier
\begin{figure}[h!]
\label{Planckdata}
\includegraphics[scale=0.6]{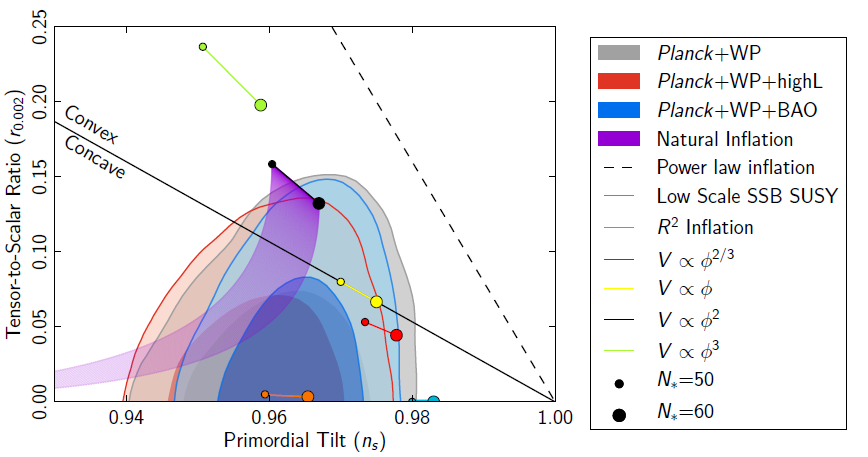}
\caption{PLANCK constraints on $n_s$ and $r$ for different models.}
\end{figure}
\FloatBarrier
\noindent
The $N_{\star}$ number represents the number of e-folds. As we can see the pure Higgs scenario with potential $\lambda\phi^4$ is far outside the data, it is not even on the diagram. To fit them the self-coupling should be much smaller than its measured in LHC, however an extra interaction between Higgs and gravity can provide a successful inflation scenario. The modified $R^2$ gravity is the only model which fits the data, however as we will show in the next paragraph it is connected to the Higgs coupled to gravity scenario. Also a recent article \cite{Freese:2014nla} shows that there is a set of parameters for which natural inflation matches the data within 1$\sigma$.
\subsection{Starobinski inflation}
\label{Starobinskyinflation}
In 1980 Starobisky proposed \cite{STAROBINSKY} a model where a pure modified gravitational action can cause non-singular evolution of the Universe, namely:
\beq
\label{Starobinskyaction}
S = \frac{1}{2}\int \sqrt{|g|}d^4x \left(M_p^2R + \frac{1}{6 M^2}R^2\right),
\eeq
where $M$ is some "mass" parameter, with value taken to fit the Planck data. Then it turned out that such a modified gravity models can give rise to inflation scenario and their FLRW equations has additional structure. For example inflation considered for Gauss-Bonnet gravity with Higgs non-minimally coupled is described in \cite{pracaAmsterdam}. See also appendix \ref{EHaction} for derivation equations of motion from $f(R)$ action and derivation of FLRW equations for this general framework. Now we will rewrite the action (\ref{Starobinskyaction}) into equivalent linear representation:
\beq
S_{l}= \frac{1}{2}\int \sqrt{|g|}d^4x \left(\frac{M_p^2}{2}R + \frac{1}{M}R\psi -3\psi^2 \right),
\eeq
then if we write equations of motion for $\psi$ we obtain:
$$
\frac{1}{M}R = 6\psi,
$$
so we get equivalency. Then if we use, as in \cite{Starobinskyreview}, a following conformal transformation:
\beq
\label{Starobinskyconformal}
g_{\mu\nu} \to e^{-\sqrt{2/3}\phi/M_p} g_{\mu\nu}= \left(1+ \frac{2\psi}{MM_p^2}\right)g_{\mu\nu},
\eeq
we get action with scalar field coupled to gravity:
\beq
S = \frac{1}{2}\int \sqrt{|g|}d^4x \left[\frac{M_p^2}{2}R +\frac{1}{2}\partial_{\mu}\phi \partial^{\mu}\phi -\frac{3}{4}M_p^4 M^2 \left(1- e^{-\sqrt{2/3}\phi/M_p}\right)\right].
\eeq
So $R^2$ term gives equivalent solutions as the evolution of scalar field with exponential type potential. According to Planck data, Starobinsky model and its decendants are the main class of models which has correct tensor to scalar ratio and scalar-tilt: 
\beq
\begin{array}{lcr}
n_s -1 \approx -\frac{2}{N}& & r \approx \frac{12}{N^2},
\end{array}
\eeq
with $N$ being the number of e-folds. For a similar calculation see paragraph \ref{HiggsInflaton}.  As we have already said, both Starobisky inflation and its descendants are of high interest nowadays. Higgs inflation, which is one of them, can be successful proposal for the inflation scenario. One can show that Starobinsky scenario provides (almost) the same observation data as the Higgs inflation. The difference in the data \cite{Starobinskyreview} is of order $10^{-3}$, between $r$ and $n_s$ in both theories. Moreover, if we get rid of kinetic term in Higgs inflation scenario, one can show their equivalence in this regime. Namely, if we take (\ref{NonminimalHiggsaction}), with $h\gg v$ and $|\partial_{\mu}h\partial^{\mu}h| \simeq 0$, whose conditions are satisfied during inflation, the action will become:
\beq
S_{H} = \int d^4x\sqrt{|g|}\left[-\frac{M^2+\xi h^2}{2}R - \frac{\lambda}{4}h^4\right].
\eeq
Then integrating out Higgs field via EL equations
$$
h^2 = \frac{\xi R}{\lambda},
$$
we obtain
\beq
S_{H} =  \int d^4x\sqrt{|g|}\left[\frac{M^2}{2}R + \frac{3\xi^2}{4\lambda} R^2\right], 
\eeq
It is the same as (up to definition of action in (\ref{NonminimalHiggsaction})) result obtained in \cite{Starobinskyreview} and (\ref{Starobinskyaction}) with
$$
M^2 = \frac{\lambda}{9\xi^2},
$$ 
Because $M \simeq 10^{-5}$ \cite{Starobinskyreview}, we get that $\xi^2 \simeq 10^{10}\lambda$ what exactly gives the range of values satisfying successful non-minimally coupled Higgs inflation. After having discussed inflation in detail we now move on to particle physics and discuss the Higgs particle and Standard Model.


\section{The Standard Model}
\label{TheSM}
\setcounter{equation}{0}
Here we will briefly describe Standard Model, using \cite{MeissnerKTP, QMbook}.
This theory is based on local gauge group: $SU(3)_{\emph{c}}\times SU(2)\times U(1)_Y$, where $c$ stays for the color charge and $Y$ for the hypercharge. The Standard Model predictions are in agreement with experiment up to very high precision. However, the theoretical investigation of SM and its failure to describe some observations, like dark matter, showed that it is not a final theory of particles. We will address those problems in paragraph \ref{SMproblems}. One has to mention that when the Standard Model was created it wasn't obvious why should it be unique, it came out later with the discovery of  gauge anomalies. Then it turned out that is the only model where these anomalies vanish i.e. it is the only possible scenario where the effective action is strictly invariant under gauge and spacetime transformations what should obviously be the case. Vanishing of anomalies also explains why the electric charge of a proton and an electron should be minus each other and the number of leptons and quarks should be equal and this is not just a coincidence \cite{MeissnerKTP}. So let us start with theoretical concepts used in the SM, namely Yang-Mills fields and Higgs mechanism. Then we will describe the components of SM and finally outline its problems and drawbacks.
\subsection{Yang-Mills theories and Higgs mechanism}
\subsubsection{Yang-Mills theories}
Quantum electrodynamics was the first successful quantum field theory, describing interactions of charged particles and photons. The lagrangian of electrodynamical part is:
\beq
\mathcal{L}_{e} = - \frac{1}{4} F_{\mu\nu}F^{\mu\nu}
\eeq
and possesses an additional symmetry: $A_{\mu} \to A_{\mu} + \partial_{\mu} \alpha$, where $\alpha$ is any scalar function. Photon is a massless particle, however in principle nothing, from the theoretical point of view, stops us from introducing the massive term $m^2 A_{\mu}A^{\mu}$ except that the above symmetry would be broken. Generalising this Lagrangian to non-abelian symmetries, we get Yang-Mills theories. The general action for the pure Yang-Mills fields is the following:
\beq
S_{YM} = \int d^4x h_{ab}F_{\mu\nu}^a F^{\mu\nu b} =\frac{1}{2g^2} \int d^4x\textrm{Tr}\left[F_{\mu\nu}F^{\mu\nu}\right],
\eeq
where $F_{\mu\nu}$ are field strength tensors:
\beq
\label{Fmn}
F_{\mu\nu} = \partial_{\mu} A_{\nu} - \partial_{\nu} A_{\mu} + \left[A_{\mu},A_{\nu}\right].
\eeq
The definitions of $A_{\mu}$ and $F_{\mu\nu}$ are: 
\beq
\begin{array}{lcr}
\label{Amua}
A_{\mu} = ig A_{\mu}^a T^a, & & F_{\mu\nu} = ig F_{\mu\nu}^a T^a;
\end{array}
\eeq  
and $F_{\mu\nu}^a = \partial_{\mu}A_{\nu}^a - \partial_{\nu}A_{\mu}^a +f^a_{bc} A_{\mu}^b A_{\nu}^c,$ has values in some algebra $\mathcal{A}$, eg. $\mathfrak{su}(2)$ or $\mathfrak{su}(3)$, where $f^a_{bc}$ are structure constants of Lie algebra and $g$ is a coupling constant. The mapping $h: \mathcal{A} \to \mathbb{R}$ is the function from that algebra to real numbers and it plays the role of a scalar product in this Lie algebra. 
For $U(1)$ group the last term in (\ref{Fmn}) vanishes, however for $\mathfrak{su}(N)$ algebras, $N>1$ this is not the case. The fields with non-abelian gauge groups are called Yang-Mills fields. For scalar electrodynamics, after using the Noether procedure, we find out that the derivatives in the lagrangian:
\beq
\mathcal{L} = -\frac{1}{4}F_{\mu\nu}F^{\mu\nu} + (D_{\mu}\phi)^{\ast}D^{\mu}\phi - \Vp,
\eeq
can be written in covariant way and are not the standard ones: $D_{\mu} = \partial_{\mu} +ieA_{\mu}$, where we have taken $e<0$ for electron. In analogy to this one can define the lagrangian with scalar or fermionic field coupled to $SU(N)$ gauge field.
\subsubsection{Higgs-mechanism}
\label{Higgsboson}
\label{Goldstone}
From experiment we know that electroweak interactions act on very short distances but if we wanted to use Yang-Mills theory to describe them, we cannot have any mass term. Solution to this problem lies is introducing another field which breaks the local symmetry and give mass to those fields \cite{MeissnerKTP}. Such situation occurs in the Standard Model where due to breaking of the SM symmetry three out of four of electroweak boson force carriers become massive, namely $W^{\pm}$, $Z^0$, while photon remains massless. Let us inspect Higgs mechanism - spontaneous breaking of local SU(2) symmetry. The Lagrangian reads
\beq
\mathcal{L} = -\frac{1}{2g^2} \textrm{Tr }\left[F_{\mu\nu}F^{\mu\nu}\right] + D_{\mu}\phi^{\dagger}D^{\mu}\phi - \frac{\lambda}{4}\left(\phi^{\dagger}\phi -v^2\right)^2,
\eeq
where $\phi$ is a vector in fundamental $SU(2)$ representation i.e. it has values in $\mathbb{C}^2$. $A_{\mu}$ is defined as in (\ref{Amua}). Then the lagrangian is invariant under the following transformation: 
\beq
\begin{array}{lr}

\phi \to U \phi, & A_{\mu} \to U A_{\mu} U^{-1} - (\partial_{\mu}U)U^{-1}, 
\end{array}
\eeq
where $U(x) \in SU(2)$. Moreover, it possesses additional global $U(1)$ symmetry. Because the potential term has minima for $|\phi| =v$ we have to choose a state of minimal energy and we choose it as
\beq
\phi_0 = \left(\begin{array}{c}  0 \\ v\end{array}\right),
\eeq
Then we make an expansion around the vacuum / lowest energy state, such that:
\beq
\label{VacHiggs}
\phi = U\left(\begin{array}{c}  0 \\ v+\rho/\sqrt{2}\end{array}\right),
\eeq
where $\rho$ is a real field. Of course for each $\phi$ we can find $U \in SU(2)$ and $\rho \in \mathbb{R}$, such that (\ref{VacHiggs}) holds. We also find such a new $\tilde{A}_{\mu}$, that for a given $U$ the following holds:
\beq
\label{AmuHiggs}
A_{\mu} = U\tilde{A}_{\mu}U^{-1} - (\partial U)U^{-1}.
\eeq
This particular choice of $\phi$ and $A_{\mu}$ gives a following Lagrangian in terms of $\rho$ and $\tilde{A}_\mu$:
\beq
\label{HiggsWmu}
\mathcal{L} = -\frac{1}{2g^2} \textrm{Tr} \tilde{F}_{\mu\nu}\tilde{F}^{\mu\nu} - \frac{1}{4}g^2(v + \rho/\sqrt{2})^2\tilde{A}_{\mu}\tilde{A}^{\mu} + \frac{1}{2} \partial_{\mu}\rho \partial^{\mu}\rho - \frac{\lambda \rho^2}{2}\left(v + \rho/(2\sqrt{2})\right)^2.
\eeq
From two complex valued field $\phi_1$ and $\phi_2$, we are left with one real field $\rho$ and three massive fields $A_{\mu}^a$, with masses:
\begin{align}
\label{massesHiggs}
m_{\tilde{A}^a}^2 &= v^2g^2/2, \\
m^2_{\rho} &= \lambda v^2. 
\end{align}
 This is an example of local symmetry breaking mechanism which is responsible for the origin of masses in Standard Model, via Higgs mechanism and chiral symmetry breaking.\\ 
\subsection{Overview of features of The Standard Model}
The Standard Model consists of fermions (quarks and leptons) and bosons (gluons,  bosons $W^{\pm}$, $Z^0$, photon and Higgs particle). According to experiment there are known 3 generations of quarks and leptons, namely: $e^i$ are: $i=1$ electron, $i=2$ muon, $i=3$ taon and $\nu^i$ denotes the proper neutrinos. On the other hand $u^i$ and $d^i$ denotes $u,s,b$ and $d,c,t$ quarks respectively. The left chiral quarks and leptons are $SU(2)$ doublets, while right chiral are siglets. Let us write the content of this model according to \cite{MeissnerKTP, QMbook}:\\
\textbf{Leptons}:
\begin{align}
&L^i = \left(\begin{array}{c} \nu_L^i \\ e_L^i \end{array} \right) &  &(Y = -1/2), \nonumber \\
&E^i = e_{R}^i & &(Y=-1).
\end{align}
\textbf{Quarks}:
\begin{align}
&Q^{ib} =  \left(\begin{array}{c} u_L^{ib} \\ 
d^{ib}_L \end{array} \right) & &(Y=1/6), \nonumber \\
&U^{ib} = u_R^{ib} & &(Y =2/3),\\
&D^{ib} = d^{ib}_R & &(Y=-1/3). \nonumber
\end{align}
and \textbf{Higgs field}:\\
\begin{align}
&\Phi =  \left(\begin{array}{c} \phi^{+} \\ \phi^0  \end{array} \right) & &(Y = 1/2).
\end{align}
Quarks and Leptons have obviously their antiparticles with opposite quantum numbers. As we can see model doesn't include right chiral neutrinos.\\
\textbf{Gauge bosons}:
\begin{itemize}
\item SU(3): $G_{\mu} = i g_3 G_{\mu}^{b}T_{b}$ 8 gluons,
\item SU(2): $W_{\mu} = i g_2W_{\mu}^a T_{a}$ 3 bosons $W^{\pm}, W^3$,
\item U(1): $B_{\mu}$ 1 boson,
\end{itemize}
\noindent where $T^b, T^a$ are generators of respectfully SU(3) and SU(2) algebras.
As we can see the Standard Model can be divided into three sectors ie. Quantum Chromodynamics (QCD), Electroweak Interactions (EW) and Higgs sector. The strong interactions (between quarks and gluons) are described by QCD with $SU(3)$ symmetry group. Leptons are not interacting with gluons, so they are not affected by it. QCD is far from being totally understood and is even today investigated and posses many surprising features, like asymptotic freedom. However, since it won't be present in the inflation scenario, we will not focus on it. The other two sectors, namely EW and Higgs sectors are related closely to each other. The electroweak lagrangian density is in the form:
\beq
\label{EWLagrangian}
\mathcal{L}_{EW} = \mathcal{L}_{g} + \mathcal{L}_{f} + \mathcal{L}_{h} + \mathcal{L}_Y,
\eeq
where:
\beq
\mathcal{L}_{g} = -\frac{1}{4} B_{\mu\nu}^2 + \frac{1}{2 g_2^2} \mathrm{Tr} \left[W_{\mu\nu}^2\right],
\eeq
with proper field strength tensors built from $B_{\mu}$ and $W_{\mu}$ respectfully (\ref{Fmn}).
The $\mathcal{L}_{f}$ part is the following:
\beq
\mathcal{L}_{f} = \bar{L}^i \gamma^{\mu}\Dmuo L^i + \bar{E}^i \gamma^{\mu} \left(\partial_{\mu} - ig_1 B_{\mu}\right) E^i + \textrm{quarks},
\eeq
The Higgs part:
\beq
\mathcal{L}_{h} =  \left(\Dmut \Phi\right)^{\dagger}\left(\Dmut\right)\Phi - \frac{\lambda}{4}\left(\Phi^{\dagger}\Phi - v^2\right)^2.
\eeq
And the Yukawa interaction term (including quarks):
\beq
\mathcal{L}_Y = \bar{L}^i\Phi Y_{ij}^E E^j + \bar{Q}^i\Phi Y_{ij}^D D^j + \bar{Q}^i\epsilon\Phi^{\ast}Y_{ij}^U U^j  + \mathrm{h.c.},
\eeq
where $\epsilon = i \sigma_2$. The $g_i$'s are coupling constants to $U(1)_Y, SU(2), SU(3)_c$, respectively. Because of $\bar{L}^i L^i =0$ relation, the masses of leptons come only from spontaneous symmetry breaking mechanism. Generally speaking Yukawa matrices are three $3 \times 3$ complex matrices, so they should posses 54 parameters, but if we redefine the fields, according to \cite{MeissnerKTP,QMbook} we obtain 13 parameters: 9 masses, 3 angles and one phase. So in total the model has 18 parameters, since we take into account 3 couplings to gauge bosons, Higgs expectation value and Higgs self-coupling. In previous paragraph we described the Higgs mechanism for a single Yang-Mills $SU(2)$ field, but generalisation for SM is straightforward. For Standard model, let us define new fields as:
\beq
\begin{array}{lcr}
A_{\mu} = \frac{g_2 B_{\mu} + g_1 W^3_{\mu}}{\sqrt{g_1^2 + g_2^2}},& & Z_{\mu} = \frac{g_2 B_{\mu} - g_1 W^3_{\mu}}{\sqrt{g_1^2 + g_2^2}}
\end{array}
\eeq
And two complex fields:
\beq
\begin{array}{lcr}
W^{\pm} = \frac{1}{\sqrt{2}}\left( W_{\mu}^1 \mp i W_{\mu}^2\right),
\end{array}
\eeq
which associated quanta are the particles: $\gamma$, $Z^0$ and $W^{\pm}$ bosons respectfully. If we provide $e$ and \emph{Weinberg angle} $\theta_W$ as:
\beq
\begin{array}{lcr}
g_1 = \frac{e}{\cos{\theta_W}},& & g_2 = \frac{e}{\sin{\theta_W}},
\end{array}
\eeq
then the electrical charge is:
\beq
Q = Y + T_3,
\eeq
where $T_3$ is the 3-rd component of weak isospin, $T_3 = \pm 1/2$ for doublets and 0 for singlets of $SU(2)$. Masses generated via Higgs mechanism (\ref{massesHiggs}) are:
\begin{align}
M_A &= 0,\\
M_Z &= \frac{\sqrt{2} e v}{\sin{2\theta_W}},\\
M_W &= \frac{ e v}{\sqrt{2}\sin{\theta_W}}, \\
M_H &= \sqrt{\lambda}v,
\end{align}
where we of course don't include the effect of running coupling constant which originates from  quantum theory, even though those effects are crucial for our understanding of the Standard Model, we will not describe them here, because they are far beyond the scope of this work \cite{MeissnerKTP, QMbook}. 
\subsection{Problems of the Standard Model}
\label{SMproblems}
The Standard Model (SM) is for sure, one of the biggest achievements in the history of knowledge of mankind. From it one can derive almost every phenomenon on the Earth. Still it has some drawbacks.\\
First of all SM doesn't incorporate gravity at all and so far there is no theory which can unify particle physics and gravity for which SM will be a low energy effective theory. There are some proposals like string theory but still they are far from being complete or having as huge predictive power as SM does. However, those are not the only arguments. There are some cosmological issues which cannot be explained by SM. First of all, there is no candidate for dark matter in this theory. Dark energy and dark matter contribute roughly $95\%$ to the observed mass-energy in the Universe. Additionally there is no mechanism which can provide a matter-antimatter asymmetry such that the relic density of matter is of the order of magnitude we observe today.\\
Other class of problems comes from the theoretical study of SM which follow only form investigating the model without confronting it with observations. First of those is the hierarchy problem. Bare parameters (written down in the lagrangian) in the model  are generally different than the ones we measure. This is due to the phenomenon of change of the coupling constants with the energy scale, it is called running and is described on the quantum level. For Higgs mass, from the naive point of view, the quantum corrections should be very huge. They are called quadratic divergences, since they are proportional to the cutoff squared. However, they are highly suppressed for this case. This means that there should be an extremely fine cancellation between the corrections or there should be a mechanism which explains it. The problem of calculating Higgs mass is obviously beyond the SM problem but it is a significant one. Another problem is called triviality. One can argue that one cannot built a consistent QFT with Higgs field. Also CP violation strong enough to create matter prevalence in the amount we see it today is still unsolved. Those and other problems have to addressed, and are one of the main theoretical physics challenges in current century. \\
There are in general three types of models (the division is not strict of course, one model can be of many types) which relies on different philosophy of building such an extension. First type are the theories where the SM structure is incorporated into another larger theory but the SM sector is preserved in the way it is. However, the theory underlying it possesses new concepts and structure. Good examples are Grand Unified Theories (GUT). They assume that the initial symmetry of the theory is some single ``big'' gauge group and on the certain large scale a spontaneous symmetry breaking scenario occurs, which causes the division into the interactions the Standard Model describes. Most popular groups are $SU(5)$ or $SO(10)$. Another one is supersymmetry concept. The main idea is that each particle has its superpartner. Due to breaking of this symmetry a  superpartner of a given particle is much heavier. Second type of theories, are those which lies on completely different fundamental assumptions and SM is only low-energy effective theory. The main representative is string theory. The third type are minimal extensions in a sense they propose only a slight extension of SM. They not only solve problems of SM but can be, in principle, valid up to the Planck scale with no new intermediate scales and give possible candidates for dark matter. Since, besides dark matter, no new phenomena in particle physics are seen and low-energy supersymmetry is absent those type of models should be analysed. The $\nu$SM is such an extension, where only right chiral neutrinos and one scalar singlet are added.  Another one is Conformal Standard Model \cite{Latosinski2015}, the name comes from the fact that conformal symmetry is broken softly \cite{CSMtwo}.


\section{Conformal Standard Model}
\label{ConformalSM}
\setcounter{equation}{0}
\subsection{Overview of the model}
In this chapter we will present and discuss the Conformal Standard Model (CSM) \cite{ Latosinski2015,MEISSNER2007}, extension of Standard Model (SM). Since the experiments show almost no deviations from SM predictions the proposed extension should preserve the structure of the model and add as least as possible. Quoting the authors: ``\emph{The SM may survive essentially as is up to the Planck scale}". They show, in the quoted articles, that their extension not only solves the problems mentioned in paragraph \ref{SMproblems}, but also enlarges the neutrino sector, point out dark matter candidates and finally gives a natural scenario for leptogenesis - resolves the problem of matter-antimatter asymmetry. So far they proposed two versions.\\
 First version had a classically unbroken conformal symmetry \cite{MEISSNER2007}. The authors incorporated in the model one additional massless scalar field ``second Higgs'' and right chiral neutrinos. Also Higgs particle supposed to be massless in their model. The spontaneous conformal symmetry breaking was achieved by the famous Coleman-Weinberg mechanism \cite{PhysRevD.7.1888}, which predicts breaking the conformal symmetry via radiative corrections. This mechanism was the originat of the masses in the their model. It turned out, however, that the mixing with the heavier scalar in the two loop efective potential become too large to fit the LHC data and because of other drawbacks of the model \cite{CSMtwo} it was abandoned and the authors proposed a different mechanism to solve the hierarchy problem, namely softly broken conformal symmetry (SBCS).\\
The assumption which underlies this concept is the following: a complete and finite fundamental theory at UV scales exist. Moreover, the cutoff $\Lambda$ of effective field theory is treated as a physical scale. Then, since $\Lambda$ is taken as a finite parameter, also `bare' parameters are finite. One should obtain a effective theory valid below $\Lambda$. There are two mechanisms which can resolve the hierarchy problem below the cutoff scale. First: one relies on a symmetry, which will ensure the cancellation of $\varpropto \Lambda^2$ terms, like supersymmetry. In the second one the fundamental theory singles out a particular, \emph{physical} cutoff, for which squared mass of the Higgs particle is much smaller than $\Lambda^2$, also at that scale all the quadratic divergences will vanish. Then from technical point of view the hierarchy problem is solved. However, there arises the question for the existence of the fundamental theory, which is still beyond our reach.\\
The authors adapted the second mechanism and inspected both SM and CSM within this scenario. For Standard Model this mechanism gives the value of $\Lambda$ far beyond the Planck scale, moreover the scalar self-coupling becomes negative near $10^{10}$ GeV, rising questions about stability of electroweak vacuum. However, for Conformal Standard Model, for cutoff of order $M_P^2$, all SBCS requirements \cite{CSMtwo} are satisfied. Since then CSM may provide a complete scenario up to this scale, and as we will point out, may solve most of the puzzling issues in particle physics and cosmology.\\
The authors argued that there should be only two energy scales left: one of order Planck scale and another one is $\mathcal{O}(1)$ Tev. Their model has to satisfy not only the requirement that CSM will reproduce SM observational data but also to ensure that there are no other scales. So they propose conditions which their model should satisfy to obtain it. They called them:
\begin{enumerate}
\item \textbf{Perturbative consistency} - absence of Landau poles up to the Planck scale $M_{Pl}$.
\item \textbf{Lower
boundedness of the RG improved one-loop effective potential}.
\item \textbf{Vacuum stability} - Electroweak vacuum should remain in the global minima in the region below Planck scale.
\end{enumerate}
The authors checked that for their model there exist such a range of parameters which satisfies all the three constraints. The new version consists in new, enlarged scalar sector (with coupling to right-chiral neutrinos), new global $SU(3)_N$ symmetry related to new scalars and right-chiral neutrinos, which is spontaneously broken giving pseudo-Goldstone in the sense of one-loop corrections, which are the terms which breaks $SU(3)$ explicitly. The authors states that their model has  additional, following advantages: 
\begin{enumerate}
\item pseudo-Goldstone bosons can be potential Dark Matter candidates, 
\item Yukawa couplings acquires the desired form for leptogenesis, see \cite{ Latosinski2015,PILAFTSIS2004303} for details,
\item some linear combination of pseudo-Goldstone bosons can be (in principle) identified with axion. 
\end{enumerate}
The authors divided their work into three pieces:
\begin{itemize}
\item Scalar sector
\item Fermionic sector
\item Pseudo-Goldstone bosons and their couplings
\end{itemize}
Following them, we will analyse each sector separately, but with emphasis on scalar sector. We will incorporate it into the inflation scenario which will be analysed in chapter \ref{CSMinflation}.
\subsection{Scalar sector}
\label{CSMscalarsector}
The complex scalar sextet $\phi_{ij} = \phi_{ji}$ is introduced, which elements are: ``\emph{blind to the SM gauge symmetry, hence sterile}''. They only interact with Higgs boson and right chiral neutrinos. The further possible extension of scalar sector is presented in next paragraph, it consist of adding a U(3) triplet $\zeta_i$. And it will be incorporated in the chapter \ref{CSMinflation} to keep unitary evolution up to the Planck scale in context of Inflation. The scalars coupled to right chiral neutrinos posses a new symmetry $SU(3)_N$ broken by Dirac-Yukawa coupling $Y^{\nu}$, but very softly $\mathcal{O}(10^{-6})$. The sextet replaces Majorana mass term by:
\beq
\langle \phi \rangle Y^M_{ij} \to y_M \langle \phi_{ij}\rangle,
\eeq
and similarly the Majorana-type Yukawa couplings. The scalar part lagrangian is:
\beq
\label{CSMlagrscalar}
\mathcal{L}_{scalar} = (D_{\mu}H)^{\dagger}(D^{\mu}H) + \mathrm{Tr}(\partial_{\mu}\phi^{\ast}\partial^{\mu}\phi) - V(H,\phi),
\eeq
and the potential is given by a formula:
\begin{align}
\label{CSMpotential}
V(H,\phi) = m^2_1 H^{\dagger}H + m_2^2 \mathrm{Tr}(\phi\phi^{\ast}) + \lambda_1(H^{\dagger}H)^2 \nonumber \\
+ \lambda_2[\TR]^2  + 2\lambda_3(H^{\dagger}H)\TR + \lambda_4\mathrm{Tr}(\phi\phi^{\ast}\phi\phi^{\ast}),
\end{align}
where all coefficients are real. The potential is invariant under:
\beq
\begin{array}{lcr}
\phi(x) \to U\phi(x)U^T, & &  U\in \textrm{U(3)}.
\end{array}
\eeq
The three different sets of conditions ensures positivity of the classical potential:
\begin{align}
\label{first}
\lambda_1, \lambda_2, \lambda_4 >0, & & \lambda_3 > -\sqrt{\lambda_1(\lambda_2 + \lambda_4/3)};\\
\label{second}
\lambda_4 <0, \lambda_1>0, \lambda_2 > -\lambda_4, & & \lambda_3 > - \sqrt{\lambda_1(\lambda_2 + \lambda_4)};\\
\label{third}
\lambda_2 <0, \lambda_1 > 0, \lambda_4 > -3\lambda_2, & & \lambda_3 > -\sqrt{\lambda_1(\lambda_2 + \lambda_4/3)}.
\end{align}
All of these conditions has to hold for all energies up to Planck scale, ie they should hold for running couplings $\lambda_i(\mu)$. The analysis made in \cite{Latosinski2015} ensures that this is exactly the case. Let us assume that the mass parameters are:
\beq
\begin{array}{lcr}
m_1^2 = - 2\lambda_1v_H^2 - 6\lambda_3v_{\phi}^2, & &  m_2^2 = -2 \lambda_3v_H^2 - (6\lambda_2 + 2\lambda_4)v_{\phi}^2,
\end{array}
\eeq
then the global minimum of (\ref{CSMpotential}) is exactly:
\beq
\begin{array}{lcccr}
\label{vacuumphi}
\langle H \rangle =\left( \begin{array}{c} 0 \\ 
v_H \end{array}\right),  & &
\langle \phi \rangle = \mathcal{U}_0\left( \begin{array}{ccc} v_{\phi} & 0 & 0 \\
0 & v_{\phi} & 0 \\
0 & 0 & v_{\phi} 
 \end{array}\right)\mathcal{U}_0^T, & & \mathcal{U}_0 \in \mathrm{U(3)},
 \end{array}
\eeq
and in addition the following condition has to be satisfied:
\beq
\label{CSMconditions}
\begin{array}{lcr}
\lambda_1\left\{\lambda_2 + \frac{\lambda_4}{3}\right\} - \lambda_3^2 >0,& & \lambda_4 >0.
\end{array}
\eeq
The  $\mathcal{U}_0$ matrix will be discussed later, in paragraph concerning pseudo-Goldstone bosons, here we will note that on the classical level one cannot determine it. Since the symmetry of vacuum state (\ref{vacuumphi}) is $SO(3)$, the Goldstone bosons manifold $\mathcal{M}$ is:
\beq
\mathcal{M} = U(3)/SO(3) \equiv U(1)_{B-L} \times SU(3)_N/SO(3),
\eeq
then we have six (pseudo-)Goldstone bosons. One of them is \emph{true} Goldstone boson, associated with global $U(1)_{B-L}$ symmetry and will be called the \emph{Majoron}. The name comes from the Majorana interaction, they could violate lepton number in double beta decay, however such a particles haven't been observed yet. After the symmetry breaking we have:
 \beq
 \label{Higgsgauge}
 H(x) = \left( \begin{array}{c} 0 \\ 
v_H + \frac{1}{\sqrt{2}}H_0(x) \end{array}\right),
\eeq
is the usual Higgs field in the unitary gauge, while for $\phi(x)$ we use a following, convinient parametrisation:
\beq
\label{phiCSM}
\phi(x) = \mathcal{U}_0e^{i\tilde{A}(x)}\left(v_{\phi} + \tilde{R}(x)\right)e^{i\tilde{A}(x)}\mathcal{U}_0^T,
\eeq
where $\tilde{A}_{ij}$ and $\tilde{R}_{ij}$ are real symmetric matrices. The trace part of 
\beq
G(x) = \mathcal{U}_0 \tilde{A}(x)\mathcal{U}_0^{\dagger},
\eeq
is (B-L) Goldstone boson (majoron) $\mathbf{a}(x)$, while the traceless part are those bosons which will be converted into pseudo-Goldstone bosons after the symmetry breaking. So we can write:
\beq
\begin{array}{lcr}
\tilde{A}_{ij}(x) = \frac{1}{2\sqrt{6}v_{\phi}}\mathbf{a}(x) \delij + A_{ij}(x), 
& &\mathrm{Tr}A(x) =0,
\end{array}
\eeq
and
\beq 
\label{ACSM}
A_{ij} = \frac{1}{v_{\phi}}G(x) = \frac{1}{4v_{\phi}}\sum_a G_a\lambda_{ij}^a,
\eeq
where the sum is over five symmetric Gell-Mann matrices, denoted as $\lambda^a$, $a\in \left\{1,2,\ldots,8\right\}$. They are traceless real $3\times3$ matrices. With the latter three Gell-Mann matrices with imaginary coefficients, they are generators of $\mathfrak{su}(3)$ algebra. They also satisfy the following property:
\beq
\mathrm{Tr}\left(\lambda_i\lambda_j\right)=2\delij.
\eeq
For $\tilde{R}_{ij}(x)$ we have the same parametrisation as for $\tilde{A}(x)$:
\beq
\label{RCSM}
\tilde{R}_{ij}(x) = \frac{1}{\sqrt{6}}r(x)\delij + \frac{1}{2}\sum_a R_a\lambda_{ij}^a(x).
\eeq
Below there are calculated some traces from CSM lagrangian (\ref{phiCSM}) in this given parametrisation :
\beq
\textrm{Tr}\phi \phi^{\ast} = \frac{1}{2}r^2(x) +\frac{1}{2} \sum_i R_i^2 + \sqrt{6}r v_{\phi} + 3 v_{\phi}^2, \\
\eeq
and:
\begin{align}
\textrm{Tr} [\partial_{\mu}\phi \partial^{\mu}\phi^{\ast}] = \textrm{Tr}[\partial_{\mu}R\partial^{\mu}R]  + 2 \textrm{Tr}\left[\partial_{\mu}\tilde{A}\cdot(v_{\phi}+\tilde{R})\cdot\partial^{\mu}\tilde{A}\cdot(v_{\phi} +\tilde{R})\right] \nonumber \\
 +2 \textrm{Tr}\left[\partial_{\mu}\tilde{A}\cdot\partial^{\mu}\tilde{A}\cdot(v_{\phi} +\tilde{R})^2\right].
\end{align}
If we consider terms only up to second powers in fields in the lagrangian (\ref{CSMlagrscalar}), to find mass modes, we obtain that, the $R_a$ modes are already mass eigenstates with eigenvalues:
\beq
M_R^2 = 4\lambda_4 v_{\phi}^2,
\eeq
and the field $r$ mixes with $H_0$ by the mixing matrix:
\beq
\label{massmatrix}
M^2 = \left(\begin{array}{cc} 4\lambda_1 v_H^2& 4\sqrt{3} v_Hv_{\phi} \\
								4\sqrt{3}v_Hv_{\phi} & 4(3\lambda_2 + \lambda_4)v_{\phi}^2\end{array}\right),
\eeq
for which we have following mass eigenstates $h_0$ and $h'$ given by the relation:
\beq
\label{massstate}
\begin{array}{lcr}
h_0 = \cos \beta H_0 + \sin \beta r, & & h' = -\sin \beta H_0 + \cos \beta r,
\end{array}
\eeq
with mixing angle $\beta$. We identify the lighter of these as observed Higgs boson with $M_{h_0} \approx 125$ GeV. Mixing will provide a second particle with the same decay channels as SM Higgs, but depending on the mass scale there could be a possibility of additional decay possibilities. 
\subsubsection{Further extended scalar sector}
\label{Further}
\setcounter{equation}{0}
In addition to complex scalar sextet, one can introduce a complex scalar triplet $\zeta_i$ (in original paper $\xi_i$, \cite{Latosinski2015}) transforming under $SU(3)_N$ as a $\mathbf{3}$ vector. The triplet is not only not coupled to SM particles (besides Higgs), but also insisting on renormalizability it mustn't couple to right-chiral neutrinos. So it could be even more sterile. Then the general renormalizable and $U(3)$-invariant potential is:
\begin{align}
\nonumber V(H, \phi, \zeta) = m_1^2 H^{\dagger}H + m_2^2 \mathrm{Tr}(\phi\phi^{\ast}) + m_3^2 \zeta^{\dagger}\zeta + (m_4\zeta^{\dagger}\phi\zeta^{\ast} + \mathrm{h.c.}) \\
\nonumber+ \lambda_1 (H^{\dagger}H)^2 + 2\lambda_3(H^{\dagger}H)\mathrm{Tr}(\phi\phi^{\ast}) + \lambda_2 [\mathrm{Tr}(\phi\phi^{\ast})]^2 + \lambda_4 \mathrm{Tr}(\phi\phi^{\ast}\phi\phi^{\ast})\\
+ \lambda_5 \zeta^{\dagger} \phi\phi^{\ast} \zeta + 2\lambda_6 H^{\dagger}H \zeta^{\dagger}\zeta + 2\lambda_7 \zeta^{\dagger}\zeta \mathrm{Tr}(\phi\phi^{\ast}) + \lambda_8 (\zeta^{\dagger}\zeta)^2,
\end{align}
with all constants being real, except for $m_4$. The potential is invariant under the following transformation:
\beq
\begin{array}{lcccccr}
\phi(x) &\to& U\phi(x)U^T,& \zeta(x)&\to&U\zeta(x), &U \in U(3).\\
\end{array}
\eeq
 With addition of $\zeta$, the expectation values of $\langle \phi_{ij} \rangle$ can be arranged in such a way that they are not proportional to the unit matrix:
\beq
\begin{array}{lcr}
\label{vacuumzeta}
\langle \zeta \rangle = \mathcal{U}_0 \left(\begin{array}{c} 0 \\ 0 \\ e^{i\alpha} v_{\zeta} \end{array}\right), & \langle H \rangle = \left( \begin{array}{c} 0 \\ v_H \end{array} \right), & \langle \phi \rangle = \mathcal{U}_0 \left( \begin{array}{ccc} v_1 & 0 & 0 \\ 0& v_1 & 0 \\ 0 & 0 & v_2 \\ \end{array} \right)\mathcal{U}_0^T,
\end{array}
\eeq
where $v_{\zeta}, v_H, v_1 v_2, (v_1 \neq v_2)$, and $\mathcal{U}_0$ are of the same origin as in the unextended scalar sector. And $\alpha$ originates from arg($m_4$). Since symmetry of (\ref{vacuumzeta}) is now $SO(2)$, then this time Goldstone bosons manifold is: 
\beq
\mathcal{M} = U(3)/ SO(2),
\eeq
with eight (pseudo-)Goldstone bosons. And $\zeta$ can parametrized as:
\beq
\label{unitaryzeta}
\zeta(x) = \mathcal{U}_0 e^{iA(x)} \tilde{\zeta}(x),
\eeq
where $A(x)$ is the same as in paragraph \ref{CSMscalarsector} and $\phi$ is parametrised as previously. The vacuum structure was analysed in the original article.
\subsection{Fermionic sector}
\label{Fermionic}
Since the right-chiral neutrinos are included to the model, it posesses 48 fundamental spin-$\frac{1}{2}$ particles, 16 fermions per family, where there are 3 families of particles with different masses, see \cite{PhysRevD.91.065029} for discussion the relation to supergravity. Since there is no experimental hints for additional fermions then any extra fermionic degrees of freedom should be heavy superpartners of the SM bosons or should be (almost) sterile. That's why the fermionic sector is supplemented only by 3 right-chiral neutrinos. Proposed Yukawa coupling in Conformal Standard Model reads:
\begin{align}
\mathcal{L}_Y = & - \left\{ Y_{ij}^EH^{\dagger}L^{i\alpha} E^{j}_{\alpha} + Y_{ij}^D H^{\dagger}Q^{i\alpha}D^{j}_{\alpha} + Y_{ij}^U H^T\varepsilon Q^{i\alpha} U^j_{\alpha} \right. \\
 &\left. + Y^{\nu}_{ij}H^T\varepsilon L^{i\alpha}N^j_{\alpha} + \frac{1}{2}y_M \phi_{ij} N^{i\alpha}N^j_{\alpha} \right\} + \textrm{h.c.},
\end{align}
where $\bar{N}^{i\alpha}$ are right-chiral neutrinos. The $i,j = 1,2,3$ are the family indices, while $\alpha$'s are $SL(2,\mathbb{C})$ indices, the authors uses slightly different notation than we used in previous chapter, see \cite{Latosinski2015,Latosinski2013}. First three terms are exactly the Yukawa couplings from SM. Right-chiral neutrinos transform under $SU(3)_N$ as:
\beq
N^i(x) \to (U^{\ast})^i_jN^j(x),
\eeq
where all other fermions are inert under this symmetry. Moreover, expectation values of sterile scalar field serve as \emph{effective} couplings of right-chiral neutrinos, which is not the case for other SM fermions. The $SU(3)_N$ is actually broken by the interaction term:
\beq
\label{Yukawa}
\mathcal{L}_Y' =  Y^{\nu}_{ij}H^T\varepsilon L^{i\alpha}N^j_{\alpha},
\eeq
which is the only interaction between the right-chiral neutrinos and SM particles. The neutrino masses emerge from spontaneous symmetry breaking, and are respectively: for light neutrinos masses as: 0.01eV, and for the heavy ones as: 1TeV.  Only if we assume that $Y^{\nu}_{ij}$ are very small, of order $\mathcal{O}(10^{-6})$, the smallness of light neutrino masses can be explained by see-saw mechanism, see \cite{Latosinski2015,Latosinski2013} and references therein. 
\subsection{Pseudo-goldstone bosons}
The Yukawa term (\ref{Yukawa}) gives rise not only to neutrino masses, but also via radiative corrections to five $A_i$ particles, which are converted to pseudo-Goldstone boson while the trace part $\mathbf{a}(x)$ remain massless. If we redefine right-chiral neutrino spinors to eliminate them from Yukawa interactions we obtain:
\beq
N_{\alpha}^i = \left(\mathcal{U}_0^{\ast}e^{-i\tilde{A}(x)}\right)\mathcal{U}_0^T\tilde{N}^j_{\alpha}(x),
\eeq
where $\mathcal{U}_0$ are taken such to ensure $\langle A \rangle = 0$ of one-loop effective potential. For remaining SM particles redefinition concern only the Majoron: $\mathbf{a}(x)$. Then the Goldstone bosons appear only via derivative couplings of the type: $\partial_{\mu}A\bar{f}\gamma^{\mu}\frac{1+\gamma^5}{2}f$ and in interaction term:
\beq
\mathcal{L}_Y' = -v_H\left(Y^{\nu}U_0^{\ast}e^{-i\tilde{A}(x)}\mathcal{U}_0^T\right)_{ij}\tilde{\nu}^{i\alpha}\tilde{N}^j_{\alpha} + \mathrm{h.c.},
\eeq
then the Majoron disappears from the interaction term and is massless. The authors analysed the renormalisibity of this interaction and investigated vanishing of logarithmic and quadratic divergences. The approximate $SU(3)_N$ symmetry ensures smallness of pseudo-Goldstone bosons masses. It also turns out that these particle are dark matter candidates, since they are stable and the effective coupling is extremely small - is of order $10^{-24}$ GeV.
\section{Conformal Standard Model Higgs as Inflaton}
Inflation scenario requires a scalar field to drive it. We can assume a fictitious, additional to the ones we know from particle physics, scalar field to do the job, but this will cause a lot of problems: how to quantise this field, how will it couple to other fields from SM (reheating), what other properties should it have. So far we recognised only one fundamental scalar field, namely the Higgs field. Then arises a question whether Higgs can serve as Inflaton. The answer is yes, moreover the spectral index and tensor perturbations amplitude for SM are in good agreement with experiment and these parameters are in 1$\sigma$ correspondence to WMAP-3 data \cite{Bezrukov}. \\
Also the extensions of Standard Model, which possess enlarged scalar sector give an opportunity to build inflation models with more than one inflaton. These proposals with two scalar fields non-minimally coupled to gravity have been discussed in recent literature, see \cite{Avgoustidis:2012yc,Two-Higgs-doublet}. In most of the cases the heavier field drops quickly to minimum leaving us with one field inflation a la Bezrukov-Shaposhnikov with effective constants. Decoupling of the heavier field can be understood by calculations in the effective field theory \cite{Avgoustidis:2012yc}. 
However, one of the drawbacks of non-minimal inflation, as it was pointed out by Burgess, Lee, Trott and Lebedev, \cite{Burgess:2010zq,Lebedev:2011aq}, is that these models suffer from problems with maintaining unitarity for large field values. To fix it, the further, unitarisation extension is required.\\
In most of the models inflaton(s) particle(s) aren't specified explicitly. Then these extension(s) of SM, which resolve the high energy particle physics problems and can give successful inflation, are of high interest. There are actually two models on the table satisfying ``minimal'' extension property and which have enlarged scalar sector. One of them is $\nu$MSM model \cite{Shaposhnikov:2006xi} proposed by Shaposnikov and collaborators. The CSM (in both old and new version) enlarges the model without incorporating low-scale supersymmetry and both of them add only three additional neutrinos to the fermionic sector and extend scalar sector property. \\
This chapter is organised in the following way. At first we will discuss the original Bezrukov-Shaposhnikov article. Then we will analyse inflation within CSM. In the following sections we will propose a natural unitarisation procedure, discuss various choices of couplings and derive a simple argument for decoupling of heavier inflation field. Since within whole chapter we will use extensively conformal transformations and notion of Jordan and Einstein frame, the relation between those two frames and quantities calculated in both of them are discussed in Appendices \ref{ConformalTrans}, \ref{JordanEinstein}.
\subsection{Higgs particle as Inflaton}
\setcounter{equation}{0}
\label{Higgsinflation}
\label{HiggsInflaton}
In this paragraph we will follow the steps described in \cite{Bezrukov}, one can also look into \cite{Higgscritical} and references therein for more details concerning pure Higgs inflation. 
Let us start with Standard Model Lagrangian with non-minimal coupling to gravity:
\beq
\label{Nonminimalcoupling}
\mathcal{L} = L_{SM} -\frac{M^2}{2}R - \xi H^{\dagger}H R,
\eeq
where $L_{SM}$ is Standard Model Lagrangian, $M$ is some mass parameter, $H$ is Higgs field and $\xi$ is the coupling to gravity. The non-minimal coupling is a key assumption, not-coupled Higgs cannot drive successful inflation and match the parameters which was discussed in section \ref{Experimental}. The authors argue that there should be some ``good'', non-zero choice of $\xi$ and $M$ such that they fit inflation scenario and particle physics. According to \cite{Bezrukov} we will consider only $\xi$ such that: $1\ll \sqrt{\xi} \lll 10^{17}$, since this will simplify formulas and in which: $M \simeq M_P$. Let us ignore gauge couplings and set unitary gauge: $H = h/\sqrt{2}e^{i\theta}$.  So we obtain the following action:
\beq
\label{NonminimalHiggsaction}
S_{H} = \int d^4x\sqrt{|g|}\left[-\frac{M_P^2+\xi h^2}{2}R + \frac{\partial_{\mu}h\partial^{\mu}h}{2}+\frac{h^2\partial_{\mu}\theta \partial^{\mu}\theta}{2} - \frac{\lambda}{4}\left(h^2 -v^2\right)^2\right],
\eeq
where $\theta$ is massless Goldstone boson and can be integrated out since it is not coupled to gravity. This action is well known and studied in literature eg. \cite{Salopek,Kaiser, Bezrukov}. Action (\ref{NonminimalHiggsaction}) is written in Jordan frame, to perform analysis we will change the frame to the one, where $R$ is canonically normalised. To obtain this we will use a conformal factor:
\beq
\label{Jordan frame}
\Omega^2 = 1 +\frac{\xi h^2}{M_P^2},
\eeq
so the transformed metric is:
\beq
\label{JordanEinstein}
g_{E\mu\nu} = \Omega^2 g_{J\mu\nu}.
\eeq
Moreover,, if we use a convenient new scalar field (\ref{hatphi}):
\beq
\label{NewScalar}
\frac{d\chi}{dh} = \sqrt{\frac{\Omega^2 +6\xi^2h^2/M_P^2}{\Omega^4}},
\eeq
we arrive at the action in the Einstein frame:
\beq
\label{inflationEinstein}
S = \int d^4x\sqrt{g_E}\left[-\frac{M^2_P}{2} R_E + \frac{\partial_{\mu}\chi \partial^{\mu}\chi}{2} - U(\chi)\right],
\eeq
where:
\beq
\label{NewPotential}
U(\chi) =\frac{1}{\Omega^4}\frac{\lambda}{4}\left(h(\chi)^2 - v^2\right)^2.
\eeq
For small field values: $h \simeq \chi$ and $\Omega^2 \simeq 1$ for both fields potential has the same initial values. However, it is not so for $h \gg M_P/\sqrt{\xi}$ (or $\chi \gg \sqrt{6}M_P$). In this limit one can solve (\ref{NewScalar}) and get:
\beq
\label{HighNewScalar}
h \simeq \frac{M_P}{\sqrt{\xi}}\exp\left(\frac{\chi}{\sqrt{6}M_P}\right),
\eeq
from (\ref{NewPotential}, \ref{HighNewScalar}) we obtain exponentially flat potential:
\beq
\label{ExpPotential}
U(\chi) = \frac{\lambda M_P^4}{\sqrt{\xi}}\left(1+ \exp\left(-\frac{2\chi}{\sqrt{6}M_P}\right)\right)^{-2}.
\eeq
The plot of the potential is given below and is taken from the original work of Bezrukov and Shaposhnikov.
\FloatBarrier
\begin{figure}[h!]
\label{Uchipotential}
\includegraphics[scale=0.4]{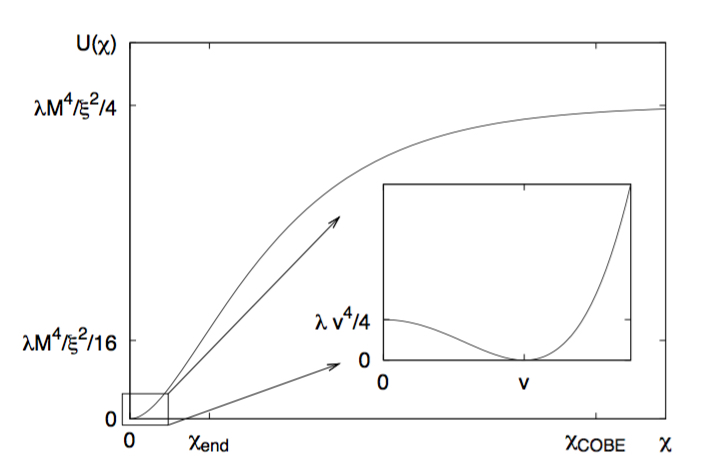}
\caption{Effective potential in the Einstein frame.}
\end{figure}
\FloatBarrier
\noindent We will analyse this potential using slow - roll approximation. We can calculate slow-roll parameters (\ref{Inflconditions}), in the limit of $h^2 \gg M^2_P/\xi \gg v^2$, as:
\begin{align}
\epsilon &= \frac{M^2_P}{2} \left(\frac{dU/d\chi}{U}\right)^2 \simeq \frac{4M^4_P}{3\xi^2h^4},\\
\eta &= M_P^2\frac{d^2U/d\chi^2}{U} \simeq \frac{4M_P^2}{3\xi h^2}.
\end{align}
Slow roll ends when $\epsilon \simeq 1$, so $h_{end}\simeq 1.07M_P/\sqrt{\xi}$. The number of e-foldings (\ref{numberofe-folds}) is given by the formula:
\beq
\label{Bezrukovefolds}
N = \int \frac{1}{M_P^2}\frac{U}{dU/dh}\left(\frac{d\chi}{dh}\right)^2 dh \simeq \frac{6}{8}\frac{h^2}{M_P^2/\xi}.
\eeq
For all values $\sqrt{\xi} \lll 10^{17}$, the $v$ parameter doesn't appear anywhere so inflation stage is not affected by its value. If one uses the relation between $\epsilon$, $\eta$ and $N$ and use (\ref{nsepsiloneta}, \ref{repsilon}), one obtains a familiar relation:
\begin{align}
n\simeq 1 - 2\eta \simeq 1 - 2/N \simeq 0.97,\\
 r =16 \epsilon \simeq 12/N^2 \simeq 0.0033,
\end{align}
where $N\simeq 60$. The authors argues \cite{Bezrukov} that their model agrees with inflation data, what we can see on Fig. 2 in this paper or on Fig. [\ref{Planckdata}]. Inserting into COBE normalisation \cite{Lyth:1998xn}: $U/\epsilon = (0.027M_P)^4$ and with $N_{COBE} \simeq 62$, we obtain that:
\beq
\xi \simeq \sqrt{\frac{\lambda}{3}}\frac{N_{COBE}}{0.027^2} \simeq 49000 \sqrt{\lambda},
\eeq
so for $\lambda \sim \mathcal{O}(1)$, $\xi \simeq 49000$ is the value for which Higgs scenario fit the data, as we have argued in paragraph concerning Starobinsky inflation \ref{Starobinskyinflation}.\\
The authors made a statement that if one could calculate this value and compare it with the data then it could provide connection between Higgs mass and amplitude of primordial perturbations. Despite the fact that Higgs-inflation can work, there is no theoretical explanation or observational evidence (besides inflation and CMB of course) for introducing such a coupling. Then this calculation would provide a theoretical explanation for this kind of scenarios. Also this statement sums up actual development of inflation scenarios that (almost) all inflation models share the property of being only effective models rather than fundamental theories. Moreover, one can speculate that giving an explanation of relation of $\xi$ and $m_H$ will be a huge step towards our understanding the Universe and quantum gravity. However, this problem seems to be unsolvable yet.\\
\subsection{Inflation scenario within Conformal Standard Model}
\label{CSMinflation}
In this and the two following paragraphs we will propose and discuss a similar scenario within Conformal Standard Model. To analyse non-minimally coupled inflation, within the CSM, we have to simplify it. We restrict ourselves only to interactions associated with $H_0(x)$ and $r(x)$ and their couplings to gravity as we usually do when concerning multi-inflation. However, one can argue that these assumptions are quite artificial even though frequently used \cite{PhysRevD.90.043505}. We will take only $r(x)$ not full $\textrm{Tr}(\phi\phi^{\ast})$ into account in the section \ref{Observables}. So, from our assumptions, only diagonal term of $R(x)$ remains. Then the kinetic term is:
\beq
\textrm{Tr} [\partial_{\mu}\phi \partial^{\mu}\phi^{\ast}] = \partial_{\mu}r\partial^{\mu}r +4r^2\partial_{\mu}\tilde{A}\partial^{\mu}\tilde{A},
\eeq
and as we can see the kinetic term associated with $\tilde{A}(x)$ decouples and will be no longer considered. For this assumption also another equality holds: 
$$
\textrm{Tr}[\phi\phi^{\ast}\phi\phi^{\ast}] = \frac{1}{3}\left(\textrm{Tr}\left[\phi\phi^{\ast}\right]\right)^2.
$$
Then the potential (\ref{CSMpotential}), using (\ref{Higgsgauge}, \ref{phiCSM}), is given by the formula:
\begin{align}
\label{VCSMI}
V_J(H,\phi) = \frac{1}{4}\lambda_1(H_0^2-v_H^2)^2 + \frac{1}{4}\lambda_p(r^2-v_{\phi}^2)^2\nonumber\\
 + \frac{1}{2}\lambda_3\left(H_0^2-v_H^2\right)\left(r^2-v_{\phi}^2\right),
\end{align}
with $\lambda_p = \lambda_2+ \lambda_4/3$. The unbounded from below conditions for this potential were discussed in section \ref{ConformalSM}. From now on we will use the following notation: $h := H_0(x)$ and $s := r(x)$. We proposed that the inflation comes from non-minimally coupled to gravity lagrangian which in Jordan frame reads:
\beq
\mathcal{L} = \frac{1}{2}\partial_{\mu}h\partial^{\mu}h + \frac{1}{2}\partial_{\mu}s \partial^{\mu}s - \frac{\left(M_P^2 +\xi_1 h^2 +  \xi_2s^2\right)}{2} R- V_J(h, s),
\eeq
with $\xi_i >0$. We will proceed in the scheme of \cite{Two-Higgs-doublet,Lebedev:2011aq}. First, to change the frame to the Einstein one we make the following conformal transformation:
\beq
\label{CSMcoupling}
\begin{array}{lcr}
\tilde{g}_{\mu\nu}= \Omega^2 g_{\mu\nu}, & & \Omega^2 = 1 + \frac{\xi_1 h^2 + \xi_2s^2}{ M_P^2}.
\end{array}
\eeq
We set $M_P=1$ to simplify the equations but it will be restored later on for slow-roll parameters analysis. Then from (\ref{Two-field}) we obtain:
\beq
\label{LagrangianE}
\mathcal{L}_{\textrm{E}} = -\frac{R}{2} +\frac{3}{4}\left[\partial_{\mu}\log(\Omega^2)\right]^2 +\frac{1}{2\Omega^2}\left[(\partial_{\mu}h)^2 + (\partial_{\mu}s)^2\right] - \frac{1}{\Omega^4}V(H_0,r).
\eeq
Since for inflation scenario we consider large fields limit, which is notabene necessary to recover canonical kinetic terms and is discussed in appendix \ref{Twofieldanalysis},  we take:
\beq
\xi_1h^2 + \xi_2s^2 \gg M_P^2 \gg v_i^2,
\eeq
If we redefine the fields as: 
\begin{align}
\chi =& \sqrt{\frac{3}{2}}\log(\xi_1 h^2 + \xi_2s^2), \\
\tau =& \frac{h}{s},
\end{align}
then the kinetic part of the lagrangian reads:
\begin{align}
\mathcal{L}_{\textrm{kin}} &= \frac{1}{2}\left(1 + \frac{1}{6}\frac{\tau^2 +1}{\xi_1\tau^2 + \xi_2}\right)(\partial_{\mu}\chi)^2 + \frac{1}{\sqrt{6}}\frac{(\xi_2 - \xi_1)\tau}{(\xi_1\tau^2 +\xi_2)^2}(\partial_{\mu}\chi)(\partial^{\mu}\tau) \\
&+ \frac{1}{2}\frac{\xi_1^2\tau^2 + \xi_2^2}{(\xi_1\tau^2 + \xi_2)^3}(\partial_{\mu}\tau)^2.
\end{align} 
We are interested in large fields and coupling regime: $\xi' = \xi_1 + \xi_2 \gg 1$. 
This is the same case as for single Higgs inflation, where $\xi \approx 49000\sqrt{\lambda}$. Then the mixing term $(\partial_{\mu}\chi)(\partial^{\mu}\tau)$ and the second term in front of $(\partial_{\mu}\chi)^2$ are suppressed by term $1/(\xi')$ so the kinetic part is:
\beq
\mathcal{L}_{\textrm{kin}} \simeq \frac{1}{2}(\partial_{\mu}\chi)^2 + \frac{1}{2}\frac{\xi_1^2\tau^2 +\xi_2^2}{(\xi_1\tau^2 + \xi_2)^3} (\partial_{\mu}\tau)^2,
\eeq  
and the potential in new variables reads:
\beq
\label{VECSM}
V_E(\tau, \chi)= U(\tau)W(\chi) = \frac{\lambda_1\tau^4 + \lambda_p+ 2\lambda_3\tau^2}{4(\xi_1\tau^2 + \xi_2)^2}\left(1 + e^{-2\chi/\sqrt{6}}\right)^{-2}.
\eeq
The calculated  minima of $U(\tau)$ are shown in the table below, with $a = \lambda_1 \xi_2 - \lambda_3 \xi_1$, $b = \lambda_p\xi_1 - \lambda_3\xi_2$. In the section \ref{taubehaviour} we will show that the ratio of fields drops eventually to stable minimum, before the inflation ends and Shaposhnikov-type evolution of $\chi$ is described in the paragraph \ref{ParametersCSM}.
\FloatBarrier
\begin{table}[h!]
\label{MinimaU}
\caption{Minimal values of the radial part of inflation potential}
\begin{tabular}{c|c|c}
$\tau_0$ values & stable minimum condition & $U_0$ \\
\hline
$\tau_0 =0$ & $a>0$ and $b<0$ & $\frac{\lambda_1}{4\xi_1^2}$, \\
$\tau_0 = + \infty$ & $a<0$ and $b>0$ & $\frac{\lambda_p}{4\xi_2^2}$, \\
$\tau_0 = \pm \sqrt{\frac{b}{a}}$ & $a>0$ and $b>0$ & $\frac{\lambda_1 \lambda_p - \lambda_3^2}{4(\lambda_1\xi_2^2 + \lambda_p \xi_1^2 - 2\lambda_3 \xi_1\xi_2)}$,\\
$\tau=0$ or $\tau_0= +\infty$ & $a<0$ and $b<0$ & $\frac{\lambda_1}{4\xi_1^2}$ or $\frac{\lambda_p}{4\xi_2^2}$.
\end{tabular}
\end{table}
\FloatBarrier
\noindent Then we have two types of scenarios. Either we have single Inflaton case: Higgs or single ``shadow'' Higgs inflation, when $\tau_0$ obtains zero or infinity value. Or we have multi-inflaton scenario, where ratio of fields goes to the value: $\tau_0 = \sqrt{\frac{b}{a}}$. Single Higgs scenarios we will describe briefly below, here we will addresss scenario when $a,b>0$. To have desired values of parameters, the following conditions has to be satisfied: 
\begin{align}
a = \lambda_1 \xi_2 - \lambda_3 \xi_1>0 \label{positiveI},\\
b = \lambda_p\xi_1 - \lambda_3\xi_2 > 0 \label{positiveII},\\
\lambda_1\lambda_p - \lambda_3^2 >0  \label{positiveIII},
\end{align}
\noindent
where the third one is required to prevent from metastability of electroweak vacuum and for a positivity of vacuum energy during inflation stage. If look at constraints for parametrisation (\ref{CSMconditions}) and constraints for couplings given by CSM (\ref{first}), the last condition is satisfied due to fact, that $V_J$ has to be unbounded from below. For $\lambda_3<0$ both: $a>0$ and $b>0$ are obviously satisfied. On the other hand the last condition is automatically satisfied for $\lambda_3>0$ since it comes from the first two.  Moreover, the choice $\lambda_3<0$ is more convinient to predict the theoretical Higgs mass the same as its observed (125 GeV) value. \\
So for the these choice of parameters (\ref{first}) the CSM model is consistent with the inflation scenario. The measurement of $\lambda_3$ is crucial to determine which of the final values of $\tau_0$ could be obtained in inflation. Assuming that the shadow Higgs would be found in LHC and the measured value of $\lambda_3$ will be less than zero then single Higgs scenario presented in \cite{Bezrukov} will be immediately falsified, even for $\xi_1 =0$ or $\xi_2 =0$. Moreover, for $\lambda_3>0$ single or mixed Higgs scenario can be realised in CSM with non-minimal couplings. So inflation in CSM can contain these two types of scenarios.
\subsection{Analysis of $U(\tau)$ potential}
\label{taubehaviour}
We would like to show that $\tau$ drops quickly to the minima so it can be integrated out while $W(\chi)$ is almost constant. The are many arguments advocating this thesis presented in the literature, see for example \cite{Avgoustidis:2012yc, Lebedev:2011aq} or \cite{Two-Higgs-doublet}, but some of them seems to be a bit heuristic rather than rely on the simple calculation presented below which can easily extended. To prove it we start with the action:
\beq
S = \int d^4x\sqrt{|g|}\left[\frac{1}{2}\partial_{\mu}\chi\partial^{\mu}\chi + \frac{1}{2}\frac{\xi_1^2\tau^2 +\xi_2^2}{(\xi_1\tau^2 + \xi_2)^3} (\partial_{\mu}\tau)(\partial^{\mu}\tau) - V_E\right],
\eeq
If we assume that: $\chi (t,\vec{x}) = \chi(t)$ and $\tau(t,\vec{x}) = \tau(t)$ and the metric is in the FLRW form, we obtain the following action:
\beq
S = \int dt a^3\left[\frac{1}{2}(\partial_t \chi)^2 + \frac{1}{2}b(\tau)(\partial_t \tau)^2 - V(\tau,\chi)\right],
\eeq
where $B(\tau) =\frac{\xi_1^2\tau^2 +\xi_2^2}{(\xi_1\tau^2 + \xi_2)^3}$. 
The Euler-Lagrange equations are:
\beq
\frac{\partial\mathcal{L}}{\partial \phi_i} - \partial_{\mu} \left(\frac{\partial \mathcal{L}}{\partial \partial_{\mu}\phi_i
}\right)=0,
\eeq
so we obtain equations for $\tau$, $\chi$ supplemented by Friedmann equation for $H$:
\begin{align}
\ddot{\tau} + 3H\dot{\tau} &= -\frac{1}{2} \frac{\partial \ln B(\tau)}{\partial \tau}\dot{\tau}^2 - \frac{1}{B(\tau)} \frac{\partial V}{\partial \tau},  \\
\ddot{\chi} +3H\dot{\chi} &= -\frac{\partial V}{\partial \chi}, \\
H^2 &= \frac{1}{6} \left(\dot{\chi}^2 + B(\tau)(\dot{\tau})^2 + 2V(\tau,\chi)\right),
\end{align}
We used the fact that $\dot{B}(\tau) = \partial_{\tau}B(\tau)\dot{\tau}$ so
after applying slow-roll conditions we obtain:
\begin{align}
 \dot{\tau} &=  - \frac{1}{B(\tau)} \frac{\partial \ln V}{\partial \tau}\sqrt{V}, \\
\dot{\chi} &= -\frac{\partial \ln V}{\partial \chi}\sqrt{V}, \\
3H^2 &=  V(\tau,\chi).
\end{align}
If we use the fact that for large values of $\chi$: $V(\tau,\chi) \simeq \frac{1}{4}U(\tau)$ we get:
\beq
\dot{\tau} = - \frac{\left(\xi_1 \tau^2 + \xi_1\right)\left[-b\tau + a\tau^3\right]}{\left(\xi_1^2\tau^2 + \xi_2^2\right)\left(\lambda_p + 2\lambda_3 \tau^2 + \lambda_1 \tau^4\right)^{1/2}},
\eeq
with $\chi \simeq \textrm{const}$. In the first order approximation: $\left(\lambda_p + 2\lambda_3 \tau^2 + \lambda_1 \tau^4\right)^{1/2} \simeq \sqrt{\lambda_1}\tau^2$ and after integration we obtain:
\beq
-t +C = \frac{\sqrt{\lambda_1}\ln\left(-b +a\tau^2\right)}{2a},
\eeq
where $C$ is constant of integration. Then we get:
\beq
\tau^2 = \frac{b}{a} + \textrm{e}^{-2at/\sqrt{\lambda_1}} \cdot \textrm{e}^{2aC/\sqrt{\lambda_1}},
\eeq
If we take $\lambda_3 <0$ then $a \gg \sqrt{\lambda_1}$ and we have quick decay to the value $b/a$. However, also for $\lambda_3 >0$ we have decay but the rate of decay cannot be estimated without knowing specific values of the parameters. And in the second order approximation:  
$$\left(\lambda_p + 2\lambda_3 \tau^2 + \lambda_1 \tau^4\right)^{1/2} \simeq \sqrt{\lambda_1}\tau^2\left(1 + \frac{\lambda_3}{\lambda_1\tau^2}\right),$$ 
so there is no explicit formula for $\tau(t)$:
\beq
-t + C =  \frac{(a+\lambda_1b)\ln\left(-b +a\tau^2\right)-2a\ln(x)}{2ab\sqrt{g}},
\eeq
However, for $b\gg 1$ and $\lambda_i \sim \mathcal{O}(1)$ we can drop the $-\ln(x)/(b\sqrt{g})$ term and obtain the same asymptotic behaviour. One can ask question, why we can use slow-roll conditions while we observe that changes of $\tau$ are rapid? It turns out that if one plugs this solution into the full equation then it can be seen that the deviation vanishes exponentially. This comes from the fact that K-G equations possess an attractor solution \cite{Senatore} because $U(\tau)$ has only one stable minimum. Then $\tau$  behaves in such a way that slow-roll conditions are satisfied. Since then locally all the solutions converge to $\tau_s$, when $s$ denotes slow roll solution which is an attractor in the space on solutions. Because $\tau$ exponentially fast drops to potential minima, one can think about it as proto-inflation step which is needed to solve the initial conditions  problems indicated in \cite{Dimopoulos:2016yep}. It has to be checked in detail whether this is the case.
\subsection{$\chi$ behaviour, parameters and observables from CSM inflation}
\label{ParametersCSM}
Since the heavy state decouples we are left with classical Bezrukov-Shaposhnikov evolution:
\beq
V(\chi) = \frac{\lambda_{eff}}{4\xi^2}W(\chi),
\eeq
with $\xi = \xi_1\tau_0^2 + \xi_2$ and $\lambda_{eff} = \lambda_1 + \lambda_p\tau_0^4 + 2\lambda_3\tau_0^2$. This potential was analysed in section \ref{Higgsinflation} and has the same shape as in Fig. [\ref{Uchipotential}]. Here we recall the general scheme. For large $\chi$ potential is flat and the inflation occurs. As the field rolls to smaller values the $\epsilon \simeq 1$ gives the end of inflation where the slow roll parameters are:
\begin{align}
\epsilon &= \frac{M_P^2}{2}\left(\frac{dW/d\chi}{W}\right)^2 \simeq \frac{4M_P^2}{3}\frac{e^{-4\chi/\sqrt{6}}}{\left(1+e^{-2\chi/\sqrt{6}}\right)^2},\\
\eta &= M_P^2 \frac{d^2W/d^2 \chi}{U} \simeq - \frac{4M_P^2}{3}e^{-2\chi/\sqrt{6}}\frac{1 - 2e^{-2\chi/\sqrt{6}}}{\left(1 + e^{-2\chi/\sqrt{6}}  \right)^2}.
\end{align}
The number of e-folds is given by (\ref{Numberefolds}, \ref{numberofe-folds}):
\beq
N = \int_e^i \frac{V(\chi)}{V'(\chi)}d\chi = \frac{3}{4}\left[ e^{2\chi_i/\sqrt{6}} - e^{2\chi_e/\sqrt{6}} + \frac{2}{\sqrt{6}}\left(\chi_i - \chi_e\right)\right],
\eeq
and the slow roll conditions are violated for $e^{2\chi_e/\sqrt{6}} \simeq 0.155$. Then the initial value of the field, for $N=60$, is given by: $e^{2\chi_e/\sqrt{6}} \simeq 80$. Hence the initial values of fields, after the decoupling stage, are given by:
\begin{align}
s_i & \simeq M_P \sqrt{\frac{4N}{3\xi}}, \\
h_i & \simeq M_P \sqrt{\frac{4N}{3\xi_1}}\sqrt{\left[1 - \frac{\xi_2}{\xi}\right]}.
\end{align}
Inserting it into COBE normalisation \cite{Lyth:1998xn}: $W/\epsilon = (0.027M_P)^4$ and with $N_{COBE} \simeq 62$, we obtain that (in analogy to Bezrukov-Shaposhnikov):
\beq
\xi \simeq \sqrt{\frac{\lambda_{eff}}{3}}\frac{N_{COBE}}{0.027^2} \simeq 49000 \sqrt{\lambda_{eff}},
\eeq
for $\lambda_{eff} \sim 1$ the coupling to gravity is: $\xi \simeq 49000 = \xi_1 \tau_0^2+ \xi_2$. Since $\tau^2 \sim v_H^2/v_{\phi}^2 = \mathcal{O}(1)$, then roughly $\xi' \simeq \xi$. The spectral tilt is (since it depends on the shape of the potential rather then its amplitude and can be expressed by $\epsilon$ and $\eta$, as we have shown in section dedicated to inhomogeneities \ref{Inhomogeneouslimit}):
\beq
n_s \simeq 1 - \frac{2}{N} \simeq 0.97,
\eeq
with tensor to scalar ratio: $r \simeq 12/N^2 \simeq 0.0033.$ Hence Conformal Standard Model with two non-coupled to gravity scalars can provide a successful inflation scenario which fits the data. Not only we have showed the parameters required to provide successful inflation scenario are in agreement with CSM predictions but moreover quantum effects of running coupling constants were addressed for both two scalar inflatons  and CSM in details, see \cite{Two-Higgs-doublet,Lebedev:2011aq} and for CSM \cite{CSMtwo,Latosinski2015}. With present measurements we find it impossible to distinguish between Starobinsky model and its descendants by measurement of $r$ and $n_s$ only. Even natural inflation discussed in  the outline \ref{Extensions} and \cite{PhysRevD.70.083512} which relies on different mechanism is hard to be falsified.\\
The scale on which the new physics should appear (unitary violation) is the $\mu_{U} \sim M_P^2/{\xi}'$, as we will show in the next paragraph. Since the particle physics couplings haven't been specified yet we cannot be sure whether unitarisation procedure is required for $h$ field. The $\phi$ sextet is sterile so there might be a suitable sort of parameters for which $h$ interactions don't break the unitary evolution, see \cite{Lee:2013nv}. On the other hand, if it is not the case we present below that addition of one sterile scalar is sufficient. Moreover in CSM there is such a natural extension, which was presented in paragraph \ref{Further}.
\subsection{Violation of unitary evolution} 
Now we will address the unitary violation and unitarisation at scale $M_P/\xi'$, in the manner of \cite{Lebedev:2011aq}. As a setup let us take initial field values $|h|\ll M_p/\xi_{1}$ and $|s| \ll M_P/\xi_2$, then up to leading order the kinetic term of lagrangian in Jordan frame (\ref{LagrangianE}) is:
\beq
\mathcal{L}_{\mathrm{kin}} \simeq \frac{1}{2}(1+ 6\xi_2^2s^2)(\partial_{\mu}s)^2 + \frac{1}{2}(1+ 6\xi_1^2h^2)(\partial_{\mu}h)^2, 
\eeq
and the canonically normalized variables are:
\beq
\begin{array}{clr}
\rho = s(1 + \xi_2^2s^2), & & \varphi=h(1+\xi_1^2h^2).
\end{array}
\eeq
Now we would like to make an expansion around the expectation values:
\beq
\begin{array}{clr}
\rho = \rho_0 + \bar{\rho}, & & \varphi=\varphi_0 + \bar{\varphi},
\end{array}
\eeq
and the same for $s= s_0 + \bar{s}$ and $h = h_0 + \bar{h}$. The fluctuations are related by: $\bar{s} \simeq (1-3\xi_2s_0^2)\bar{\rho} -3\xi_2^2s_0\bar{\rho}^2$ and $\bar{h} \simeq (1-3\xi_2h_0^2)\bar{\varphi} -3\xi_2^2h_0\bar{\varphi}^2$. Now we will consider the interaction of Higgs with the gauge bosons (\ref{HiggsWmu}):
\beq
\mathcal{L}_{\textrm{gauge}} = \frac{1}{2}g^2h^2W_{\mu}^+W^{\mu-},
\eeq
where we used Standard Model $W^{\pm}_{\mu}$ fields, and $g$ is the coupling. Since the terms of order $\xi_1h^2$ and $\xi_2s^2$, compared to $\xi^2_1h^2$ and $\xi^2_2s^2$, are negligible in this case for canonical variables we obtain:
\beq
\mathcal{L}_{\textrm{gauge}} = \frac{1}{2}g^2\varphi_0^2\left(1 + 2a\frac{\bar{\varphi}}{\varphi_0} + b\frac{\bar{\varphi}^2}{\varphi_0^2}\right)W_{\mu}^+W^{\mu-},
\eeq
with $a=1-3\xi_1\varphi^2$ and $b=1-12\xi_1\varphi^2$, where we also put: $h_0 \approx \varphi_0$. Since the normalisation of the $WW$ scattering is non-canonical, one can show that the amplitude grows with energy: $\mathcal{A}(WW \to WW) \sim \xi_1^2E^2$ and unitarity is broken at $M_P/\xi_1$ energy scale \cite{ Burgess:2010zq,Lebedev:2011aq,Lee:2013nv}. The unitarity is also spoiled in the scalar interaction by 6-point interactions at the same scale. 
\subsubsection{$\zeta$ field unitarization}
Since the CSM model shouldn't provide any new mass scale up to Planck scale one can introduce a new heavy scalar particle to restore unitary evolution. In CSM this is quite natural, as it was pointed in Appendix A of \cite{Latosinski2015} and many times within this thesis. If one takes a particular lagrangian in Jordan frame:
\beq
\mathcal{L}_J = -\frac{1}{2} (M_P^2+\xi_{3}\sigma^2 + \tilde{\xi}_1 h^2 + \tilde{\xi}_2 s^2)R + \mathcal{L}_{\mathrm{kin}} - \frac{1}{4}\kappa(\sigma^2 - \Lambda^2 - \alpha h^2 - \beta s^2)^2 - V_J(h,s),
\eeq
with $\Lambda = 1/\sqrt{\xi_3}$ and $\sigma \equiv \tilde{\zeta}$, where we took the analog of unitary parametrisation (\ref{unitaryzeta}) for $\zeta$. If we take $\tilde{\xi}_1,\tilde{\xi}_2\ll \xi_3$; $\Lambda \gg v_i$ then in low energy limit the $\sigma$-field can be integrated out, by minimalizing the potential:
\beq
\sigma^2 = \Lambda^2 + \alpha h^2 + \beta s^2,
\eeq
and the effective action is the one we have inspected before with effective couplings: $\xi_1 = \tilde{\xi}_1 + \alpha \xi_{3}$, $\xi_2 = \tilde{\xi}_2 + \beta \xi_{3}$. Let us briefly inspect the inflationary case (when $\sigma \gg \Lambda$). Then in the Einstein frame the kinetic part reads:
\begin{align}
\mathcal{L}_{\textrm{kin}} =& \frac{3}{4} \left[ \partial_{\mu} \ln(\xi_{3}\sigma^2 + \tilde{\xi}_1 h^2 + \tilde{\xi}_2 s^2)\right]^2\\ 
&+ \frac{1}{2(\xi_{3}\sigma^2 + \tilde{\xi}_1 h^2 + \tilde{\xi}_2 s^2)}\cdot
\left[(\partial_{\mu}\sigma)^2 + (\partial_{\mu}h)^2 + (\partial_{\mu}s)^2\right],
\end{align}
with the following definitions:
\beq
\begin{array}{lcr}
\chi = \sqrt{\frac{3}{2}}\ln(\xi_3\sigma^2,)& \tau_h = \frac{h}{\sigma}, & \tau_s = \frac{s}{\sigma},
\end{array}
\eeq
To the leading term in $1/\xi_3$ the kinetic part is:
\beq
\mathcal{L}_{\textrm{kin}} = \frac{1}{2}(\partial_{\mu}\chi)^2 + \frac{1}{2\xi_3}(\partial_{\mu}\tau_h)^2,
\eeq
with mixing terms suppressed. For large $\sigma$ and $\xi_3$ the potential term becomes:
\beq
V \simeq \frac{1}{4\xi_3^2}\left[\kappa(1-\alpha\tau_h^2 - \beta\tau_s^2)^2 + \lambda_1\tau_h^4+ \lambda_p\tau_s^4+ \lambda_3\tau_h^2\tau_s^2\right]. 
\eeq
\noindent Therefore this model possesses the same properties as the original inflation scenario does, this time however two fields drop to minima. One can check that the same conditions for couplings in both models (recalling that $\xi_1 \simeq \alpha \xi_3$, $\xi_2 \simeq \beta \xi_2$) need to hold \cite{Lebedev:2011aq}. Since the shape of the potential is preserved:
\beq
U(\chi) = \frac{\lambda_{\textrm{eff}}}{4\xi_3^2}\left(1+ \exp\left(-\frac{2\chi}{\sqrt{6}}\right)\right)^{-2},
\eeq
the observational predictions ($r$, $n_s$) aren't spoiled.
The $\lambda_{\textrm{eff}}$ is obtained by minimalization of $V$ for large $\chi$. One can take $\tilde{\xi}_{1,2} \sim \mathcal{O}(1)$ and the unitary constraints are satisfied up to the Planck scale. Moreover, large vacuum energy of the Higgs field which contributes from dropping into minimum can successfully take part in the reheating process \cite{Lee:2013nv}. One can also take only $\zeta$ and $H_0$ as inflatons and then inspect inflation conditions in the manner presented here.\\
One can think also, in the case of addition a heavy sterile scalar $\zeta$, about a larger fraction of induced gravity scenario, with proper $\xi_3$ obviously. Then inflation, as a theory, would be a bridge between regime of quantum gravity and particle physics. The addition of $\zeta$ scalar triplet also opens the possibility of $U(1)$ symmetry breaking in lagrangian and drive inflation by additional pseudo-Goldstone boson in the manner of pure natural inflation (NI) or as as an additional field coupled to gravity, like in Gong, Lee and Kang article \cite{Two-Higgs-doublet}. We will briefly discuss both possibilities in the outline and extensions section.
\subsection{Possible modifications of coupling to gravity}
\label{Observables}
In the previous paragraph we have investigated importance of possible extension of Conformal Standard Model. In this chapter we would like to discuss possible modifications of previously analysed scenario by modifying the non-minimal coupling. The proposed in this work coupling to gravity:
\beq
\label{omegaone}
\Omega^2 = \frac{M^2 + \xi_1h^2 + \xi_2r^2}{2},
\eeq
might look quite strange even though our assumption that only diagonal terms are important is obviously justified for (\ref{omegaone}) (because we are considering couplings to gravity with values much bigger than one, while the other couplings are of order one). Since the $\textrm{Tr}(\phi\phi^{\ast})$ is the full quadratic term in the model and at first sight looks more natural:
\beq
\Omega'^2 = \frac{M^2 + \xi_1h^2 +\xi_2\textrm{Tr}(\phi\phi^{\ast})}{2},
\eeq
there are two reasons why we didn't inspect that scenario. First one is simplicity of the model. For the full $\mathrm{Tr}(\phi\phi^{\ast})$ there would be no physical reason to drop the non-diagonal interactions for $\phi$. Without doing so, the analysis seems to be impossible even if we treat all $R_i$'s as the one effective field. The second reason is the following: if we take mixing matrix eigenstates as inflatons (\ref{massstate}):
\beq
\begin{array}{lcr}
h_0 = \cos (\beta) H_0 + \sin (\beta) r,& & h' = -\sin (\beta) H_0 + \cos (\beta) r,
\end{array}
\eeq
with mixing angle $\beta$, and propose the coupling as:
\beq
\Omega^2 = \frac{M_P^2 + (\xi_1'h_0^2+\xi_2'h'^2)}{2}.
\eeq
Then we can find such $\xi_1'$, $\xi_2'$, and by redefining $\lambda_i$'s and/or $v_i$ if necessary, that for a given $\beta$ we can get such a set of parameters for $H_0$ and $r$ which will fit the stable minimum of $V(s,\chi)$ and give the same observational data. There is also another (natural) way of defining the coupling: 
\beq
\Omega^2_M = \frac{M_P^2 + \xi(h_0^2+h'^2)}{2},
\eeq
then we obtain the following relation: $\xi_1=\xi\cos^2\beta,$ $\xi_2 = \xi \sin^2\beta$. Since the $\beta \approx 0$ then the lighter eigenstate has dominant role and for this case the heavy state decouples even faster and can be better understood using simpler arguments \cite{Two-Higgs-doublet}. So we obtain (almost) single Higgs scenario for this coupling. However, for this proposal there is no set parameters which can prevent from spoiling unitarity without introducing a new degree of freedom.\\
\section{Summary and outlook}
\label{Summary}
\setcounter{equation}{0}
\subsection{Summary}
In this thesis we have analysed the inflation scenario, where the mixed mass states of Higgs and ``shadow'' Higgs where considered as inflatons. The decoupling of heavier state was broadly discussed. It turned out that for large set of parameters the presented mechanism gives successful inflation scenario. The stable minima of the potential were identified and all three cases, ie $\tau_0 = \sqrt{\frac{b}{a}}$, $\tau_0 =0$, $\tau_0=+\infty$, were discussed. The consistency with CSM and observational data was checked. One has to stress that the presented model doesn't posses a unique prediction for inflation. However, it can be potentially falsified on the particle physics side. Moreover, the following modification (addition of induced gravity couplings) doesn't spoil the gravity observations so far. So this model together with CSM provides, in principle, a unified scenario for all cosmological and particle physics issues excluding dark energy and quantum gravity, up to the Planck scale. Such a unified description seemingly hasn't been discussed before. It was also argued that such inflation model is valid up to the Planck scale since there is a natural extension of CSM, by additional scalar triplet, while many other models suffer from incomplete description concerning only inflation. So the power of the presented scenario lies on the fact that it there is one unified description for broad range of cosmological observables. The further extensions of presented scenario and other mechanisms of inflation within CSM are discussed in the next subsection. 
\subsection{Outlook and further work}
\label{Extensions}
One can ask whether Conformal Standard Model can provide a successful inflation itself, without induced coupling to gravity or coupling $\xi\ll\xi_{eff}$. To find out, one has to explore the features of the model. There are some possibilities which could be investigated:\\
\textbf{K-inflation}\\
As it was pointed out by Picon, Damour and Mukhanov inflation may not only originate from a specific type of potential but can be also driven by higher kinetic terms. Let us consider a lagrangian:
\beq
S = \int d^4 x \sqrt{|g|} \frac{M_P^2}{2}R + p(X,\varphi),
\eeq
with $X$ defined in [\ref{Inhomogeneouslimit}] as $X = \guv\partial^{\mu}\varphi\partial^{\nu}\varphi$. If we impose that when $X \to 0$ then $p(X, \varphi)$ vanishes we can expand matter lagrangian in the powers of $X$ near $X \approx 0$. One of the motivations to study such a theory is low energy effective action for string theory or models possessing effective action with two scalar fields with quartic potentials coupled to each other as was pointed out in \cite{Avgoustidis:2012yc}. It was discussed \cite{ArmendarizPicon:1999rj,GARRIGA} that such a lagrangian of type:
\beq
\mathcal{L} = K(\varphi)X + L(\varphi)X^2,
\eeq
 can lead to a successful inflation scenario with graceful exit. They also calculated spectral-tilt and tensor-to-scalar ratio. As we have showed in section \ref{Inhomogeneouslimit} these quantities in general depend on $c_s^2$ i.e. the speed of sound. It is not equal to one for higher kinetic terms and can provide in general proper values of $r$ and $n_s$. Since the multiinflation model effective actions can give rise to such a lagrangian it is worth studying within CSM.\\
\textbf{Natural inflation}\\ Another mechanism for successful inflation scenario is called natural inflation \cite{PhysRevD.70.083512}. The Inflaton in this model is pseudo-Goldstone boson(s) with potential of form: $V(\phi) = \Lambda^4\left[1 \pm \cos(\phi/f)\right].$ The Goldstone bosons become pseudo-Goldstone bosons (PGB) when there is an explicit symmetry breaking and the only symmetry remaining is the shift symmetry: $\phi \to \phi + \textrm{const}$. The flatness of the potential and slow-roll conditions  is obtained due to those remaining symmetries. This scenario is called natural since it satisfies naturalness condition proposed by van 't Hooft: \emph{a small parameter $\alpha$ is called natural if, in the limit $\alpha \to 0$, the symmetry of the system increases.} Here the naturalness corresponds to:
\beq
\chi = \Delta V/(\Delta \phi)^4 = \frac{\textrm{height}}{\textrm{width}^4} \leq 10^{-6},
\eeq
ie the ratio between height of the potential to the width should be quite small, however most of particle physics potentials give: $\chi = \mathcal{O}(1)$. So for the QCD axion, where:
\beq
\Lambda/f^4 \sim 10^{-64},
\eeq
 which the authors took as the inflaton initially, with self-coupling $\chi$. Moreover, Freese and Kinney advocates in 2014 article, that Natural Inflation seems to be consistent with Planck CMB data. Conformal Standard Model posses, in new version five pseudo-Goldstone bosons, which can potentially drive inflation. Also with new field $\zeta$ one can think of interaction which will break a global symmetry and provide a pseudo-Goldstone boson as an additional inflaton ``helping'' non-minimal coupling inflation.\\
\textbf{Conformal anomaly driven inflation}\\
In classical conformal field theory the energy-momentum tensor is invariant under rescalings, ie it's traceless. However, quantum effects can give rise to non-zero expectation value of $\hat{T}_{\mu\nu}$ for these theories and that quantity is called the conformal anomaly. The origin and its general analytic form is broadly studied using effective action. Then classical conformal invariance is broken at the  quantum level:
\beq
\gvu \langle T_{\mu\nu}\rangle \neq 0,
\eeq
moreover the anomalous trace is of geometric origin and independent of the state. It can be written as:
\beq
T_{\mu}^{\mu} = \frac{1}{180(4\pi)^2}\left(c_sC^2 + a_s E_4\right),
\eeq
where
\begin{align}
C^2 &= R_{\mu\nu\rho\sigma}R^{\mu\nu\rho\sigma} - 2 R_{\mu\nu}R^{\mu\nu} + \frac{1}{3}R^2,\\
E_4 &=  R_{\mu\nu\rho\sigma}R^{\mu\nu\rho\sigma} -4 R_{\mu\nu}R^{\mu\nu} + R^2,
\end{align}
and $a_s$ and $c_s$ depends on the field(s) spin which are involved. So, with gravity included, we want to solve the following equation:
\beq
R_{\mu\nu} + \frac{1}{2}Rg_{\mu\nu} = 8\pi G\langle T_{\mu\nu}\rangle,
\eeq 
and as we can see it can have a drastic effect on the Einstein field equation, like in \cite{Godazgar:2016swl}. For maximally symmetric spacetimes trace anomaly acts exactly as cosmological constant for these spacetimes. Then it can give rise to de-Sitter solution to the Einstein equations. Also there is a graceful exit \cite{Hawking:2000bb,STAROBINSKY} and there exists a natural solution which decays into matter-dominated FLRW universe. The authors of \cite{Hawking:2000bb} show that this can be achieved for  $N=4$ super Yang-Mills theory and they calculated the inflation parameters and number of e-folds --  it turned out that this approach is within the current experimental data for some modified gravity theories \cite{PhysRevD.90.043505}. Then one can apply their methods for Conformal Standard Model, calculate trace-anomaly and see whether it can give Inflation scenario, either with non-minimal coupling(s) or without. Moreover, vanishing of the $T_{\mu\nu}$ can be related to $N\geq 5$ super-gravities with which CSM seems to be connected, see \cite{Meissner:2016onk}. \\\\
\noindent 
We would like to stress that all the components are already in the model. For all three possibilities one has to check whether there is a suitable choice of parameters in CSM such that the Inflation scenario can occur and moreover they can fit the CMB data. If it is so, it would be an argument in favor of the CSM model, since this model would provide an inflation scenario within itself, without assuming further assumptions or conditions like large non-minimal coupling to gravity. There could be also more than one mechanism driving the inflation.\\
Other similar extensions of SM, like $\nu$MSM \cite{Shaposhnikov:2006xi}, can also provide an inflation phase. However, for this particular model inflation with non-minimal couplings hasn't been yet fully analysed, as far as the author is informed. The features of both models, when concerning inflation scenario, seem to be very similar. So the analysis performed in this paper can be easily repeated for $\nu$MSM, however that model lacks an additional, natural extension which can preserve unitarity. \\
There is also one more problem with all those models, the initial conditions problem. Namely the $R^2$ (Starobisky) Inflation type scenarios, like for example Higgs coupled to gravity \cite{Bezrukov}, matches the data, but they have problems close to the Planck scale. The quantum fluctuations in the Planck scale are so large that can stop the expansion so the Universe will stay in this phase. The authors \cite{Dimopoulos:2016yep} proposed a scenario where there is a proto-inflation field which provides $\ddot{a}>0$ and homogenise the Universe and then Starobinsky inflation takes place. This scenario might be promising due to resolving both problems of fine tuning in high energies and matching the data. This initial conditions problem was addressed in \cite{Dimopoulos:2016yep} but could also be resolved by decoupling of heavy scalar degree(s) of freedom, like $\zeta$, $\tau$, which should be investigated. These decoupling can also give high average value of the energy and then resolve the problems of reheating.\\
Besides proposing and investigating the CSM-inflation scenario, this thesis had a second purpose from very beginning. The inflation programme is far from complete and the literature of the subject is vast but a bit messy, in a sense there are no standard convention. The author decided to find, among broad and diverse literature, the most general derivations, arguments and reasonings concerning that part of cosmology and put them together. This is why the chapters \ref{Mainideas}-\ref{inflation}, appendices and parts of \ref{Higgsinflation} are written in a very general manner, to cover possibly all the scenarios and mechanisms discussed. The author hopes that these chapters could serve as an entry level manual for researchers interested in inflation. It is especially dedicated to those who aren't familiar with the formalism or even with general relativity. The manual should be obviously supplemented by detailed discussion of renormalisation procedure and effective action which are tools broadly used and needed to investigate inflation scenario fully and which are absent in this thesis. The chapters concerning Standard Model and Conformal Standard Model are very brief and discuss only issues required to study inflation scenario within the framework of particle physics. Such a selection of material and structure of text helped to understand these issues at least one person, the author. \\
Thesis is based on article \cite{Kwapisz:2017vjt}.


\appendix
\setcounter{equation}{0}
\renewcommand{\theequation}{\Alph{section}.\arabic{equation}}
\section{General Relativity appendix}
\label{GRappendix}
\subsection{Tensor calculus}
\label{Covariant}
\textbf{Differentiation}
\begin{itemize}
\item$\phi_{,\nu} := \partial_{\nu}\phi$, means standard derivative,
\item$\phi_{;\nu}$, means covariant derivative.
\end{itemize} 
Covariant derivative for scalar field is equal to the standard one: $\phi_{,\nu}=\phi_{;\nu}$. For a vector field: 
$$
V_{\mu;\nu} =\nabla_{\nu} V_{\mu} = \partial_{\nu}V_{\mu} +\Gamma^{\sigma}_{\mu\nu}V_{\sigma},
$$
and contravariant:
$$ 
V^{\mu}_{;\nu} =\partial_{\nu}V^{\mu} - \Gamma^{\mu}_{\nu\sigma}V^{\sigma},
$$
where connection coefficients are: $\Gamma^{\mu}_{\nu\sigma}$ defined in (\ref{Christoffel}, \ref{Christoffel2}). One can extend it to other types of fields of course.\\
\noindent
\textbf{Christoffel symbols}\\
Christoffel symbols of first kind are defined as follows:
\beq
\label{Christoffel}
\Gamma_{\mu\nu\rho} = \frac{1}{2} \left[ \partial_{\rho} g_{\sigma\nu} + \partial_{\nu} g_{\sigma\rho} - \partial_{\mu} g_{\rho\nu}\right].
\eeq
And of second kind (broadly used in General Relativity) as:
\beq
\label{Christoffel2}
\Gamma^{\sigma}_{\nu\rho} = g^{\mu\sigma} \Gamma_{\sigma\nu\rho}.
\eeq
\textbf{Riemann tensor}.\\
Riemann curvature tensor in coordinates is defined as \cite{Gravity}:
\beq
\label{Riemanntensor}
R^{\alpha}_{\beta\gamma\delta} = \partial_{\gamma}\Gamma^{\alpha}_{\beta\delta} - \partial_{\delta}\Gamma^{\alpha}_{\beta\gamma} + \Gamma^{\alpha}_{\mu\gamma}\Gamma^{\mu}_{\beta\delta} - \Gamma^{\alpha}_{\mu\delta}\Gamma^{\mu}_{\beta\gamma}.
\eeq
It has 256 componets, but many of then are the same / vanishes by symmetry. The easiest form to show those is full covariant version: $R_{\alpha\beta\gamma\delta}=g_{\sigma\alpha}R^{\sigma}_{\beta\gamma\delta}$. Then we have the following symmetries \cite{Fieldtheory}:
\begin{align} 
R_{\alpha\beta\gamma\delta} &= - R_{\beta\alpha\gamma\delta},\\
R_{\alpha\beta\gamma\delta} &= - R_{\alpha\beta\delta\gamma},\\
R_{\alpha\beta\gamma\delta} &= R_{\gamma\delta\alpha\beta},\\
R_{\alpha[\beta\gamma\delta]} &= R_{\alpha\beta\gamma\delta} + R_{\alpha\gamma\delta\beta} + R_{\alpha\delta\beta\gamma} = 0,
\end{align}
where the final one is called First Bianchi identity. The second Bianchi identity reads as:
\beq
 R_{\alpha\beta[\gamma\delta;\epsilon]} =0.
\eeq
We define Ricci tensor as:
\beq
\label{RicciTensor}
R_{\mu\nu} = R^{\alpha}_{\mu\alpha\nu}, 
\eeq
which is symmetric in its indices and is also only non-vanishing contraction of Riemann tensor. We can also define Ricci scalar curvature as:
\beq
\label{RicciScalar}
R = g^{\mu\nu}R_{\mu\nu} = \sum_{i=0}^3 R_i^i.
\eeq
And finally we introduce Einstein tensor as:
\beq
\label{EinsteinTensor}
G_{\mu\nu} = R_{\mu\nu} - \frac{1}{2}\guv R,
\eeq
which is the left hand side of Einstein equations.
\subsection{Conformal transformations}
\label{ConformalTrans}
Let $\Omega^2(x)$ be a positive function with no critical point, then we call a following transformation \emph{conformal mapping} on Lorentz manifold $\mathcal{M}$:
\begin{align}
\hat{g}_{\mu\nu} = \Omega^2(x)\guv,\\
\hat{g}^{\mu\nu} = \Omega^{-2}(x)g^{\mu\nu},\\
\sqrt{\hat{|g|}} = \Omega^D \sqrt{|g|}, 
\end{align}
where $D$ is the dimension of spacetime. The $\Omega^2$ function may also depend on other fields described in theory like in (\ref{Starobinskyconformal}). 
Then connection coefficients also transforms \cite{pracaAmsterdam, EinsteinJordan} as:
\beq
\hat{\Gamma}^{\sigma}_{\mu\nu} = \Gamma^{\sigma}_{\mu\nu} + \left[ \delta_{\nu}^{\sigma} \partial_{\mu}+ \delta_{\mu}^{\sigma}\partial_{\nu} - \guv g^{\sigma\rho}\partial_{\rho}\right] \ln{\Omega}.
\eeq
And the Ricci tensor is as following (\ref{RicciTensor}):
\begin{align}
\hat{R}_{\mu\nu}= R_{\mu\nu} + \left[(D-2)\partial_{\mu}\partial_{\nu} - g_{\mu\nu}\square\right] \ln \Omega +\\
 +(D-2)\left[ \partial_{\mu}\ln \Omega \partial_{\nu} \ln \Omega - g_{\mu\nu} g^{\alpha\beta}(\partial_{\alpha} \Omega)(\partial_{\beta}\ln \Omega)\right]. \nonumber
\end{align}
And finally the Ricci scalar (\ref{RicciScalar}):
\beq
\hat{R} =\Omega^{-2} \left[R -\frac{2(D-1)}{\Omega}\square\Omega - \frac{(D-4)(D-1)}{\Omega^2}g^{\alpha\beta}(\partial_{\alpha}\Omega)(\partial_{\beta}\Omega)\right],
\eeq
where $\square$ is D'Alembert operator: $\square = \guv \partial^{\mu}\partial^{\nu}$.
\subsection{Geodesic equation }
\label{Geodesic}
Since the square of infinitesimal length is given by the formula:
$$
ds^2 = g_{\mu\nu}dx^{\mu}dx^{\nu}
$$
So length of the curve from point $\lambda_1$ to $\lambda_2$ is:
\beq
\label{actioncurve}
S =  \int^{\lambda_2}_{\lambda_1} ds = \int^{\lambda_2}_{\lambda_1} \sqrt{g_{\mu\nu}(x(\lambda))\frac{dx^{\mu}}{d\lambda}\frac{dx^{\nu}}{d\lambda}}d\lambda,
\eeq
and we assume that particles travel on geodesics, namely on the curves which length is extremal. So if we calculate the variation of (\ref{actioncurve}):
$$
\delta S = \int \delta\left(\frac{\left(g_{\mu\nu}dx^{\mu}dx^{\nu}\right)}{2\sqrt{g_{\mu\nu}\frac{dx^{\mu}}{d\lambda}\frac{dx^{\nu}}{d\lambda}}}\right)d\lambda,
$$
then we will reparametrize with proper time $\tau$ coordinates to get $g_{\mu\nu}\frac{dx^{\mu}}{d\tau}\frac{x^{\nu}}{d\tau} = 1$, so we are left with equation:
$$
0=\int \delta\left(g_{\mu\nu}\frac{dx^{\mu}}{d\tau}\frac{dx^{\nu}}{d\tau}\right) d\tau.
$$
From this we obtain:
$$
0=2 \int d \tau \delta x^{\nu}\left[\guv\frac{d^2x^{\mu}}{d\tau^2} + \frac{1}{2} \frac{dx^{\rho}}{d\tau}\frac{dx^{\nu}}{d\tau}\left( \partial_{\rho} g_{\mu\nu} + \partial_{\nu} g_{\mu\rho} - \partial_{\mu} g_{\rho\nu}\right)\right].
$$
Because for every $\delta x^{\mu}$ the integral has to be eual $0$, we deduce that following equation has to be satisfied:
\beq
\label{geodesic equation}
\frac{d^2x^{\mu}}{ds^2} + \Gamma^{\mu}_{\nu\rho}\frac{dx^{\rho}}{ds}\frac{dx^{\nu}}{ds} =0,
\eeq
where $\Gamma^{\mu}_{\nu\rho}$  is Christoffel symbol of second kind (\ref{Christoffel2}).
\subsection{Einstein-Hilbert action}
\label{EHaction}
Einstein equations relates metric with energy-momentum tensor. They can be derived from the following action using stationary action principle:
\beq
\label{Einsteinaction}
S = \int \left(\frac{1}{2\kappa} R + \mathcal{L}_m \right) \sqrt{|g|}d^4x,
\eeq
where $\kappa = 8\pi G c^{-4}$. And Einstein equations are:
\beq 
\label{Einstein}
\Ruv- \frac{1}{2} R \guv = \Tuv. 
\eeq
Since by \emph{Lovelock theorem} the Einstein-Hilbert action isn't the only fundamental proper action built from metric tensor and can be further generalised, since then there are many extensions. One natural extension is general $f(R)$ gravity where action is given by:
\beq
\label{EH_action}
S = \frac{1}{2\kappa} \int d^4x \sqrt{|g|}f(R) + S_m,
\eeq
So Starobinsky action is among them where $f(R)$ is given as:
\beq
f(R)=M_P^2\left(R+ \beta R^2\right),
\eeq
with $\beta = \frac{1}{6M_P^2 M^2}$. We obtain the following derivatives:
\beq
\begin{array}{lcr}
\dot{f}(R) = \dot{R} + 2\beta R\dot{R}, &f_{,R} =1+ 2\beta R & \dot{f}_{,R} = 2\beta \dot{R}.
\end{array}
\eeq
For general $f(R)$ action we will show how to derive equations for metric tensor. For this case:
\begin{align}
0&= \delta S = \frac{1}{2\kappa} \delta \int d^4x\sqrt{|g|} f(R) + \delta S_m \nonumber \\
&= \frac{1}{2\kappa} \int d^4x \sqrt{|g|}\frac{\guv}{2}f(R)\delta \gvu + f_{,R}\delta R  + \delta S_m,
\end{align}
where $f_{,R} = \frac{d f(R)}{d R}$.  After an easy calculation \cite{pracaAmsterdam} we obtain:
\beq
\delta R = R_{\mu\nu}\delta \gvu + \left(\guv \nabla^2 - \nabla_{\mu}\nabla_{\nu}\right) \delta \gvu, 
\eeq
then if we integrate by parts we will obtain the following gravitational part of the integral:
\beq
\int d^4x \sqrt{g} \delta \gvu \left[ \frac{\guv}{2}f(R) + f_{,R}R_{\mu\nu} + \guv \nabla^2f_{,R} - \nabla_{\mu}\nabla_{\nu} f_{,R}\right].
\eeq
Since the variation is arbitrary and energy momentum tensor is given by the known formula:
\beq
S_m = \frac{1}{2}\int d^4x T_{\mu\nu} \delta \gvu, 
\eeq
we obtain the following $f(R)$ equations:
\beq
\label{fREinstein}
\frac{\guv}{2}f(R) + f_{,R}R_{\mu\nu} + \guv \left[\nabla^2 - \nabla_{\mu}\nabla_{\nu}\right] f_{,R}  = \kappa T_{\mu\nu},
\eeq
and for $f(R) = R$ the standard relation is obtained. From now we put $\kappa = 1$. Hence taking the trace gives us:
\beq
\label{tracefR}
f_{,R}R + 3\nabla^2f(R) + 2f(R) = T,
\eeq
where $T = \gvu T_{\mu\nu}$. The derivation of FLRW equations for the $f(R)$ gravity is known and can be done, for example like in \cite{pracaAmsterdam}.
The resulting equation for $00$ component (\ref{fREinstein}) is:
\beq
f_{,R}R_{00} + \frac{1}{2}f(R)g_{00} - \partial^2_0 f_{,R} + g_{00}\nabla_{\mu}^2 f_{,R} = g_{00}T^0_0,
\eeq
where we have for FLRW metric:
\beq
g_{00}\nabla^2 f_{,R} = 3H\dot{f}_R + \ddot{f}_R,
\eeq
so we get:
\beq
f_{,R}R_{00} + \frac{1}{2}f(R)g_{00} +3H\dot{f}_R  = T_{00},
\eeq
so, with help of (\ref{Riccizero}, \ref{RicciHubble}), we get the second Friedmann equation:
\beq
f_{,R} \left(3H^2 - \frac{1}{2}R\right) + \frac{1}{2}f(R) +3H\dot{f}_{,R} = \varepsilon,
\eeq
while with the help of trace equation (\ref{tracefR}) we get the first one:
\beq
-6f_{,R}(\dot{H}+ 2H^2) + 3\ddot{f}_{,R} + 2f(R) = (\varepsilon - 3p),
\eeq
for $f(R) = R$ one can conceive himself that they are the same as (\ref{IFriedeq}, \ref{IIFried}). In general to solve equations $f(R)$ has to be specified. Still, even for Starobinsky action $f(R)$ FLRW equations posses very complicated structure, however for inflation case not all terms are relevant.
\subsection{Cauchy surfaces and Killing vectors}
\subsubsection{Killing vectors}
Symmetries are essential for solving equations, especially so complicated as Einstein Equations. If metric in some coordinate basis is independent from a coordinate $\beta$, then: $g_{\mu\nu;\beta} = 0$. Then any curve $x^{\alpha} = c^{\alpha}(\lambda)$ can be shifted in the $\beta$ direction, and let $\Delta x^{\beta} =\varepsilon$. Let us compare length (from $\lambda_1$ to $\lambda_2$)  of two curves \cite{Gravity}.\\
\emph{Non-shifted:}
$$
L = \int_{\lambda_1}^{\lambda_2} \left[g_{\mu\nu}(x(\lambda))\frac{dx^{\mu}}{d\lambda}\frac{dx^{\nu}}{d\lambda}\right]^{1/2}d\lambda,
$$
\emph{shifted:} 
$$
L(\epsilon) = \int_{\lambda_1}^{\lambda_2} \left[\left\{g_{\mu\nu}(x(\lambda))+ \varepsilon \frac{\partial g_{\mu\nu}}{\partial x^{\beta}}\right\}\frac{dx^{\mu}}{d\lambda}\frac{dx^{\nu}}{d\lambda}\right]^{1/2}d\lambda,
$$
but $g_{\mu\nu;\beta} = 0$, so $dL/d\varepsilon = 0$. Let us define Killing vector as:
\beq
\xi = d/d\varepsilon = \frac{\partial}{\partial x^{\beta}} 
\eeq
One can show that it satisfies Killing equation:
\beq
\label{Killingeq}
\xi_{\mu;\nu} + \xi_{\nu;\mu} = 0.
\eeq
From (\ref{geodesic equation}) and (\ref{Killingeq}) we can deduce a theorem:
\begin{thm}Scalar product of Killing vector with a vector tangent to a geodesic is constant along this geodesic, namely: $\textbf{P} = \frac{d}{d\lambda}$
\beq
\textbf{P}^{\beta} := P \cdot \xi = \emph{constans}
\eeq
\end{thm}
\noindent Killing  vectors are very useful in solving geodesic equations \cite{Gravity}, but also we can use them to inspect global structure of spacetime when a Killing vector is globally defined or clasify solutions to Klein-Gordon equation in curved spacetime, we briefly discuss in [\ref{QFTcurved}].\\
\subsubsection{Cauchy surfaces and hyperbolicity}
Worldlines of particles are modeled by causal curves:
\begin{itemize}
\item \textbf{Future / Past development} called also as domain of dependence or causal domain of the point $p$ is a set, which:
\beq
\label{development}
J^{\pm} (p) = \left\{q \in \mathcal{M}:  \textrm{ exists future/past causal }\gamma (t): \textrm{ } \gamma(0) = p, \textrm { } \gamma(1) = q\right\},
\eeq 
we can define in analogy future / past development for a set $S$: $J^{\pm}(S)$, such that every  curves intersects $S$.
\item \textbf{Future / Past  lightcone} from the point $p$:
\beq
\label{Lightcone}
V^{\pm} (p) = \left\{q \in \mathcal{M}: q\in J^{\pm} \textrm{ and } \gamma(t) \textrm{ is future/past directed null geodesic}\right\}
\eeq
\end{itemize}
We define full domain of dependence as:
\beq
J(S) = J^{+}(S) \cup J^{-}(S)
\eeq
\begin{defi} A spacetime satisfies the \textbf{causality conditions} when there is no closed causal curves. For instance $\mathcal{M}$ mustn't be compact\cite{BB_2009} to satisfy those conditions.
\end{defi}
\noindent Now we would like to introduce \textbf{Cauchy Surfaces}, which enable us to provide the description of simultaneity. Let us consider a spacetime $(\mathcal{M},g)$, time-orientation is chosen \cite{Wojtek,BlackHoles,Wald}.
\begin{thm}
\label{subman} Let S be a (nonempty) closed achronal set with edge $(S) = \emptyset$. Then S is a three-dimensional, embedded, $C^0$ submanifold of $\mathcal{M}$, where $\mathcal{M}$ is spacetime manifold.
\end{thm}
\begin{defi}\textbf{Cauchy surface}
\label{Cauchysurface}
$\Sigma$ is a closed achronal set for which $J(\Sigma) = \mathcal{M}$.
\end{defi}
\noindent Obviously edge of Cauchy surface is empty, thus by theorem [\ref{subman}] Cauchy surface is an embedded $C^0$ submanifold of $\mathcal{M}$. One may think about $\Sigma$ as representing ``instant of time'' surface in spacetime.
\begin{defi}
\label{hiper}We say that spacetime $(\mathcal{M},g)$ is \textbf{globally hiberbolic}, when it posses Cauchy surface.
\end{defi}
\noindent
The following theorem \cite{BB_2009} shows that spacetimes with Cauchy surface are special:
\begin{thm}
\label{globhiper}
These conditions are equivalent:
\begin{enumerate}
\item spacetime $\mathcal{M}$ is globally hiperbolic,
\item $\mathcal{M} \approx \mathbb{R} \times \Sigma_t$ and every $\Sigma_t$ is a Cauchy surface,
\item there are no causal closed curves and $\forall_{p,q\in M} J^+(q)\cap J^-(p)$ are compact.
\end{enumerate} 
\end{thm}
\noindent Many common spacetimes posses Cauchy surface, for example: Minkowski spacetime, spacetime with FLRW metric, Rindler wedge in Minkowski. Spacetime which is not globally hiperbolic is anti-deSitter spacetime, this quantity is related to AdS/CFT correspondence \cite{AdS/CFT}.
Global hiperbolicity will be a crucial property, when we want to quantise a theory on curved spacetime. It will guarantee us uniqness of Green function and proper definition of the theory, for further details see chapter [\ref{QFTcurved}].
\begin{defi} We say that metric $\mathrm{(g)}$ is \textbf{stationary}, when it has a future directed time Killing vector, so the general form of this metric is\cite{Wojtek}:
$$
ds^2 = \alpha^2(y)dt^2 - \omega_i dt dx^i - h_{ij}(y)dx^idx^j,
$$
where $\alpha^2(y)>0$, $h_{ij}$ has signature $(+,+,+)$. If we can choose $\Sigma_t$ such that $\omega_i =0$, we say that this metric is static.
\label{Stationary metric}
\end{defi}
\noindent For stationary metric we can construct Hamiltonian which will be strictly positive and not depending on time, which is the case in section \ref{QFTcurved}.

\section{Quantum field theory in curved space-time}
\label{QFTcurved}
\setcounter{equation}{0}
\subsection{General Introduction}
Noninteracting quantum field theory in Minkowski spacetime, assuming there is no gravity at all, can be solved exactly. We just simply write down free classical lagrangian and quantise it. Sometimes its not very trivial, for example for electromagnetic field, but in the end free theory is quantised and can be solved exactly. Interactions cause much more trouble, but still can calculate some Feynman diagrams to get perturbative expansion or use other tools like symmetries or path integral formulation. When we consider noninteracting QFT on classical gravity background, things are slightly different. In that case when we write Euler-Lagrange equations, the solutions depends on the metric and are non-trivial ones, often without analytical solutions. Moreover, even the simplest field to describe, ie noninteracting scalar field gives interesting physical predictions like Unruh effect or Hawking radiation. Even though inflation is analysed on the classical level, however the inhomogeinities needs to be quantised. So inflation is one among the motivations for the study of QFT on curved background. \\
Let us start with the action \cite{MukiWin} of scalar field in curved spacetime:
\beq
\label{KGaction}
S = \int d^4x \sqrt{|g|} \left(\frac{1}{2}g^{\mu\nu}\phi_{,\mu}\phi_{,\nu} - V(\phi)\right),
\eeq
using E-L equations we obtain Klein-Gordon equation in a form:
\beq
\label{KGeq}
g^{\mu\nu}\phi_{,\mu\nu} + \frac{1}{\sqrt{|g|} }\left(g^{\mu\nu}\sqrt{|g|}\right)_{,\mu}\phi_{,\nu} - \frac{\partial V}{\partial \phi} = 0.
\eeq
In case of our study we would like to study FLRW (\ref{FLRWmet}) metric, then (\ref{KGeq}) takes the form:
\beq
\label{KGFLRW}
\left(\frac{1}{a^3}\partial_t (a^3\partial_t) - a^{-2}\Delta_h\right) \phi(t,y) = \frac{\partial V}{\partial \phi}.
\eeq
We have canonical structure, namely: $\pi(t,y)=a^3\sqrt{|h|}\dot{\phi}$ is the canonical conjugated momenta and the Poisson equation:
$$
\begin{array}{c}
\{\pi(t,y),\phi(t,y')\} = \delta_{y}(y')
\end{array}
$$
And the Hamilton equation of motion are:
\begin{align}
\dot{\pi}  &= \left(a\Delta_h\phi - \frac{V(\phi)}{\phi}a^3\right)\sqrt{|h|},\\
\dot{\phi} &= \frac{1}{a^3\sqrt{|h|}}\pi.
\end{align}
So we have the hamiltonian:
\beq
\label{QFTChamiltonian}
H_t = \int \frac{1}{2}\left(a^{-3}\sqrt{|h|}^{-1}\pi^2 + a\sqrt{|h|}h^{ij} \phi_{,i}\phi_{,j} + a^3 \sqrt{|h|}V(\phi)\right).
\eeq
From now on for simplicity we will assume $V(\phi) = M^2\phi^2$ to have free theory of massive scalar real field. We will rewrite action in conformal coordinates \cite{MukiWin}, namely: $dt = a(\eta)d\eta$ and $\chi(\eta)= a(\eta) \phi$:
\beq
S= \frac{1}{2} \int d^3\mathbf{x} d\eta \left((\chi')^2 - (\nabla \chi)^2 - m_{eff}^2(\eta)\chi^2\right),
\eeq
where: 
$$
m_{eff}^2(\eta) = M^2a^2 - \frac{a''}{a},
$$
and Klein - Gordon equation is then:
\beq
\chi'' - \Delta \chi + m_{eff}^2\chi = 0.
\eeq
We can expand $\chi$ using Fourier modes:
$$
\chi(\mathbf{x},\eta) = \int \frac{d^3\mathbf{k}}{(2\pi)^{3/2}} \chi_{\mathbf{k}}(\eta)e^{i\mathbf{\mathbf{k} \cdot \mathbf{x}}},
$$
then for each mode $\chi_k$ we obtain K-G:
\beq
\label{KGFourier}
\chi_{\mathbf{k}}'' + \left[{\mathbf{k}}^2 + m^2a(\eta) -\frac{a''}{a}\right]\chi_{\mathbf{k}} = \chi_{\mathbf{k}}'' + \omega^2_{\mathbf{k}}(\eta)\chi_k = 0,
\eeq
so the general solution can be written as \cite{MukiWin}:
\beq
\chi_{\mathbf{k}}(\eta) = \frac{1}{\sqrt{2}}\left[a_{\mathbf{k}}^-v^{\ast}_{\mathbf{k}}(\eta) + a^+_{-k}v_{\mathbf{k}}(\eta)\right].
\eeq
Of course due to $\chi$ being real we have a relation: $a_{k}^+ = (a_k^-)^{\ast}$.
So far it looks like in Minkowski case, despite of the dependence of $v(\eta)$ on $\eta$, which will cause lots of problems while quantising the theory. So here is the first trouble. How to define vacuum? In Minkowski vacuum was the state which was $SO(1,3)$ invariant. In case of curved spacetime there is no state which is invariant under change of reference frame, but nevertheless we will quantise this theory and we will addresss this problem in next paragraph \ref{Vacuumstate}. Let us introduce commutation relation, namely:
\beq
\label{commutation}
[a,b] = i \{a,b\}
\eeq
where ${a,b}$ is Poisson bracket. 
So we obtain such equal time commutation relations:
$$
[\hat{\chi}(\mathbf{x},\eta), \hat{\pi}(\mathbf{y},\eta)] = i \delta(\mathbf{x - y}),\textbf{     } [\hat{\chi}(\mathbf{x},\eta), \hat{\chi}(\mathbf{y},\eta)] = 0, 
\textbf{     }[\hat{\pi}(\mathbf{x},\eta), \hat{\pi}(\mathbf{y},\eta)] = 0.
$$
We define field operator as:
\beq
\label{Fieldmodes}
\hat{\chi}(\mathbf{x},\eta) = \intkpi \frac{1}{\sqrt{2}} \left(\eikx v_\mathbf{k}^{\ast}(\eta)\ami +\emikx v_{\mathbf{k}}(\eta)\apl\right),
\eeq
where $v_k$ obeys the relation:
$$
v_{\mathbf{k}}'' + \omega_{\mathbf{k}}^2(\eta) v_{\mathbf{k}}= 0
$$
So we have standard comutation relations:
\beq
[\ami,\aplp] = \delta(\mathbf{\mathbf{k}-\mathbf{k}'}), \textrm{    } [\ami,\amip] = 0, \textrm{    } [\apl,\aplp] = 0.
\eeq
So far so good. Let us for a moment switch to physical time $t$. We would like to introduce vacuum as state annihilated by $\ami$, but here arise a problem. For which $t$ should we take $\ami$. We will start with definition of vacuum for a given Cauchy surface as:
\beq
\ami(\eta) |0\rangle_{t} = 0.
\eeq
Let us now construct a unitary transformation between time $t$ and $t'$ so we would like to compare vacuums for both times. This unitary transformation is so called Bogoliubov transformation and its a intertwining map between Fock space in time $t$ and $t'$. Let us change notation a little bit: $a_k^{\pm}(t)= A_{\mathbf{k},t}^{\pm}$, for a given $\mathbf{k}$. In general we have:
\beq
\left( \begin{array}{c}
 A_{\mathbf{k},t}^{+} \\
A_{\mathbf{k},t}^{-}  \\
\end{array} \right) =
\left( \begin{array}{cc}
p & q  \\
r & s \\
\end{array} \right) \left( \begin{array}{c}
 A_{\mathbf{k},t'}^{+} \\
A_{\mathbf{k},t'}^{-}  \\
\end{array} \right) 
\eeq
the transformations have to satisfy 3 conditions (which are conditions on coefficients of transformation):
\begin{itemize}
\item reality: $$A_{k,t}^{-}=(A_{\mathbf{k},t}^{+})^{\ast}= \bar{q}A_{\mathbf{k},t'}^{+} + \bar{p} A_{\mathbf{k},t'}^{-}\Rightarrow  \bar{q} =r \textrm{ and } \bar{p}=s$$
\item preservation of commutator: $$1=[\hat{A}_{\mathbf{k},t}^{-},\hat{A}_{\mathbf{k},t}^{+}] = (|p|^2-|q|^2)$$ 
\item Anihilation of vacuum:
$$
0=\hat{A}_{\mathbf{k},t}^{-}|0\rangle_t = \left(\bar{q}\hat{A}_{\mathbf{k},t'}^{+} + \bar{p} \hat{A}_{\mathbf{k},t'}^{-}\right)|0\rangle_t,
$$
\end{itemize}
as we can see that in general vacuum state for $t$ is not a vacuum state for $t'$, since any of the restrictions gives the condition that $\hat{A}_{\mathbf{k},t}^{-}$ annihilates vacuum for any time. Let us introduce number of particles operator for a given momentum $\mathbf{k}$ operator in time $t'$ as:
\beq
\hat{N}_{k,t'} = \hat{A}_{\mathbf{k},t'}^{+} \hat{A}_{\mathbf{k},t'}^{-}.
\eeq
One can calculate expectation value on the vacuum state at time $t$ \cite{Wojtek}:
$$
\langle 0|_t\hat{A}_{k,t'}^{+} \hat{A}_{\mathbf{k},t'}^{-} \vac = \cav (\bar{p}\hat{A}_{\mathbf{k},t}^{+} -q\hat{A}_{\mathbf{k},t}^{-})(-\bar{q}\hat{A}_{\mathbf{k},t'}^{+} +p\hat{A}_{k,t}^{-})\vac = |q|^2,
$$
so expectation value of $N_{t'}$ on $\vac$ in general is not zero. This simple calculation presents that in curved-spacetime the intuitive definition of vacuum as a state with average particle number identically zero is completely false. Then the concept of number of particles is not well defined for whole spacetime. Still for any given time we can define vacuum. Maybe we can find some unitary transformation between each vacuum?

\subsection{Vacuum state}
\label{Vacuumstate}
There are many ways to define vacuum. Here we will present one, so called \textbf{instantaneous vacuum}, which is related to a notion of vacuum as a minimal energy state. Let us select a  moment of time $\eta = \eta_0$. Then let us define \textbf{instantaneous vacuum} $|0\rangle_{\eta}$ as state of minimum energy for a hamiltonian $\hat{H}(\eta)$ at the time $\eta_0$. Then we have to minimalize ${}_{(v)}\langle 0 | \hat{H}(\eta_0)\vacuum_{(v)}$ to find $v_{\mathbf{k}}(\eta_0)$.
Using mode expansion we obtain following hamiltonian (\ref{QFTChamiltonian}, \ref{Fieldmodes}) \cite{MukiWin}:
\beq
\hat{H}(\eta) = \frac{1}{4}\int d^3 \mathbf{k}\left[\hat{a}_{\mathbf{k}}^{-}\hat{a}_{\mathbf{-k}}^{-}F^{\ast}_{\mathbf{k}} + \hat{a}_{\mathbf{k}}^{+} \hat{a}_{\mathbf{-k}}^{+}F_{\mathbf{k}} + \left(2\hat{a}_{\mathbf{k}}^{+}\hat{a}_{\mathbf{k}}^{-} + \delta^3(0)\right)E_{\mathbf{k}}\right],
\eeq
where $F_{\mathbf{k}}$ and $E_{\mathbf{k}}$ are defined by:
\begin{align}
E_{\mathbf{k}} &:= |v_{\mathbf{k}}'|^2 + \omega_k^2|v_{\mathbf{k}}|^2, \\
F_{\mathbf{k}} &:= v_{\mathbf{k}}'{}^2 + \omega_k^2v_{\mathbf{k}}^2.
\end{align}
Then the expectation value of Hamiltonian is:
\beq
{}_{(v)}\langle 0 | \hat{H}(\eta_0)\vacuum_{(v)} = \frac{1}{4} \delta^3(0)\varepsilon = \frac{1}{4} \delta^3(0) \int d^3\textbf{k} E_{\mathbf{k}}|_{\eta= \eta_0},
\eeq
where $\delta^3(0)$ comes from the fact that the total volume of space is infinite \cite{MukiWin}. Now we have to determine $v_{\mathbf{k}}(\eta)$ for each mode separetely. After some calculations we obtain that:
\beq
\begin{array}{lcr}
v_{\mathbf{k}}(\eta_0) = \frac{1}{\sqrt{\omega_k(\eta_0)}},& & v_{\mathbf{k}}'(\eta_0) = i \omega_k v_{\mathbf{k}}(\eta_0),
\end{array}
\eeq
 however this approach works only for $\omega_k^2(\eta_0)>0$, otherwise the instantaneous lowest-energy vacuum state cannot be found. Generally speaking the notion of vacuum in curved spacetimes is a problematic issue. For some class of spacetimes number of particles for time $\eta'$ is infinite for $|0\rangle_{\eta}$ vacuum. Then one can introduce so called \textbf{adiabatic vacuum}. However, for \textbf{stationary spacetimes}  the hamiltotian is time independent and we have space of solutions defined by Killing vector transformation:
\beq
T_t\phi(s) = \phi(s-t),
\eeq
and uniqueness of vacuum comes from the positivity of hamiltonian. For compact Cauchy surface we have very convenient and easy to check condition: 
\begin{thm} If $M>0$ in Hamiltonian with $V(\phi) = M\phi^2$, then exists unique stationary Fock vacuum for which creation/anihilation observables are obtained with respect to time evolution hamiltonian. 
\end{thm}
\noindent For non compact Cauchy surface case is much more delicate and is far beyond our investigation \cite{Wojtek, MukiWin}. However, for QFT in curved spacetime the most convenient way to define the theory is using two point functions as basic objects:
$$
\lambda (s,s') = \langle\phi(s),\phi(s')\rangle.
$$
This kind of construction of Hilbert space using two-point functions is called GNS construction. In these approach one can avoid many difficulties associated with canonical quantisation of fields. However, as we have said, we are not going into details.\\
The Inflatons are in general massive particles either by having explicit mass term or by Coleman-Weinberg mechanism \cite{PhysRevD.7.1888}. The inflation scenario is described by classical scalar field / fields which causes the accelerated expansion of the Universe. The quantum perturbations of the scalar field induce the  scalar and tensorial gravitational perturbations. The tensorial modes is associated to gravitational waves and one has to quantise them to obtain the proper picture. Even thoughgh the quantum gravity theory is far from being complete and understand, inflation scenario predicts that there should be such a theory to describe the origin of perturbations, which evolve into galaxies and eventually our planet.\\
\subsection{WKB method}
In section \ref{Slow-roll} we have discussed inflation equations, which turns out to be 1D Klein-Gordon equation in curved spacetime with constraint - Friedmann equation (\ref{IIFriedeq}). For the equations of this kind exists so called WKB approximation \cite{QMbook}. In our case we will expand solution in Planck constant $\hbar$. The equation we are analysing in the Klein-Gordon equation in FLRW background for massive scalar field (\ref{KGeq}).
If we decompose it in by the Fourier transform and rewrite (\ref{KGFourier}) with $\hbar$ and using $\phi$ we obtain:
\beq
-\hbar^2a^{-3}\partial_t \left(a^3 \partial_t \phi_m \right) -\left( a^{-2} \lambda(m) +M^2 \right) \phi_m = 0
\eeq
where $m$ states for are Fourier decomposition coefficients and $\lambda(m)$ are coefficients related to those. We postulate solution up to order $k+1$ in $\hbar$: $\tilde{\phi}_m = e^{i\tilde{S}}$, with $\tilde{S} = \frac{1}{\hbar}S +S_0 + \hbar S_1+ \ldots + h^{n+1}S_{n+1}$. This approximation works best for large values of $m$. Lets calculate the outcome \cite{Wojtek} up to some order $k$: 
\begin{align}
\nonumber
KG \tilde{\phi}_m = \left\{\dot{S}^2 + 2\sum_{k \neq 0} \hbar^{k+1}\dot{S}\dot{S}_k + \sum_k \hbar^{2k+2}\dot{S}^2_k + \sum_{k<j}\hbar^{k+j+2}\dot{S}_k\dot{S}_j \right. \\
\nonumber
 \left.- i(\hbar \ddot{S} + \hbar^2 \ddot{S}_o + \ldots) - (a^{-2}\lambda(m)+M^2)\right\}e^{i\tilde{S}} = \mathcal{O}(\hbar^{k+3}),
\end{align}
and expand in $\hbar$ to calculate $S, S_0$ and $S_i$:
\begin{align}
\hbar^0:& \dot{S}^2 - \left(a^{-2}\lambda(m) +M^2\right) =0 \\
\hbar^1:& 2\dot{S}\dot{S}_0 - 3i\frac{\dot{a}}{a}\dot{S}_0 - i\ddot{S} =0 \\
\hbar^2:& \ldots \nonumber
\end{align}
For $S$ we have two independent solutions:
\beq
S^{\pm} = \pm \int \sqrt{a^{-2}\lambda(m) +M^2},
\eeq
 from which we will denote full solutions as $\tilde{\phi}^{\pm}_m = e^{i\tilde{S}^{\pm}}$. Moreover, one can proof that:
\begin{itemize}
\item $(S_{k}^+)' = - (S_{k}^-)'$ for odd $k$,
\item $(S_{k}^+)' = (S_{k}^-)'$ for even $k$.
\end{itemize}
Now we have to analyse Poisson brackets of solutions:
$$
G(t,t') = -i\hbar\left[\tilde{\phi}^{-}_m(t)\tilde{\phi}^{+}_m(t')- \tilde{\phi}^{+}_m(t)\tilde{\phi}^{-}_m(t')\right]
$$
Then:
\beq
1= 2a^3(t)G(t,t')_{,t,t=t'}
\eeq
From the condition above we can find algebraic relations for $S_{2k}$, in particular for $S_0$:
$$
S_0 = i \frac{1}{2}  \ln(2a^3\dot{S})
$$
In this paragraph we presented the most common way of solving Klein-Gordon equation in curved spacetime. This method is used, for example, to describe tensorial perturbations caused by inflation.


\section{Einstein and Jordan frames}
\label{JordanEinstein}
As we have seen in  section \ref{Starobinskyinflation}, theories of gravity with non-minimally coupled scalar fields and $R^2$ gravity theory are related to each other via conformal transformation. In this chapter we will provide a systematic derivation of conformal mapping which is used in broad parts of text, especially in sections \ref{HiggsInflaton}, \ref{CSMinflation}. Einstein frame is a frame where $R$ is canonically coupled, ie $f_E(R)= \frac{M_P^2}{2} R$. The Jordan frame is the frame where scalar(s) field possess canonical kinetic term(s). For some Lagrangians there are such transformations of fields and coordinates which gives both canonical $R$ coupling and kinetic terms in canonical form. Due to their physical equivalence \cite{EinsteinJordan}, we can perform calculations in any of those frames. So at first, basing on \cite{Multiconformal}, we will analyse one field coupled to gravity. Then we will show that generally there is no transformation which gives Lagrangian in Einstein frame and standard kinetic terms for multiple fields and show what conditions needs to be imposed to assure existence of such transformation. 
\subsection{One field analysis}
Let us now take a general scalar field action coupled non-minimally to gravity:
\beq
S = \int d^Dx\sqrt{|g|}\left[-f(\phi)R + \frac{\partial_{\mu}\phi\partial^{\mu}\phi}{2} - U(\phi)\right]
\eeq
Firstly we will investigate the $f(\phi)R$ part. After a conformal mapping we obtain:
\begin{align}
\int d^D x \sqrt{|g|}f(\phi)R = \int d^Dx \frac{\sqrt{|g|}}{\Omega^D} \left[ \Omega^2 \hat{R} +\frac{2(D-1)}{\Omega} \square \Omega\right.\nonumber\\
\left. + \frac{(D-1)(D-4)}{\Omega^2} \gvu \nabla_{\mu}\Omega\nabla_{\nu} \Omega \right].
\end{align}
To go to Einstein frame we use following transformation:
\beq
\label{Omegaf}
\Omega^{D-2}(x) = 2f(\phi).
\eeq
After using the fact that $\partial_{\mu} = \hat{\partial}_{\mu} = \nabla_{\mu}$ we finally obtain \cite{Multiconformal}:
\beq
\int d^D x \sqrt{|g|}f(\phi)R = \int d^Dx \sqrt{|g|} \frac{1}{2}\left[\hat{R} -(D-1)(D-2)\frac{1}{\Omega^2} \hat{g}^{\mu\nu} \hat{\nabla}_{\mu}\Omega\hat{\nabla}_{\nu}\Omega \right].
\eeq
Now we consider scalar field part, where: 
\beq
\label{hatV}
\hat{V} = \frac{V(\phi)}{\Omega^D},
\eeq 
and we have:
\begin{align}
\label{scalarpart}
\int d^Dx \sqrt{|g|} \left[ \frac{1}{2} \gvu \nabla_\mu\phi\nabla_{\nu}\phi - V(\phi)\right] = \nonumber\\ 
\int d^Dx \sqrt{|\hat{g}|}\left[  \frac{1}{2\Omega^{D-2}} \hat{g}^{\mu\nu} \hat{\nabla}_{\mu}\phi\hat{\nabla}_{\nu}\phi - \hat{V}\right].
\end{align}
Then the full action in the transformed frame is the following:
\beq
S = \int d^Dx\sqrt{|g|}\left[ -\frac{1}{2}\hat{R} + \frac{1}{2} \frac{(D-1)}{(D-2)} \frac{1}{f^2}\hat{g}^{\mu\nu} \hat{\nabla}_{\mu}f \hat{\nabla}_{\nu} f + \frac{1}{4f} \hat{g}^{\mu\nu} \hat{\nabla}_{\mu}\phi\hat{\nabla}_{\nu}\phi - \hat{V}\right].
\eeq
In one field case we can find a transformation: $\phi \to \hat{\phi}$ such that:
\beq
\frac{1}{2} \hat{g}^{\mu\nu} \hat{\nabla}_{\mu}\hat{\phi} \hat{\nabla}_{\nu} \hat{\phi} =  \frac{1}{2} \frac{(D-1)}{(D-2)} \frac{1}{f^2}\hat{g}^{\mu\nu} \hat{\nabla}_{\mu}f \hat{\nabla}_{\nu} f + \frac{1}{4f} \hat{g}^{\mu\nu} \hat{\nabla}_{\mu}\phi\hat{\nabla}_{\nu}\phi,
\eeq
moreover we can relate them in a direct way:
\beq
\label{hatphi}
\frac{d\hat{\phi}}{d\phi} = F(\phi) = \sqrt{\frac{1}{2f^2(\phi)}}\sqrt{f(\phi) + \frac{2(D-1)}{D-2}[f'(\phi)]^2},
\eeq
then the action in rescaled variables has both canonical Einstein-Hilbert action and the canonical term for the scalar field. If we will take the potential in the Jordan frame as: \beq
U(\phi) = M^2(f-1)^2,
\eeq
which is the one that guaranties the flat plateau, so the slow roll. Then for the Einstein frame we get (for $D=4$):
\beq
\hat{U} = M^2\frac{(f^2 -1)^2}{4f^2} = M^2 \left(1- \frac{1}{f}\right)^2.
\eeq
Then one can also calculate the number of e-folds, spectral index / tilt or tensor to scalar ratio in a closed form:
\beq
N= \int Hdt \simeq \int \frac{V}{V_{,\phi}}\left(\frac{d\hat{\phi}}{d\phi}\right)^2d\phi = \int \frac{f-1}{2 f f'} \left(f+ \frac{3}{2}f'^2\right)d\phi.
\eeq
 Moreover, if $f'^2 \gg f$, then: 
\beq
\label{numberofe-folds}
 N \simeq 3(f-\log f)/4,
 \eeq
then for particular $f$ one can provide a relation between $\epsilon$, $\eta$ and number of e-folds. 
This is exactly the case for Higgs particle nonminimally coupled to gravity as Inflaton, where: $f(\phi) = 1 + \xi h^2$ and $F(\phi)$ can be calculated, see (\ref{NewScalar}). 
\subsection{Two field analysis}
\label{Twofieldanalysis}
Here we will perform a general study of two scalar fields coupled to gravity, extension to more fields can be found in \cite{Multiconformal}. Let $\phi^1, \phi^2$ be scalar fields coupled to gravity, for $D$ dimensions we have the following action:
\beq
\int d^Dx \sqrt{|g|}\left[-f(\phi^1,\phi^2)R + \frac{1}{2} \delij \gvu \nabla_{\mu}\phi^i\nabla_{\nu}\phi^j - V(\phi^1, \phi^2)\right].
\eeq
Let us know perform conformal transformation: $\guv \to \hat{g}_{\mu\nu}$ with the term $\Omega^2(x)$. The steps of the transformation for the gravitational part are exactly the same and gives, using (\ref{Omegaf}):
\beq
\int d^Dx\sqrt{|g|} f(\phi^i)R = \int d^Dx\sqrt{|\hat{g}|} \left[ \frac{1}{2} \hat{R} - \frac{1}{2} \frac{(D-1)}{(D-2)} \frac{1}{f^2} \hat{g}^{\mu\nu} \hat{\nabla}_{\mu}f\hat{\nabla}_{\nu}f\right], 
\eeq
where: 
\beq
\hat{\nabla}_{\mu}f = \left(\hat{\nabla}_{\mu}\phi^i\right)f_{,i}.
\eeq
The scalar part transforms similarly to (\ref{scalarpart}), using (\ref{hatV}) we find:
\begin{align}
\label{Multifieldtransform}
\intd\left[\frac{1}{2}\delij \gvu \nabla_{\mu}\phi^i\nabla_{\nu}\phi^j - V(\phi^1, \phi^2)\right] = \nonumber \\
\int d^Dx\sqrt{|\hat{g}|} \left[\frac{1}{4f} \delij \hat{g}^{\mu\nu}\hat{\nabla}_{\mu}\phi^i\hat{\nabla}_{\nu}\phi^j - \hat{V}\right].
\end{align}
So finally the action becomes:
\beq
\label{Two-field}
\int d^Dx\sqrt{|\hat{g}|} \left[ -\frac{1}{2} \hat{R} +\frac{1}{2} \frac{(D-1)}{(D-2)} \frac{1}{f^2} \hat{g}^{\mu\nu} \hat{\nabla}_{\mu}f\hat{\nabla}_{\nu}f + \frac{1}{4f} \delij \hat{g}^{\mu\nu}\hat{\nabla}_{\mu}\phi^i\hat{\nabla}_{\nu}\phi^j - \hat{V}\right].
\eeq
There arises a question whether there is such a field transformation that gives canonical kinetic term structure. Let us rewrite \cite{Multiconformal} the action in terms of metric in field space $\mathcal{G}_{ij}$. From now on we will also skip hat notation:
\beq
\intd \left[-\frac{1}{2} R + \frac{1}{2} \mathcal{G}_{ij} \gvu \nabla_{\mu} \phi^i \nabla_{\nu} \phi^j - \hat{V}\right],
\eeq
with
\beq
\mathcal{G}_{ij} = \frac{1}{2f} \delij + \frac{(D-1)}{(D-2)} \frac{1}{f^2} f_{,i}f_{,j}.
\eeq
The necessary condition for the conformal transformation: $\mathcal{G}_{ij} \to \hat{\mathcal{G}}_{\ij} = \delij$ to exist (which gives Minkowski spacetime in fields space) is that all the Riemann tensors coefficients vanish. Let us now focus on $\hat{R} = \hat{\mathcal{G}}^{\ij} \hat{R}_{ikj}^k$. We know that if $\hat{R} \neq 0$, then $R_{jkl}^i \neq 0$. Therefore its sufficient to demonstrate that $\hat{R} \neq 0$ for $\mathcal{G}_{ij}$. Let us first rescale $\mathcal{G}_{\ij}$: 
\beq
\mathcal{G}_{\ij} \to \tilde{\mathcal{G}}_{\ij} = 2f \mathcal{G}_{\ij} = \delij + \frac{2(D-1)}{D-2} \frac{1}{f}f_{,i}f_{,j},
\eeq
then $\tilde{R}$ takes the following form:
\beq
\label{overlineR}
\tilde{R} = \frac{2(D-1)(D-2)}{L(\phi)} [2ff_{11}f_{22} - f^2_{,1}f_{,22} - f^2_{,2}f_{,11} - 2f_{,12}(ff_{,12} - f_{,1}f_{,2})],
\eeq
where:
\beq
\label{Lphi}
L(\phi^i) = \left[ (D-2)f + 2(D-1)\sum_i f_{,i}^2\right]^2.
\eeq
Moreover, we know \cite{Multiconformal} that for two fields (in two dimensional) case: $\tilde{R}^i_{jkl} \varpropto \tilde{R}$. Since then, if and only if Riemann tensor vanishes, one can find such a transformation. Below we will show that for most typical $f$ it is impossible. If we denote $\phi^1 = \phi$ and $\phi^2 = \chi$ and take $f(\phi, \chi)$ as:
\beq
f(\phi, \chi) = \frac{1}{2}\left[M^{D-2} + \xi_{\phi}\phi^2 + \xi_{\chi}\chi^2\right],
\eeq 
then according to (\ref{overlineR}, \ref{Lphi}) we obtain:
\beq
\label{overR}
L(\phi,\chi)\tilde{R} = 2(D-1)(D-2)\xi_{\phi}\xi_{\chi}M^{D-2},
\eeq
therefore we see that, for $D>2$ only if $M=0$ one can find such a conformal transformation that would bring both the gravitational and kinetic terms into canonical form. This implies a further condition on $\xi_i$, namely:
\beq
 \frac{\xi_{\phi}\phi^2 + \xi_{\chi}\chi^2}{M^{D-2}} \ggg 1,
\eeq
to obtain the canonical Einstein and kinetic term.
\section*{Bibliography, list of figures and tables}
\nocite{*} 
\addcontentsline{toc}{section}{The Bibliography}
\bibliography{mybibfile.bib}{}
\bibliographystyle{siam}
\listoffigures
\addcontentsline{toc}{section}{List of Figures}
\listoftables
\addcontentsline{toc}{section}{List of Tables}
\end{document}